\documentclass[12pt,a4paper]{article}
\usepackage{graphicx}
\usepackage{subeqnarray}
\begin{document}
%
%                         _________________ 
%                         |               |   
%                         |               |
%                         |  VERSIONE  D  |
%                         |               |
%                         _________________
%
%                      
%
\def\astrobj#1{#1}
\newenvironment{lefteqnarray}{\arraycolsep=0pt\begin{eqnarray}}
{\end{eqnarray}\protect\aftergroup\ignorespaces}
\newenvironment{lefteqnarray*}{\arraycolsep=0pt\begin{eqnarray*}}
{\end{eqnarray*}\protect\aftergroup\ignorespaces}
\newenvironment{leftsubeqnarray}{\arraycolsep=0pt\begin{subeqnarray}}
{\end{subeqnarray}\protect\aftergroup\ignorespaces}
\newcommand{\diff}{{\rm\,d}}
\newcommand{\pprime}{{\prime\prime}}
\newcommand{\ppprime}{{\prime\prime\prime}}
\newcommand{\szeta}{\mskip 3mu /\mskip-10mu \zeta}
\newcommand{\FC}{\mskip 0mu {\rm F}\mskip-10mu{\rm C}}
\newcommand{\appleq}{\stackrel{<}{\sim}}
\newcommand{\appgeq}{\stackrel{>}{\sim}}
\newcommand{\Int}{\mathop{\rm Int}\nolimits}
\newcommand{\Nint}{\mathop{\rm Nint}\nolimits}
\newcommand{\sgn}{\mathop{\rm sgn}\nolimits}
\newcommand{\range}{{\rm -}}
%newcommand{\erf}{\mathop{\rm erf}\nolimits}
%\newcommand{\psfc}{\mathop{\rm psfc}\nolimits}
%\newcommand{\Psf}{\mathop{\rm psf}\nolimits}
\newcommand{\displayfrac}[2]{\frac{\displaystyle #1}{\displaystyle #2}}
\newcommand{\mmatrix}[1]{\left|\left|\matrix{#1}\right|\right|}
\def\astrobj#1{#1}
%\begin{titlepage}
%\setcounter{page}{0}
%\headnote{Astron.~Nachr.~000 (2001) 0, 000--000}
%\makeheadline
%
\title{Bivariate least squares linear regression: \\
towards a unified analytic formalism. \\
I. Functional models}
\author{{R.~Caimmi}\footnote{
{\it Astronomy Department, Padua Univ., Vicolo Osservatorio 3/2,
I-35122 Padova, Italy}
email: roberto.caimmi@unipd.it~~~
fax: 39-049-8278212}
%, {E.~Milanese}\footnote{
%{\it Astronomy Department, Padua Univ., Vicolo Osservatorio 3/2,
%I-35122 Padova, Italy}
%email: elena.milanese@studenti.unipd.it~~~
%fax: 39-049-8278212}
\phantom{agga}}
%
%\medskip
%\small{Dipartimento di Astronomia}}
%
%\date{Received..................................................
%Accepted..................................................}
\maketitle
\begin{quotation}
\section*{}
\begin{Large}
\begin{center}
%\summary

Abstract

\end{center}
\end{Large}
\begin{small}

\noindent\noindent
Concerning bivariate least squares linear regression, the
classical approach pursued for functional
models in earlier attempts (York, 1966; 1969)
is reviewed using a new formalism in terms
of deviation (matrix) traces
which, for homoscedastic data, reduce to usual
quantities leaving aside an
unessential (but dimensional) multiplicative factor.
Within the framework of classical error models,
the dependent variable relates to the independent
variable according to the usual additive model.
The classes of linear models considered are
regression lines
%
%slope and intercept
%estimators, and related variance estimators,
%are expressed
%
in the general case of
correlated errors in $X$ and in $Y$ for
heteroscedastic data, and in the opposite limiting
situations of (i) uncorrelated errors in
$X$ and in $Y$, and (ii) completely
correlated errors in $X$ and in $Y$.
The special case of (C) generalized
orthogonal regression
%
%errors in $X$
%equally weighted and equally correlated
%with respect to errors in $Y$ i.e.
%constant variance ratio and correlation
%coefficient for all data points,
%
is considered in detail together with
well known subcases, namely: (Y) errors
in $X$ negligible (ideally null) with
respect to errors in $Y$; (X) errors
in $Y$ negligible (ideally null) with
respect to errors in $X$; (O) genuine orthogonal
regression; (R) reduced major-axis regression.
In the limit of homoscedastic data, the
results determined for functional models
are compared with their counterparts
related to extreme structural models i.e. the
instrumental scatter is negligible
(ideally null) with respect to the
intrinsic scatter (Isobe et al., 1990;
Feigelson and Babu, 1992).   While regression
line slope and intercept estimators
for functional and structural models
necessarily coincide, the contrary
holds for related variance estimators
even if the residuals obey a Gaussian
distribution, with the exception of Y
models.   An example of astronomical
application is considered, concerning
the [O/H]-[Fe/H] empirical relations
deduced from five samples related to
different stars and/or different methods
of oxygen abundance determination.
For selected samples and assigned
methods, different regression models
yield consistent results within the
errors $(\mp\sigma)$ for both
heteroscedastic and homoscedastic data.
Conversely, samples related to different
methods produce discrepant results,
due to the presence of (still undetected)
systematic errors, which implies no
definitive statement can be made at
present.   A comparison is also made
between different expressions of
regression line slope and intercept
variance estimators, where fractional
discrepancies are found to be not
exceeding a few percent, which grows
up to about 20\% in presence of large
dispersion data.   An extension of the
formalism to structural models
is left to a forthcoming paper.

\noindent
{\it keywords - 
galaxies: evolution - stars: formation; evolution - methods: data analysis -
methods: statistical.}

pacs codes: 98.62.-g; 97.10.Cv; 02.50.-r
%END
%\end{titlepage}
\end{small}
\end{quotation}

\section{Introduction} \label{intro}

\noindent\noindent

Linear regression is a fundamental and frequently
used statistical tool in almost all branches of
science, such as astronomy, biology, chemistry,
geology, physics, and statistics of course; for
a full discussion refer to a classical paper
(Isobe et al., 1990; hereafter quoted as Ial90).
In spite of its apparent simplicity, the task
of drawing the ``best'' straight line through
data on a Cartesian plot is difficult and
controversial.   The problem is twofold:
regression line slope and intercept estimators
are expressed involving minimizing or maximizing
some function of the data;
on the other hand, regression line slope and
intercept variance estimators are expressed
requiring knowledge of the error distributions
of the data.

The complexity mainly arises from the
occurrence of intrinsic dispersion,
which could be related to a non
Gaussian distribution, in
addition to the dispersion related to
the measurement processes (hereafter
quoted as instrumental dispersion),
which necessarily implies a Gaussian
distribution.   An increasing difficulty
is encountered in more exotic situations,
such as truncated regression, where a
variable is assumed to be truncated
below or above a threshold, and censored
regression, where several data are assumed
to be undetected at various sensitivity
levels.   For further details of astronomical
interest, refer to a classical paper
(Feigelson and Babu, 1992; erratum, 2011;
hereafter quoted together as FB92) and, in
general, to specific texts on the subject
(e.g., Klein and Moeschberger, 2005).

In statistics, problems where the true
points lie precisely on an expected
line are called functional regression
models, while problems where the true
points are (intrinsically) scattered
about an expected line are called
structural regression models.
Accordingly, functional regression
models may be conceived as structural
regression models where the intrinsic
dispersion is negligible (ideally null)
with respect to the instrumental
dispersion.   A distinction between
functional and structural  modelling
is currently preferred, where the former
can be affected by intrinsic scatter
but with no or only minimal assumptions
on related distributions, while the
latter implies (usually parametric)
models are placed on the above mentioned
distributions.   For further details
refer to specific textbooks (e.g.,
Carroll et al., 2006, Chap.\,2, \S 2.1).
In addition, models
where the instrumental dispersion
is the same from point to point for
each variable, are called homoscedastic
models, while models where the
instrumental dispersion is (in general)
different from point to point, are called
heteroscedastic models.   Similarly,
related data are denoted as homoscedastic
and heteroscedastic, respectively.

Bivariate least squares linear regression
related to heteroscedastic
functional models with uncorrelated and
correlated errors, following
Gaussian distributions, were
analysed and formulated in two classical
papers (York, 1966; 1969; hereafter quoted
as Y66 and Y69, respectively).   Bivariate
least squares linear regression
related to extreme structural
models, where the instrumental dispersion
is negligible (ideally null) with respect
to intrinsic dispersion, was exhaustively
treated in a classical paper (Ial90).

An extension to homoscedastic functional
and structural models was performed in
a subsequent paper (FB92), yielding the
same expression of regression line slope
and intercept estimators,
%and relatedvariance estimators,
provided the
instrumental dispersion in the former
case coincides with the intrinsic
dispersion (assumed to be dominant)
in the latter case, for
each variable, and the residuals
follow a Gaussian distribution.

Further extension to homoscedastic
structural models where instrumental
and intrinsic dispersion are of the
same order, was carried in a later
paper (Akritas and Bershday, 1996;
hereafter quoted as AB96).
Heteroscedastic structural models
with instrumental dispersion
negligible with respect to intrinsic
dispersion in one variable, were also
presented (AB96).

The above mentioned
papers provide the simplest description
of linear regression.   More sophisticated
attempts imply additional effects such
as truncated and censored regression
(e.g., FB92), analytical methods such
as correction of the observed moments
of the data (e.g., Fuller, 1987; AB96;
Freedman et al., 2004), minimization of
an effective $\chi^2$ statistic (e.g.,
Clutton-Brock, 1967; Barker and Diana,
1974; Press et al., 1992; Tremaine et
al., 2002), assuming a probability
distribution for the true independent
variable values (e.g., Schafer, 1987,
2001; Roy and Banerjee, 2006),
computational methods such as bootstrap
and jackknife (e.g., FB92), matrix
formalism (e.g., Schwarzemberg-Czerny,
1995; Branham, 2001), simultaneous
adjustement (e.g., Pourbaix, 1998), and
Bayesan approach (e.g., Zellner, 1971;
Gull, 1989; Dellaportas and Stephens,
1995; Carroll et al., 1999; Scheines et
al., 1999; Kelly, 2007).

The last
investigation is particularly relevant
in that it is the first example, in
the astronomical literature, where
linear regression is considered
following the modern (since about
half a century ago) approach based
on likelihoods rather than the old
(up to about a century ago) least-squares
approach.   More specifically, a
hierarchical measurement error
model is set up therein, the
complicated likelihood is written
down, and a variety of minimum
least-squares and Bayesan solutions
are shown, which can treat functional,
structural, multivariate, truncated
and censored mesaurement error
regression problems.

Even in dealing with the simplest
homoscedastic (or heteroscedastic)
functional and structural models,
still no unified analytic formalism has been
developed (to the knowledge of the
author) where (i) structural
heteroscedastic models with
instrumental and intrinsic
dispersion of comparable order
in both variables, are considered;
(ii) previous results are recovered
in the limit of dominant instrumental
dispersion; and (iii) previous results
are recovered in the limit of dominant
intrinsic dispersion.   A related
formulation may be useful also for
computational methods, in the sense
that both the general case and
limiting situations can be described
by a single numerical code.

The current paper aims at making a
first step towards a unified analytic formalism
of bivariate least squares linear regression
involving functional models.
More specifically, earlier attempts
shall be reviewed and reformulated
by definition and use of deviation
(matrix) traces, within the framework
of classical error models where
the dependent variable relates to the independent
variable according to the usual additive model.

Homoscedastic
and heteroscedastic functional models
are presented in section \ref{fumo},
basing on two classical papers (Y66;
Y69).
%Structural models where
%instrumental dispersion is negligible
%with respect to intrinsic dispersion
%in both variables, are outlined in
%section \ref{stmo} grounding on a
%classical paper (Ial90).   Structural
%models with instrumental and intrinsic
%dispersion of comparable order are
%described in section \ref{stst} on
%the basis of a classical paper (AB96).
An example of astronomical application
is outlined in section \ref{apfm}.
The discussion is performed in section
\ref{disc}.   Finally, the conclusion
is shown in section \ref{conc}.
Some points are developed with more
detail in the Appendix.   An extension
of the formalism to structural models is
left to a forthcoming paper.

\section{Least-squares fitting of a straight line} \label{fumo}

\subsection{General considerations} \label{geco}

\noindent\noindent

Attention shall be restricted to the classical
problem of least-squares fitting of a straight
line, where both variables are measured with
errors and the true points lie on the unknown
regression line i.e. functional models (e.g.,
Y66; Y69), which can be considered as structural
models in the limit of negligible (ideally null)
intrinsic scatter.   In general, the dependent
variable, $y$, relates to the independent variable,
$x$, according to the usual additive model (e.g.,
AB96; Carroll et al., 2006, Chap.\,1, \S 1.2,
Chap.\,3, \S 3.2.1; Kelly, 2007; Buonaccorsi, 2010,
Chap.\,4, \S 4.3):
\begin{equation}
\label{eq:riS}
y_{{\rm S}i}=ax_{{\rm S}i}+b+\epsilon_i~~;\qquad1\le i\le n~~;
\end{equation}
where ${\sf P}_{{\rm S}i}^\ast\equiv(x_{{\rm S}i},
y_{{\rm S}i})$ are the actual points whose coordinates
are affected by no instrumental error and $\epsilon_i$
is a random variable with null expectation value
representing the intrinsic scatter in
$(x_{{\rm S}i},y_{{\rm S}i})$ about the regression line%
\footnote{The Italian convenction shall
be adopted here, according to which the
slope and the intercept of a straight
line on the Cartesian plane, are denoted
as $a$, $b$, respectively.}.

Due to the occurrence of instrumental
errors, the observed points, ${\sf P}_i
\equiv(X_i,Y_i)$, are evaluated in
place of the actual points, ${\sf P}_{{\rm S}i}^\ast$.
The coordinates of observed and actual
points are assumed to be related as:
\begin{leftsubeqnarray}
\slabel{eq:erma}
&& X_i=x_{{\rm S}i}+(\xi_{{\rm F}_x})_i~~;\qquad1\le i\le n~~; \\
\slabel{eq:ermb}
&& Y_i=y_{{\rm S}i}+(\xi_{{\rm F}_y})_i~~;\qquad1\le i\le n~~;
\label{seq:erm}
\end{leftsubeqnarray}
where $(\xi_{{\rm F}_x})_i$, $(\xi_{{\rm F}_y})_i$,
are the instrumental errors on $x_{{\rm S}i}$
and $y_{{\rm S}i}$, respectively, assumed to
be normally distributed with null expectation
values and known variances,
$(\sigma_{xx})_i=[(\sigma_x)_i]^2$,
$(\sigma_{yy})_i=[(\sigma_y)_i]^2$, and
covariance,$(\sigma_{xy})_i=(\sigma_{yx})_i$.   The terms
``independent variable'' and ``dependent variable''
are purely conventional when the model is
symmetric in $x$ and in $y$ provided $a\ne0$.
For a vanishing intrinsic scatter, $\epsilon_i\to0$,
$1\le i\le n$, actual points lie on the unknown
regression line whereas adjusted points, $\hat{\sf P}_i
\equiv(x_i,y_i)$, lie on the estimated
regression line:
\begin{equation}
\label{eq:esl}
y_i=\hat{a}x_i+\hat{b}~~;\qquad1\le i\le n~~;
\end{equation}
where, in general, estimators are denoted
by hats.   For further details refer to
earlier attempts (York, 1967; Y69).

In the case under discussion, the regression
estimator minimizes the sum (over the $n$
observations) of squared residuals (e.g.,
Y69), or statistical distances of the
observed points, ${\sf P}_i\equiv(X_i,Y_i)$,
from the estimated line in the unknown
parameters, $a, b, x_1, ..., x_n$ (e.g.,
Fuller, 1987, Chap.\,1, \S 1.3.3).   Under
restrictive assumptions, the regression
estimator is the functional maximum
likelihood estimator (e.g., Carroll et al.,
2006, Chap.\,3, \S 3.4.2).

To the knowledge of the author, only
classical error models are considered for
astronomical applications, and for this
reason different error models such as
Berkson models and mixture error models
(e.g., Carroll et al., 2006, Chap.\,3, Sect.\,3.2)
shall not be dealt with in the current
attempt.   From this point on, investigation
shall be limited to functional models and
least-squares regression estimators for the
following reasons.   First, they are important
models in their own right, furnishing an
approximation to real world situations.
Second, a careful examination of these
simple models helps for understanding
the theoretical underpinnings of methods
for other models of greater complexity
such as hierarchical models (e.g., Kelly, 2007).

\subsection{Functional models} \label{funm}

\noindent\noindent

With regard to functional models, bivariate
least squares linear regression were analysed in two
classical papers (Y66; Y69).   The same
line of thought shall be followed here
and the sole changes shall be concerned
with the formalism, as clearly indicated.
The general case shall first be presented,
while special cases shall be deduced later
as limiting situations.

In the light of the model outlined in
subsection \ref{geco} in absence of
intrinsic scatter, Eqs.\,(\ref{seq:erm})
and (\ref{eq:esl}), the actual points,
${\sf P}_{{\rm S}i}^\ast\equiv(x_{{\rm S}i},
y_{{\rm S}i})$ coincide with the true points,
${\sf P}_i^\ast\equiv(x_i^\ast, y_i^\ast)$,
whose coordinates lie on the unknown regression line:
%
%Let the variables of interest be denoted
%as $(x_i, y_i)$ and related measurements
%(or observed data) as $(X_i, Y_i)$, $1\le
%i\le n$, and let the variance
%and the covariance of distributions
%depending on $X_i, Y_i,$ be denoted
%as $(\sigma_{xx})_i=[(\sigma_x)_i]^2$;
%$(\sigma_{yy})_i=[(\sigma_y)_i]^2$;
%$(\sigma_{xy})_i$; respectively.
%the actual points, $\hat{\sf P}_i\equiv
%(x_i, y_i)$, $1\le i\le n$, lie on the
%regression line:
%
\begin{equation}
\label{eq:rlf}
y_i^\ast=ax_i^\ast+b~~;\qquad1\le i\le n~~;
\end{equation}
while the observed points, ${\sf P}_i
\equiv(X_i, Y_i)$, are
scattered with respect to the regression
line.
%The parameters, $a$ and $b$,
%are still to be determined and for this
%reason are not indicated as estimators
%as done in Eq.\,(\ref{eq:esl}).

The coordinates, $(x_i, y_i)$, may be
conceived as the adjusted values of
related observations, $(X_i, Y_i)$,
on the calculated regression line
(Y66; Y69), Eq.\,(\ref{eq:esl}) and,
in addition, as estimators of the
coordinates, $(x_i^\ast, y_i^\ast)$,
on the true regression line determined
in absence of mesaurement errors,
Eq.\,(\ref{eq:rlf}).
The line of adjustment, $\overline
{{\sf P}_i\hat{{\sf P}}_i}$ (e.g.,
Y69), may be conceived as an estimator
of the statistical distance, $\overline
{{\sf P}_i{\sf P}_i^\ast}$ (e.g.,
Fuller, 1987, Chap.\,1, \S 1.3.3),
where $\hat{\sf P}_i(x_i, y_i)$ is the
adjusted point on the estimated
regression line and
${\sf P}_i^\ast(x_i^\ast,
y_i^\ast)$ is the true point on the
true regression line.

The squared weighted residuals are defined
as (Y69):
\begin{leftsubeqnarray}
\slabel{eq:R2ga}
&& (\widetilde{R}_i)^2=\frac{w_{x_i}(X_i-x_i)^2+w_{y_i}(Y_i-y_i)^2-2r_i
\sqrt{w_{x_i}w_{y_i}}(X_i-x_i)(Y_i-y_i)}{1-r_i^2}~~;\qquad \\
\slabel{eq:R2gb}
&& r_i=\frac{(\sigma_{xy})_i}{[(\sigma_{xx})_i(\sigma_{yy})_i]^{1/2}}~~;
\qquad\vert r_i\vert\le1~~;\qquad1\le i\le n~~;
\label{seq:R2g}
\end{leftsubeqnarray}
where $w_{x_i}$, $w_{y_i}$, are the
weights of the various measurements
(or observations) and $r_i$ the correlation
coefficients.   An equivalent formulation
in matrix formalism can be found in specific
textbooks, where weighted true residuals are
conceived as ``statistical distances'' from
data points to related points on the
regression line [e.g., Fuller, 1987,
Chap.\,1, \S 1.3.3, Eq.\,(1.3.16)].

In the limit of uncorrelated errors,
%$r_i=0$, $1\le i\le n$,
Eq.\,(\ref{seq:R2g})
reduces to (Y66):
\begin{leftsubeqnarray}
\slabel{eq:R20a}
&& (\widetilde{R}_i)^2=w_{x_i}(X_i-x_i)^2+w_{y_i}(Y_i-y_i)^2~~; \\
\slabel{eq:R20b}
&& r_i=0~~;\qquad1\le i\le n~~;
\label{seq:R20}
\end{leftsubeqnarray}
where the covariances are necessarily
null, $(\sigma_{xy})_i=0$, $1\le i\le n$.

In the limit of perfectly correlated
errors, $r_i=\sgn(r_i)$, $1\le i\le n$,
it can be seen that the following relation
holds:
\begin{equation}
\label{eq:r1}
\frac{Y_i-y_i}{X_i-x_i}=\left(\frac{w_{x_i}}{w_{y_i}}\right)^{1/2}\sgn(r_i)~~;
\end{equation}
where $\sgn$ is the sign function%
\footnote{The sign function is defined
as $\sgn(x)=\vert x\vert/x$, $x\ne0$;
$\sgn(x)=0$, $x=0$.}.
Accordingly, Eq.\,(\ref{eq:R2ga}) reduces to:
\begin{lefteqnarray*}
&& (\widetilde{R}_i)^2=\lim_{r_i\to\sgn(r_i)}\left[\frac{w_{x_i}(X_i-x_i)^2+
w_{x_i}(X_i-x_i)^2\sgn^2(r_i)}{1-r_i^2}\right. \\
&& \phantom{(\widetilde{R}_i)^2=\lim_{r_i\to\sgn(r_i)}~}\left.
-\frac{2r_i\sgn(r_i)w_{x_i}(X_i-x_i)^2}{1-r_i^2}\right] \\
&& \phantom{(\widetilde{R}_i)^2}=w_{x_i}(X_i-x_i)^2\lim_{r_i\to\sgn(r_i)}
\frac{2-2r_i\sgn(r_i)}{1-r_i^2\sgn^2(r_i)} \\
&& \phantom{(\widetilde{R}_i)^2}=w_{x_i}(X_i-x_i)^2\lim_{r_i\to\sgn(r_i)}
\frac2{1+r_i\sgn(r_i)}~~;
\end{lefteqnarray*}
which, owing to Eq.\,(\ref{eq:r1}),
takes the form:
\begin{leftsubeqnarray}
\slabel{eq:R21a}
&& (\widetilde{R}_i)^2=w_{x_i}(X_i-x_i)^2=w_{y_i}(Y_i-y_i)^2~~; \\
\slabel{eq:R21b}
&& r_i=\sgn(r_i)~~;\qquad1\le i\le n~~;
\label{seq:R21}
\end{leftsubeqnarray}
where the covariances are necessarily
equal to the (positive or negative) 
square root of the variance product,
$(\sigma_{xy})_i=[(\sigma_{xx})_i
(\sigma_{yy})_i]^{1/2}$, $1\le i\le n$.

Turning back to the general case, let
the squared residual matrix be defined as:
\begin{eqnarray}
\label{eq:MR}
&& {\rm M}_{\widetilde{R}}=\mmatrix{\widetilde{R}_i\widetilde{R}_j \cr} ~~;
\qquad1\le i\le n~~;\qquad1\le j\le n~~;
\end{eqnarray}
which is a square matrix of order, $n$.
Let the related trace:
\begin{equation}
\label{eq:TR}
T_{\widetilde{R}}=\sum_{i=1}^n(\widetilde{R}_i)^2~~;
\end{equation}
be defined as the squared residual trace.
The regression estimator is that minimizing
the squared residual trace (Y66; Y69):
\begin{lefteqnarray}
\label{eq:ssR}
&& T_{\widetilde{R}}(x_1, x_2, ..., x_n, y_1, y_2, ..., y_n)=\sum_{i=1}^n
(\widetilde{R}_i)^2\nonumber \\
&& =\sum_{i=1}^n\frac{w_{x_i}(X_i-x_i)^2+w_{y_i}(Y_i-y_i)^2-2r_i
\sqrt{w_{x_i}w_{y_i}}(X_i-x_i)(Y_i-y_i)}{1-r_i^2}~~;
\end{lefteqnarray}
with the constraint expressed by
Eq.\,(\ref{eq:esl}) where the coefficients,
$\hat{a}$, $\hat{b}$, are still to be
determined, and for this reason are denoted
as $a$, $b$, respectively.   If the values,
$(x_1, x_2, ..., x_n, y_1, y_2,$
$..., y_n, a, b)$,
relate to a constrained extremum
point, then the following relations
must necessarily hold (Y66; Y69):
\begin{lefteqnarray}
\label{eq:dF0}
&& \delta T_{\widetilde{R}}=-2\sum_{i=1}^n\left[\frac{w_{x_i}(X_i-x_i)
\delta x_i+w_{y_i}(Y_i-y_i)\delta y_i}{1-r_i^2}\right. \nonumber \\
&& \phantom{\delta F=-2\sum_{i=1}^n~}\left.
-\frac{r_i\sqrt{w_{x_i}w_{y_i}}[(Y_i-y_i)\delta x_i+(X_i-x_i)\delta
y_i]}{1-r_i^2}\right]=0~~; \\
\label{eq:lr0}
&& \delta y_i-a\delta x_i-x_i\delta a-\delta b=0~~;\qquad1\le i\le n~~;
\end{lefteqnarray}
where Eq.\,(\ref{eq:lr0}) is also
valid after inserting on both sides
a multiplier (to be specified later),
$\lambda_i$, $1\le i\le n$.   The sum
of the ensuing relations yields:
\begin{equation}
\label{eq:srl0}
\sum_{i=1}^n\lambda_i\delta y_i-a\sum_{i=1}^n\lambda_i\delta x_i-\sum_{i=1}^n
\lambda_ix_i\delta a-\sum_{i=1}^n\lambda_i\delta b=0~~;
\end{equation}
and the sum of the left-hand sides
of Eqs.\,(\ref{eq:dF0}) and (\ref{eq:srl0})
produces:
\begin{lefteqnarray}
\label{eq:sFr0}
&& \sum_{i=1}^n\left[\frac{w_{x_i}(X_i-x_i)-r_i\sqrt{w_{x_i}w_{y_i}}(Y_i-y_i)}
{1-r_i^2}-a\lambda_i\right]\delta x_i-\sum_{i=1}^n\lambda_i\delta b \nonumber
\\
&& +\sum_{i=1}^n\left[\frac{w_{y_i}(Y_i-y_i)-r_i\sqrt{w_{x_i}w_{y_i}}(X_i-x_i)}
{1-r_i^2}+\lambda_i\right]\delta y_i-\sum_{i=1}^n\lambda_ix_i\delta a=0~~;
\qquad
\end{lefteqnarray}
which implies each coefficient is
null, as:
\begin{leftsubeqnarray}
\slabel{eq:cF0a}
&& \frac{w_{x_i}}{1-r_i^2}(X_i-x_i)-\frac{r_i\sqrt{w_{x_i}w_{y_i}}}{1-r_i^2}
(Y_i-y_i)-a\lambda_i=0~~; \\
\slabel{eq:cF0b}
&& \frac{w_{y_i}}{1-r_i^2}(Y_i-y_i)-\frac{r_i\sqrt{w_{x_i}w_{y_i}}}{1-r_i^2}
(X_i-x_i)+\lambda_i=0~~;
\label{seq:cF0}
\end{leftsubeqnarray}
\begin{lefteqnarray}
\label{eq:sla}
&& \sum_{i=1}^n\lambda_i=0~~; \\
\label{eq:slx}
&& \sum_{i=1}^n\lambda_ix_i=0~~;
\end{lefteqnarray}
for $1\le i\le n$.

The combination 
of Eqs.\,(\ref{eq:cF0a}) and (\ref{eq:cF0b})
yields:
\begin{leftsubeqnarray}
\slabel{eq:XYa}
&& X_i-x_i=\left(a-\frac{r_i}{\Omega_i}\right)\frac{\lambda_i}{w_{x_i}}~~; \\
\slabel{eq:XYb}
&& Y_i-y_i=\frac{r_i}{\Omega_i}\left(a-\frac1{r_i\Omega_i}\right)\frac
{\lambda_i}{w_{x_i}}~~;
\label{seq:XY}
\end{leftsubeqnarray}
\begin{equation}
\label{eq:Omi}
\Omega_i=\sqrt\frac{w_{y_i}}{w_{x_i}}~~;\qquad1\le i\le n~~;
\end{equation}
and the combination 
of Eqs.\,(\ref{eq:esl}) and (\ref{seq:XY})
produces:
\begin{lefteqnarray}
\label{eq:lai}
&& \lambda_i=W_i(aX_i+b-Y_i)~~;\qquad1\le i\le n~~; \\
\label{eq:Wi}
&& W_i=\frac{w_{x_i}\Omega_i^2}{1+a^2\Omega_i^2-2ar_i\Omega_i}~~;\qquad
1\le i\le n~~;
\end{lefteqnarray}
finally, the substitution of 
Eq.\,(\ref{eq:lai}) into (\ref{eq:sla})
and (\ref{eq:slx}) yields, in the latter
case after some algebra:
\begin{lefteqnarray}
\label{eq:slaW}
&& \sum_{i=1}^nW_i(aX_i+b-Y_i)=0~~; \\
\label{eq:slxW}
&& \sum_{i=1}^nW_iX_i(aX_i+b-Y_i)-a\sum_{i=1}^n\frac{W_i^2}{w_{x_i}}
(aX_i+b-Y_i)^2 \nonumber \\
&& +\sum_{i=1}^n\frac{W_i^2}{w_{x_i}}\frac{r_i}{\Omega_i}
(aX_i+b-Y_i)^2=0~~;
\end{lefteqnarray}
where the regression line slope and intercept
estimators, $\hat{a}$ and $\hat{b}$, are found
solving the system of
%the estimating equations
%(e.g., Buonaccorsi, 2010, Chap.\,6, \S 6.3),
Eqs.\,(\ref{eq:slaW}) and (\ref{eq:slxW}).

In terms of the weighted means:
\begin{leftsubeqnarray}
\slabel{eq:wZa}
&& \widetilde{Z}=\displayfrac{\sum_{i=1}^nW_iZ_i}{\sum_{i=1}^nW_i}~~;\qquad
Z=X,Y~~; \\
\slabel{eq:wZb}
&& \sum_{i=1}^nW_i(Z_i-\widetilde{Z})=0~~;\qquad Z=X,Y~~;
\label{seq:wZ}
\end{leftsubeqnarray}
the intercept is expressed by
casting Eq.\,(\ref{eq:slaW})
under the equivalent form:
\begin{equation}
\label{eq:bar}
b=\widetilde{Y}-a\widetilde{X}~~;
\end{equation}
the point, $\widetilde{\sf P}
\equiv(\widetilde{X},\widetilde{Y})$,
clearly lies on the regression line,
and can be considered as the
``barycentre'' of the data,
${\sf P}_i\equiv(X_i,Y_i)$, $1\le i
\le n$ (Y66; Y69).

Using Eq.\,(\ref{eq:bar}), the
following relation holds:
\begin{equation}
\label{eq:DXY}
aX_i+b-Y_i=a(X_i-\widetilde{X})-(Y_i-\widetilde{Y})~~;\qquad1\le i\le n~~;
\end{equation}
in terms of the deviations from
the weighted mean, $(X_i-\widetilde{X})$
and $(Y_i-\widetilde{Y})$.   Using
Eqs.\,(\ref{eq:slaW}), (\ref{eq:bar})
and (\ref{eq:DXY}), together with the
identities, $Z_i=(Z_i-\widetilde{Z})+
\widetilde{Z}$, $1\le i\le n$, $Z=X, Y$,
the following relations are found after
some algebra:
\begin{lefteqnarray}
\label{eq:SW}
&& \sum_{i=1}^nW_iX_i(aX_i+b-Y_i)=a\sum_{i=1}^nW_i(X_i-\widetilde{X})^2-
\sum_{i=1}^nW_i(X_i-\widetilde{X})(Y_i-\widetilde{Y});\quad \\
\label{eq:SV}
&& \sum_{i=1}^nV_i(aX_i+b-Y_i)^2=a^2\sum_{i=1}^nV_i(X_i-\widetilde{X})^2+
\sum_{i=1}^nV_i(Y_i-\widetilde{Y})^2 \nonumber \\
&& \phantom{\sum_{i=1}^nV_iX_i(aX_i+b-Y_i)^2=}
-2a\sum_{i=1}^nV_i(X_i-\widetilde{X})(Y_i-\widetilde{Y})~~; \\
\label{eq:SU}
&& \sum_{i=1}^nU_i(aX_i+b-Y_i)^2=a^2\sum_{i=1}^nU_i(X_i-\widetilde{X})^2+
\sum_{i=1}^nU_i(Y_i-\widetilde{Y})^2 \nonumber \\
&& \phantom{\sum_{i=1}^nU_iX_i(aX_i+b-Y_i)^2=}
-2a\sum_{i=1}^nU_i(X_i-\widetilde{X})(Y_i-\widetilde{Y})~~; \\
\label{eq:Vi}
&& V_i=\frac{W_i^2}{w_{x_i}}=\frac{w_{x_i}\Omega_i^4}{(1+a^2\Omega_i^2-2ar_i
\Omega_i)^2}~~;\qquad1\le i\le n~~; \\
\label{eq:Ui}
&& U_i=\frac{W_i^2}{w_{x_i}}\frac{r_i}{\Omega_i}=\frac{w_{x_i}r_i\Omega_i^3}
{(1+a^2\Omega_i^2-2ar_i
\Omega_i)^2}~~;\qquad1\le i\le n~~;
\end{lefteqnarray}
and the substitution of
Eqs.\,(\ref{eq:SW}), (\ref{eq:SV}),
(\ref{eq:SU}), into (\ref{eq:slxW})
yields (Y69):
\begin{lefteqnarray}
\label{eq:pce}
&& a^3\sum_{i=1}^nV_i(X_i-\widetilde{X})^2-a^2\left[2
\sum_{i=1}^nV_i(X_i-\widetilde{X})(Y_i-\widetilde{Y})+
\sum_{i=1}^nU_i(X_i-\widetilde{X})^2\right] \nonumber \\
&& -a\left[\sum_{i=1}^nW_i(X_i-\widetilde{X})^2-
\sum_{i=1}^nV_i(Y_i-\widetilde{Y})^2-2
\sum_{i=1}^nU_i(X_i-\widetilde{X})(Y_i-\widetilde{Y})\right] \nonumber \\
&& +\left[\sum_{i=1}^nW_i(X_i-\widetilde{X})(Y_i-\widetilde{Y})-
\sum_{i=1}^nU_i(Y_i-\widetilde{Y})^2\right]=0~~;
\end{lefteqnarray}
where the resulting terms have been
ordered in decreasing powers of the
slope, $a$, and the coefficients,
$W_i$, $V_i$, $U_i$, $1\le i\le n$,
depend in turn on the slope, as
shown by 
Eqs.\,(\ref{eq:Wi}), (\ref{eq:Vi}),
(\ref{eq:Ui}), respectively.   Then
Eq.\,(\ref{eq:pce}) is a pseudo cubic
equation which can be iteratively
solved with the desired degree of
precision, provided 
$W_i$, $V_i$, $U_i$, $1\le i\le n$,
are weakly dependent on $a$.

With the aim of getting a more compact
formalism, let the weighted deviation
matrices be defined as:
\begin{equation}
\label{eq:MQpq}
{\rm M}_{\widetilde{Q}_{pq}}=\mmatrix{\sqrt{Q_iQ_j}(X_i-\widetilde{X})^p
(Y_j-\widetilde{Y})^q \cr}~;~~1\le i\le n~;~~1\le j\le n~;\qquad
\end{equation}
which are square matrices of order, $n$.
Let the related traces:
\begin{equation}
\label{eq:wgc}
\widetilde{Q}_{pq}=\sum_{i=1}^nQ_i(X_i-\widetilde{X})^p(Y_i-\widetilde{Y})^q
~~;\qquad Q=W, V, U, P~~;
\end{equation}
be defined as the weighted deviation traces.
Pure and mixed traces occur for $p=0$ and/or
$q=0$; $p>0$ and $q>0$; respectively.
The special cases, $(p,q)=(2,0),(0,2);
(1,1);$ relate to expressions used in
earlier attempts for both weighted (Y66;
Y69) and unweighted (Ial90; FB92) residuals.
The special case, $(p,q)=(0,0)$, yields the
product, $n\overline{Q}$.   With this notation,
Eq.\,(\ref{eq:pce}) reads:
\begin{equation}
\label{eq:pcc}
\widetilde{V}_{20}a^3-(2\widetilde{V}_{11}+\widetilde{U}_{20})a^2-
(\widetilde{W}_{20}-\widetilde{V}_{02}-2\widetilde{U}_{11})a+
(\widetilde{W}_{11}-\widetilde{U}_{02})=0~~;
\end{equation}
and the solutions can be determined
along the following steps.
\begin{description}
\item[(1)]
Estimate the slope, $a^{(1)}$, of the
regression line, and calculate related  
deviation traces,
$\widetilde{Q}_{pq}^{(1)}$, appearing in
Eq.\,(\ref{eq:pcc}).
\item[(2)]
Solve the corresponding cubic equation, and
select the solution of interest, $a^{(2)}$.
\item[(3)]
Check the inequality, $\vert a^{(i)}/a^{(i-1)}-1
\vert<\upsilon$, where $i=2$ and $\upsilon$ is
the desired degree of precision.
\item[(4)]
If the above inequality is not satisfied,
return to (1) using $a^{(2)}$ instead of
$a^{(1)}$, or in general $a^{(i+1)}$ instead
of $a^{(i)}$.
\item[(5)]
If the above inequality is satisfied for
$i=n$, then $\hat{a}=a^{(n)}$ is the
regression line slope estimator.   As pointed out in
the parent paper (Y66) and confirmed by 
the author, in all cases considered three
real solutions are found, where the one
of interest coincides with the third
appearing in standard formulae.   For
further details refer to Appendix
\ref{a:psec}.
\end{description}
The pseudo cubic, Eq.\,(\ref{eq:pcc}),
may be reduced algebrically to a pseudo
quadratic equation (Y69) which, in terms
of deviation traces,
expressed by Eq.\,(\ref{eq:wgc})
via Eqs.\,(\ref{eq:Vi}) and (\ref{eq:Ui}),
may be cast under the form:
\begin{lefteqnarray}
\label{eq:pqc}
&& (\widetilde{V}_{11}-\widetilde{U}_{20})a^2+
(\widetilde{P}_{20}-\widetilde{V}_{02})a-
(\widetilde{P}_{11}-\widetilde{U}_{02})=0~~; \\
\label{eq:Pi}
&& P_i=\frac{W_i^2}{w_{y_i}}=\frac{W_i^2}{w_{x_i}\Omega_i^2}=
\frac{w_{x_i}\Omega_i^2}{(1+a^2\Omega_i^2-2ar_i
\Omega_i)^2}~~;\qquad1\le i\le n~~;
\end{lefteqnarray}
where no explanation and no quotation
are provided in the parent paper (Y69)
on how Eq.\,(\ref{eq:pqc}) can be
deduced from Eq.\,(\ref{eq:pcc}).
Surely, the solution of interest
must necessarily be chosen among
the three of the pseudo cubic or
the two of the pseudo quadratic.

At this stage, the regression line slope
and intercept estimators, $\hat{a}$, and
$\hat{b}$, can be determined via
Eqs.\,(\ref{eq:pcc}) or (\ref{eq:pqc})
and (\ref{eq:bar}), respectively.
%
%the mathematical problem
%has been completely solved by evaluating
%the slope and the intercept of the
%regression line related to the minimum
%condition expressed by Eqs.\,(\ref{eq:dF0})
%and (\ref{eq:lr0}), and to the considered
%data set.   Different data sets would yield
%different regression lines, which makes the
%statistical problem implying the evaluation
%of the variances of the slope and the intercept,
%to be conceived as random variables.   In
%this view, calculated values of slope,
%intercept, and related variances,
%usually denoted as $\hat{a}$, $\hat{b}$,
%$(\hat{\sigma}_{\hat{a}})^2$,
%$(\hat{\sigma}_{\hat{b}})^2$, are
%estimators of the corresponding true values,
%$a$, $b$, $(\sigma_a)^2$, $(\sigma_b)^2$,
%respectively.   Accordingly,
%Eqs.\,(\ref{eq:bar}), (\ref{eq:pcc}),
%and (\ref{eq:pqc}) should be rewritten with
%$\hat{a}$, $\hat{b}$, in place of $a$, $b$,
%respectively.
%
An exact expression of the regression line
slope and intercept variance estimators may
be calculated using the method of partial
differentiation, along the following steps (Y69).
\begin{description}% 
\item[(1)]
Cast the pseudo quadratic equation,
Eq.\,(\ref{eq:pqc}), into the implicit
form:
\begin{leftsubeqnarray}
\slabel{eq:phia}
&& \phi(X_i,Y_i,\hat{a})=A(X_i,Y_i,\hat{a})(\hat{a})^2+B(X_i,Y_i,\hat{a})
\hat{a}+C(X_i,Y_i,\hat{a})=0~;\qquad \\
\slabel{eq:phib}
&& A(X_i,Y_i,\hat{a})=\widetilde{V}_{11}-\widetilde{U}_{20}~~; \\
\slabel{eq:phic}
&& B(X_i,Y_i,\hat{a})=\widetilde{P}_{20}-\widetilde{V}_{02}~~; \\
\slabel{eq:phid}
&& C(X_i,Y_i,\hat{a})=\widetilde{U}_{02}-\widetilde{P}_{11}~~;
\label{seq:phi}
\end{leftsubeqnarray}
where $Z_i$ stands for $Z_1$, $Z_2$,
..., $Z_n$, and $Z=X,Y$.
\item[(2)]
Write the quadratic error propagation formula
related to weighted and correlated
measurements:
\begin{lefteqnarray}
\label{eq:prph}
&& \left(\frac{\partial\phi}{\partial\hat{a}}\hat{\sigma}_{\hat{a}}^\prime
\right)^2=\sum_{i=1}^n\left[\left(\frac{\partial\phi}{\partial X_i}\right)^2
\frac
1{w_{x_i}}+\left(\frac{\partial\phi}{\partial Y_i}\right)^2\frac1{w_{y_i}}+
\frac{2r_i}{\sqrt{w_{x_i}w_{y_i}}}\frac{\partial\phi}{\partial X_i}\frac
{\partial\phi}{\partial Y_i}\right].\qquad
\end{lefteqnarray}
\item[(3)]
Calculate the explicit expression of the partial
derivatives, $\partial\phi/\partial\hat{a}$,
$\partial\phi/\partial X_i$, $\partial\phi/\partial
Y_i$, $1\le i\le n$, using Eqs.\,(\ref{eq:Vi}),
(\ref{eq:Ui}), (\ref{eq:wgc}), (\ref{eq:Pi}),
and (\ref{seq:phi}).
\item[(4)]
Calculate the regression line slope variance
estimator, $(\hat{\sigma}_{\hat{a}})^2$, using
Eq.\,(\ref{eq:prph}) and multiplying by the
squared residual trace, using
Eq.\,(\ref{eq:ssR}), and dividing by $(n-2)$, as:
\begin{equation}
\label{eq:vara}
(\hat{\sigma}_{\hat{a}})^2=\frac{(\hat{\sigma}_{\hat{a}}^\prime)^2}{n-2}
T_{\widetilde{R}}~~.
\end{equation}
\item[(5)]
Cast the regression line intercept estimator,
expressed by Eq.\,(\ref{eq:bar}), into the
implicit form:
\begin{equation}
\label{eq:stb}
\hat{b}=\psi(X_i,Y_i)=\widetilde{Y}-\hat{a}(X_i,Y_i)\widetilde{X}~~;
\end{equation}
where $Z_i$ stands for $Z_1$, $Z_2$,
..., $Z_n$, and $Z=X,Y$.
\item[(6)]
Write the quadratic error propagation formula
related to weighted and correlated
measurements:
\begin{lefteqnarray}
\label{eq:prps}
&& \left(\hat{\sigma}_{\hat{b}}^\prime
\right)^2=\sum_{i=1}^n\left[\left(\frac{\partial\psi}{\partial X_i}\right)^2
\frac
1{w_{x_i}}+\left(\frac{\partial\psi}{\partial Y_i}\right)^2\frac1{w_{y_i}}+
\frac{2r_i}{\sqrt{w_{x_i}w_{y_i}}}\frac{\partial\psi}{\partial X_i}\frac
{\partial\psi}{\partial Y_i}\right].\qquad
\end{lefteqnarray}
\item[(7)]
Calculate the explicit expression of the partial
derivatives,
$\partial\psi/\partial X_i$, $\partial\psi/\partial
Y_i$, $1\le i\le n$, using Eqs.\,(\ref{seq:wZ}),
%(\ref{eq:lai}),
(\ref{eq:stb}), and the theorem on the derivative
of a function of a function:
\begin{equation}
\label{eq:tdff}
\frac{\partial\phi}{\partial Z_i}=\frac{\partial\phi}{\partial\hat{a}}
\frac{\partial\hat{a}}{\partial Z_i}~~;\qquad Z=X,Y~~;\qquad 1\le i\le n~~.
\end{equation}
\item[(8)]
Calculate the regression line intercept variance
estimator, $(\hat{\sigma}_{\hat{b}})^2$, using
Eq.\,(\ref{eq:prps}) and multiplying by the
squared residual trace, using
Eq.\,(\ref{eq:ssR}), and dividing by $(n-2)$, as:
\begin{equation}
\label{eq:varb}
(\hat{\sigma}_{\hat{b}})^2=\frac{(\hat{\sigma}_{\hat{b}}^\prime)^2}{n-2}
T_{\widetilde{R}}~~.
\end{equation}
\end{description}
The calculation of the partial derivatives,
$\partial\phi/\partial\hat{a}$,
$\partial\phi/\partial X_i$, $\partial\phi/\partial
Y_i$, and
$\partial\psi/\partial X_i$, $\partial\psi/\partial
Y_i$, $1\le i\le n$, is performed in Appendix
\ref{a:pade}.

Reasonable approximate values of the regression
line slope and intercept variance estimators,
are expressed as (Y66; Y69)%
\footnote{ The numerator of the fraction in
Eq.\,(\ref{eq:vabp1}) has been omitted and
put equal to unity in the parent paper (Y69)
due to a printing error, as it can be argued
by considering the physical dimensions on
both sides, or by comparison with its
counterpart in an earlier attempt (Y66).}:
\begin{lefteqnarray}
\label{eq:vaap1}
&& \left(\hat{\sigma}_{\hat{a}}\right)^2=\frac1{n-2}\displayfrac{T_
{\widetilde{R}}}{\sum_{i=1}^nW_i(X_i-\widetilde{X})^2}~~; \\
\label{eq:vabp1}
&& \left(\hat{\sigma}_{\hat{b}}\right)^2=\left(\hat{\sigma}_{\hat{a}}\right)
^2\displayfrac{\sum_{i=1}^nW_i(X_i)^2}{\sum_{i=1}^nW_i}~~;
\end{lefteqnarray}
which, using Eqs.\,(\ref{seq:wZ}), (\ref{eq:wgc}),
and determining the explicit expression of the
squared residual trace, may be
cast under the equivalent form:
\begin{lefteqnarray}
\label{eq:vaap2}
&& \left(\hat{\sigma}_{\hat{a}}\right)^2=\frac1{n-2}\left[(\hat{a})^2+
\frac{\widetilde{W}_{02}}{\widetilde{W}_{20}}-2\hat{a}\frac{\widetilde{W}_
{11}}{\widetilde{W}_{20}}\right]~~; \\
\label{eq:vabp2}
&& \left(\hat{\sigma}_{\hat{b}}\right)^2=\left(\hat{\sigma}_{\hat{a}}\right)
^2\widetilde{(X^2)}~~;
\end{lefteqnarray}
for a formal demonstration of
Eq.\,(\ref{eq:vaap2}) refer to
Appendix \ref{a:resu}.

Relevant and useful special cases
shall be discussed in the following subsections.

\subsection{Uncorrelated errors in $X$ and in $Y$} \label{uner}

\noindent\noindent

In the limit of uncorrelated errors,
$(\sigma_{xy})_i=0$, $r_i=0$, $1\le i\le n$, 
Eqs.\,(\ref{eq:Wi}), (\ref{eq:Vi}),
(\ref{eq:Ui}), and (\ref{eq:Pi}) reduce to:
\begin{lefteqnarray}
\label{eq:Wiu}
&& W_i=\frac{w_{x_i}\Omega_i^2}{1+a^2\Omega_i^2}~~;\qquad1\le i\le n~~; \\
\label{eq:Viu}
&& V_i=\frac{w_{x_i}\Omega_i^4}{(1+a^2\Omega_i^2)^2}~~;\qquad1\le i\le n~~; \\
\label{eq:Uiu}
&& U_i=0~~;\qquad1\le i\le n~~; \\
\label{eq:Piu}
&& P_i=\frac{w_{x_i}\Omega_i^2}{(1+a^2\Omega_i^2)^2}~~;\qquad1\le i\le n~~;
\end{lefteqnarray}
accordingly, the pseudo cubic,
Eq.\,(\ref{eq:pcc}), reduces to:
\begin{equation}
\label{eq:pccu}
\widetilde{V}_{20}a^3-2\widetilde{V}_{11}a^2-(\widetilde{W}_{20}-
\widetilde{V}_{02})a+\widetilde{W}_{11}=0~~;
\end{equation}
and the pseudo quadratic,
Eq.\,(\ref{eq:pqc}), reduces to:
\begin{equation}
\label{eq:pqcu}
\widetilde{V}_{11}a^2+(\widetilde{P}_{20}-
\widetilde{V}_{02})a-\widetilde{P}_{11}=0~~;
\end{equation}
where $\widetilde{U}_{pq}=0$ via
Eqs.\,(\ref{eq:wgc}) and (\ref{eq:Uiu}).

Following the same procedure outlined in
the general case, the regression line slope
and intercept estimators, $\hat{a}$ and
$\hat{b}$, and the regression line slope
and intercept variance estimators,
$(\hat{\sigma}_{\hat{a}})^2$ and
$(\hat{\sigma}_{\hat{b}})^2$, can be
determined.   For further details
refer to the parent papers (Y66; Y69).
A pictorial illustration of the method
may be found in an additional paper
(York, 1967).   For a different but
equivalent approach refer to an independent
investigation (McIntyre et al., 1966).
An earlier method implying approximate
expressions is outlined in a pioneering
attempt (Deming, 1943).

\subsection{Completely correlated errors in $X$ and in $Y$} \label{coer}

\noindent\noindent

In the limit of completely correlated errors,
$(\sigma_{xy})_i=[(\sigma_{xx})_i(\sigma_{yy})
_i]^{1/2}$ (where the positive and the negative
root relate to correlation and anticorrelation,
respectively), $r_i=\sgn(r_i)$, $1\le i\le n$,
Eqs.\,(\ref{eq:XYa}) and (\ref{eq:XYb}) reduce to:
\begin{leftsubeqnarray}
\slabel{eq:XYtca}
&& X_i-x_i=[a\Omega_i-\sgn(r_i)]\frac{\Omega_i\lambda_i}{w_{y_i}}~~; \\
\slabel{eq:XYtcb}
&& Y_i-y_i=\sgn(r_i)[a\Omega_i-\sgn(r_i)]\frac{\lambda_i}{w_{y_i}}~~;\
\label{seq:XYtc}
\end{leftsubeqnarray}
and the combination
of Eqs.\,(\ref{eq:XYtca}) and (\ref{eq:XYtcb})
yields:
\begin{equation}
\label{eq:sXY}
\frac{Y_i-y_i}{X_i-x_i}=\frac{\sgn(r_i)}{\Omega_i}~~;
\end{equation}
which is equivalent to Eq.\,(\ref{eq:r1})
via Eq.\,(\ref{eq:Omi}).   Accordingly,
Eq.\,(\ref{eq:ssR}) reduces to:
\begin{equation}
\label{eq:ssRtc}
T_{\widetilde{R}}=\sum_{i=1}^n
w_{x_i}(X_i-x_i)^2=\sum_{i=1}^nw_{y_i}(Y_i-y_i)^2~~;
\end{equation}
and the repetition of the procedure
used in the general case yields
again Eqs.\,(\ref{eq:lai})-(\ref{eq:slxW}),
(\ref{eq:bar})-(\ref{eq:pce}),
(\ref{eq:pcc})-(\ref{eq:Pi}), where
$r_i=\sgn(r_i)$, $1\le i\le n$, when
necessary.   In particular,
Eqs.\,(\ref{eq:Wi}), (\ref{eq:Vi}),
(\ref{eq:Ui}), and (\ref{eq:Pi})
reduce to:
\begin{lefteqnarray}
\label{eq:Witc}
&& W_i=\frac{w_{x_i}\Omega_i^2}{[a\Omega_i-\sgn(r_i)]^2}~~;\qquad1\le i\le n
~~; \\
\label{eq:Vitc}
&& V_i=\frac{w_{x_i}\Omega_i^4}{[a\Omega_i-\sgn(r_i)]^4}~~;\qquad1\le i\le n
~~; \\
\label{eq:Uitc}
&& U_i=\frac{w_{x_i}\Omega_i^3\sgn(r_i)}{[a\Omega_i-\sgn(r_i)]^4}~~;\qquad1
\le i\le n~~; \\
\label{eq:Pitc}
&& P_i=\frac{w_{x_i}\Omega_i^2}{[a\Omega_i-\sgn(r_i)]^4}~~;\qquad1\le i\le n~~;
\end{lefteqnarray}
while the pseudo cubic and the
pseudo quadratic maintain the
formal expression of the general
case, Eqs.\,(\ref{eq:pcc}) and
(\ref{eq:pqc}), respectively.
For a pictorial illustration of the
method refer to the parent paper (Y69).

\subsection{Errors in $X$ negligible with respect to errors in $Y$}
\label{XnrY}

\noindent\noindent

In the limit of errors in $X$ negligible with
respect to errors in $Y$, $a^2(\sigma_{xx})_i
\ll(\sigma_{yy})_i$, $a(\sigma_{xy})_i
\ll(\sigma_{yy})_i$, $1\le i\le n$.
Ideally, $(\sigma_{xx})_i\to0$, $(\sigma_{xy})_i\to0$,
$1\le i\le n$,
which implies $r_i\to0$, $w_{x_i}\to+\infty$,
$\Omega_i\to0$, $1\le i\le n$.
Accordingly, the errors in $X$ and in $Y$ are
uncorrelated.

In the limit, $w_{x_i}\to+\infty$, $1\le i\le n$,
Eqs.\,(\ref{eq:Wiu})-(\ref{eq:Piu}), with due
account taken of Eq.\,(\ref{eq:Omi}), reduce to
(Y66):
\begin{lefteqnarray}
\label{eq:WiY}
&& W_i=w_{y_i}~~;\qquad1\le i\le n~~; \\
\label{eq:ViY}
&& V_i=0~~;\qquad1\le i\le n~~; \\
\label{eq:UiY}
&& U_i=0~~;\qquad1\le i\le n~~; \\
\label{eq:PiY}
&& P_i=w_{y_i}~~;\qquad1\le i\le n~~;
\end{lefteqnarray}
accordingly, the pseudo cubic, Eq.\,(\ref{eq:pccu}),
and the pseudo quadratic, Eq.\,(\ref{eq:pqcu}), by
use of Eq.\,(\ref{eq:wgc}) reduce to:
\begin{lefteqnarray}
\label{eq:pqcY}
&& (\widetilde{w_y})_{20}a-(\widetilde{w_y})_{11}=0~~;
\end{lefteqnarray}
and the regression line slope estimator
reads:
\begin{equation}
\label{eq:aY}
\hat{a}_{\rm Y}=\frac{(\widetilde{w_y})_{11}}{(\widetilde{w_y})_{20}}~~;
\end{equation}
finally, the substitution of Eq.\,(\ref{eq:aY})
into (\ref{eq:bar}) yields the regression line
intercept estimator, as:
\begin{equation}
\label{eq:bY}
\hat{b}_{\rm Y}=\widetilde{Y}-\hat{a}_{\rm Y}\widetilde{X}~~;
\end{equation}
where the index, Y, stands for WLS(Y$|$X) i.e.
weighted least square regression, or in particular
OLS(Y$|$X) i.e.
ordinary least square regression, of the dependent
variable, $Y$, against the independent variable,
$X$ (Ial90).

With regard to the regression line slope and
intercept variance estimators, $(\hat{\sigma}_
{\hat{a}_{\rm Y}}^\prime)^2$ and $(\hat{\sigma}_
{\hat{b}_{\rm Y}}^\prime)^2$, with no
account taken of the scatter of the
data points, ${\sf P}_i\equiv(X_i, Y_i)$,
about the regression line, in the case
under discussion Eqs.\,(\ref{eq:prph})
and (\ref{eq:prps}) reduce to:
\begin{lefteqnarray}
\label{eq:prphY}
&& \left(\frac{\partial\phi}{\partial\hat{a}_{\rm Y}}\right)^2(\hat{\sigma}_
{\hat{a}_{\rm Y}}^\prime)^2=\sum_{i=1}^n\left(\frac{\partial\phi}{\partial
Y_i}\right)^2\frac1{w_{y_i}}~~; \\
\label{eq:prpsY}
&& (\hat{\sigma}_{\hat{b}_{\rm Y}}^\prime)^2=\sum_{i=1}^n\left(\frac{\partial
\psi}{\partial Y_i}\right)^2\frac1{w_{y_i}}~~;
\end{lefteqnarray}
and the substitution of Eqs.\,(\ref{eq:pdfiYb}), 
%(\ref{eq:pdZ}),
(\ref{eq:pdfsY}), and 
(\ref{eq:dppYb}) into (\ref{eq:prphY}) and
(\ref{eq:prpsY}), respectively, yields:
\begin{lefteqnarray}
\label{eq:prphY1}
&& (\hat{\sigma}_{\hat{a}_{\rm Y}}^\prime)^2=\frac1{(\widetilde{w_y})_{20}}
~~; \\
\label{eq:prpsY1}
&& (\hat{\sigma}_{\hat{b}_{\rm Y}}^\prime)^2=(\hat{\sigma}_{\hat{a}_{\rm Y}}^
\prime)^2\left[\frac{(\widetilde{w_y})_{20}}{(\widetilde{w_y})_{00}}+
(\widetilde{X})^2\right]
=(\hat{\sigma}_{\hat{a}_{\rm Y}}^\prime)^2(\widetilde{X^2})~~;
\end{lefteqnarray}
where Eqs.\,(\ref{seq:wZ}) and (\ref{eq:wgc})
have also been used.

The squared residual trace,
expressed by Eq.\,(\ref{eq:sR23}), in
the case under consideration via
Eqs.\,(\ref{eq:wgc}) and (\ref{eq:WiY})
reduces to:
\begin{lefteqnarray}
\label{eq:sR2Y}
&& T_{\widetilde{R}}=(\hat{a}_{\rm Y})^2(\widetilde{w_y})_{20}+
(\widetilde{w_y})_{02}-2\hat{a}_{\rm Y}(\widetilde{w_y})_{11}~~;
\end{lefteqnarray}
and concerning the regression line
slope and intercept variance estimators,
$(\hat{\sigma}_{\hat{a}_{\rm Y}})^2$ and
$(\hat{\sigma}_{\hat{b}_{\rm Y}})^2$,
with due account taken of the scatter
of the data points, ${\sf P}_i\equiv
(X_i,Y_i)$, about the regression line,
in the case under discussion
Eqs.\,(\ref{eq:vara}) and (\ref{eq:varb}),
by use of (\ref{eq:WiY}), (\ref{eq:prphY1}),
(\ref{eq:prpsY1}) and (\ref{eq:sR2Y})
reduce to:
\begin{lefteqnarray}
\label{eq:varaY}
&& (\hat{\sigma}_{\hat{a}_{\rm Y}})^2=\frac1{n-2}\left[(\hat{a}_{\rm Y})^2+
\frac{(\widetilde{w_y})_{02}}{(\widetilde{w_y})_{20}}-2\hat{a}_{\rm Y}\frac
{(\widetilde{w_y})_{11}}{(\widetilde{w_y})_{20}}\right]~~; \\
\label{eq:varbY}
&& (\hat{\sigma}_{\hat{b}_{\rm Y}})^2=(\hat{\sigma}_{\hat{a}_{\rm Y}})^2\left[
\frac{(\widetilde{w_y})_{20}}{(\widetilde{w_y})_{00}}+(\widetilde{X})^2\right]
=(\hat{\sigma}_{\hat{a}_{\rm Y}})^2(\widetilde{X^2})~~;
\end{lefteqnarray}
where Eqs.\,(\ref{seq:wZ}) and (\ref{eq:wgc})
have also been used.
It can be seen that, in the case under
consideration, the exact expression of the
regression line slope and intercept estimators, 
Eqs.\,(\ref{eq:varaY}) and (\ref{eq:varbY}),
coincide with the
approximate expression in the general case,
Eqs.\,(\ref{eq:vaap2}) and (\ref{eq:vabp2}),
respectively.

The expression of the regression line slope
and intercept estimators and related variance
estimators, Eqs.\,(\ref{eq:aY}), (\ref{eq:bY}),
(\ref{eq:varaY}), (\ref{eq:varbY}), coincide
with their counterparts determined for
WLS(Y$|$X) models in a recent attempt
[Lavagnini and Magno, 2007, Eqs.\,(17)-(21)
therein].

The substitution of Eq.\,(\ref{eq:aY})
into (\ref{eq:varaY}) yields:
\begin{lefteqnarray}
\label{eq:varaY2}
&& (\hat{\sigma}_{\hat{a}_{\rm Y}})^2=\frac{(\hat{a}_{\rm Y})^2}{n-2}\frac
{D_{w_y}}{[(\widetilde{w_y})_{11}]^2}~~; \\
\label{eq:Dwy}
&& D_{w_y}=(\widetilde{w_y})_{02}(\widetilde{w_y})_{20}-[(\widetilde{w_y})_
{11}]^2~~;
\end{lefteqnarray}
where $D_{w_y}$ is the determinant of a
(weighted) deviation trace matrix, defined
as a weighted deviation determinant.

Under the restriction of homoscedastic data,
$w_{x_i}=w_x$, $w_{y_i}=w_y$, $1\le i\le n$, which
implies $Q_i=Q$, $Q=W,V,U,P$, via
Eqs.\,(\ref{eq:Wi}), (\ref{eq:Vi}), (\ref{eq:Ui}),
(\ref{eq:Pi}), Eqs.\,(\ref{seq:wZ}) and (\ref{eq:wgc})
reduce to:
\begin{leftsubeqnarray}
\slabel{eq:Zpa}
&& \widetilde{Z}=\overline{Z}~~;\qquad Z=X,Y~~; \\
\label{eq:Zpb}
&& \sum_{i=1}^n(Z_i-\overline{Z})=0~~;\qquad Z=X,Y~~;
\label{seq:Zp}
\end{leftsubeqnarray}
\begin{lefteqnarray}
\label{eq:Qun}
&& \widetilde{Q}_{pq}=QS_{pq}~~;\qquad Q=W,V,U,P~~; \\
\label{eq:Spq}
&& S_{pq}=\sum_{i=1}^n(X_i-\overline{X})^p(Y_i-\overline{Y})^q~~;
\end{lefteqnarray}
where $S_{pq}$ are the (unweighted)
pure ($p=0$ and/or $q=0$)
and mixed ($p>0$ and $q>0$) deviation
traces, and $S_{00}=n$.

In the special case under discussion,
Eqs.\,(\ref{eq:aY}), (\ref{eq:bY}),
(\ref{eq:varaY}), (\ref{eq:varbY}),
(\ref{eq:varaY2}) and (\ref{eq:Dwy})
reduce to:
\begin{lefteqnarray}
\label{eq:aYu}
&& \hat{a}_{\rm Y}=\frac{S_{11}}{S_{20}}~~; \\
\label{eq:bYu}
&& \hat{b}_{\rm Y}=\overline{Y}-\hat{a}_{\rm Y}\overline{X}~~; \\
\label{eq:vaYu}
&& (\hat{\sigma}_{\hat{a}_{\rm Y}})^2=\frac1{n-2}\left[(\hat{a}_{\rm Y})^2+
\frac{S_{02}}{S_{20}}-2\hat{a}_{\rm Y}\frac{S_{11}}{S_{20}}\right]~~; \\
\label{eq:vbYu}
&& (\hat{\sigma}_{\hat{b}_{\rm Y}})^2=(\hat{\sigma}_{\hat{a}_{\rm Y}})^2
\left[\frac1nS_{20}+(\overline{X})^2\right]
=(\hat{\sigma}_{\hat{a}_{\rm Y}})^2(\overline{X^2})~~; \\
\label{eq:vaYu2}
&& (\hat{\sigma}_{\hat{a}_{\rm Y}})^2=\frac{(\hat{a}_{\rm Y})^2}{n-2}\frac
{D_{\rm S}}{(S_{11})^2}~~; \\
\label{eq:DS}
&& D_{\rm S}=S_{02}S_{20}-(S_{11})^2~~;
\end{lefteqnarray}
where $D_{\rm S}$ is the determinant of
the (unweighted) deviation trace matrix, defined
as the (unweighted) deviation determinant.

The expression of the regression
line slope variance estimator,
Eq.\,(\ref{eq:vaYu}), coincides
with earlier results known in
literature in the special case
of normal residuals [e.g., FB92,
Eq.\,(4) therein, in the limit
$c^2=(\sigma_{yy})_{\rm zzz}/
(\sigma_{xx})_{\rm zzz}\to+\infty$,
where zzz = ins, int, denote instrumental
and intrinsic scatter, respectively].
For a formal demonstration, see
Appendix \ref{a:eqFB}.   Then the
expressions of the regression
line slope and intercept variance
estimators, $(\hat{\sigma}_{\hat{a}_{\rm Y}})^2$
and $(\hat{\sigma}_{\hat{b}_{\rm Y}})^2$,
reported above, hold provided
the residuals obey a Gaussian
distribution, as expected for
functional models (Y66; Y69).

The expression of the regression line slope
and intercept estimators and related variance
estimators, Eqs.\,(\ref{eq:aYu}), (\ref{eq:bYu}),
(\ref{eq:vaYu}), (\ref{eq:vbYu}), coincide
with their counterparts determined for
OLS(Y$|$X) models in a recent attempt
[Lavagnini and Magno, 2007, Eqs.\,(3)-(7)
therein].

\subsection{Errors in $Y$ negligible with respect to errors in $X$}
\label{YnrX}

\noindent\noindent

In the limit of errors in $Y$ negligible with
respect to errors in $X$, $(\sigma_{yy})_i
\ll a^2(\sigma_{xx})_i$, $(\sigma_{xy})_i
\ll a(\sigma_{xx})_i$, $1\le i\le n$.
Ideally, $(\sigma_{yy})_i
\to0$, $(\sigma_{xy})_i\to0$, $1\le i\le n$,
which implies $r_i\to0$, $w_{y_i}\to+\infty$,
$\Omega_i\to+\infty$, $1\le i\le n$.
Accordingly, the errors in $X$ and in $Y$ are
uncorrelated.

The model under discussion can be related to,
but not confused with,
the inverse regression, which has a large
associate literature (e.g., Miller, 1966; Garden
et al., 1980; Osborne, 1991; Brown, 1993; Lavagnini
and Magno, 2007).   More specifically, the inverse
regression consists in the obtainement of a
variable, $x$, from an instrumental response,
$y$, with the confidence interval for the true
value of $x$ (e.g., Brownlee, 1960; Lavagnini and
Magno, 2007).   A statistical calibration problem
is a kind of inverse prediction, a problem of
retrospection, and some authors call it inverse
regression rather than calibration: it is probably
best explained by considering a typical univariate
calibration problem (Osborne, 1991).

In the limit, $w_{y_i}\to+\infty$, $1\le i\le n$,
Eqs.\,(\ref{eq:Wiu})-(\ref{eq:Piu}), with due
account taken of Eq.\,(\ref{eq:Omi}), reduce to
(Y66):
\begin{lefteqnarray}
\label{eq:WiX}
&& W_i=a^{-2}w_{x_i}~~;\qquad1\le i\le n~~; \\
\label{eq:ViX}
&& V_i=a^{-4}w_{x_i}~~;\qquad1\le i\le n~~; \\
\label{eq:UiX}
&& U_i=0~~;\qquad1\le i\le n~~; \\
\label{eq:PiX}
&& P_i=0~~;\qquad1\le i\le n~~;
\end{lefteqnarray}
accordingly, the pseudo cubic, Eq.\,(\ref{eq:pccu}),
and the pseudo quadratic, Eq.\,(\ref{eq:pqcu}), by
use of Eq.\,(\ref{eq:wgc}) reduce to:
\begin{lefteqnarray}
\label{eq:pqcX}
&& (\widetilde{w_x})_{11}a-(\widetilde{w_x})_{02}=0~~;
\end{lefteqnarray}
and the regression line slope estimator
reads:
\begin{equation}
\label{eq:aX}
\hat{a}_{\rm X}=\frac{(\widetilde{w_x})_{02}}{(\widetilde{w_x})_{11}}~~;
\end{equation}
finally, the substitution of Eq.\,(\ref{eq:aX})
into (\ref{eq:bar}) yields the regression line
intercept estimator, as:
\begin{equation}
\label{eq:bX}
\hat{b}_{\rm X}=\widetilde{Y}-\hat{a}_{\rm X}\widetilde{X}~~;
\end{equation}
where the index, X, stands for WLS(X$|$Y) i.e.
weighted least square regression, or in particular
OLS(X$|$Y) i.e.
ordinary least square regression, of the independent
variable, $X$, against the dependent variable,
$Y$ (Ial90).

With regard to the regression line slope and
intercept variance estimators, $(\hat{\sigma}_
{\hat{a}_{\rm X}}^\prime)^2$ and $(\hat{\sigma}_
{\hat{b}_{\rm X}}^\prime)^2$, with no
account taken of the scatter of the
data points, ${\sf P}_i\equiv(X_i, Y_i)$,
about the regression line, in the case
under discussion Eqs.\,(\ref{eq:prph})
and (\ref{eq:prps}) reduce to:
\begin{lefteqnarray}
\label{eq:prphX}
&& \left(\frac{\partial\phi}{\partial\hat{a}_{\rm X}}\right)^2(\hat{\sigma}_
{\hat{a}_{\rm X}}^\prime)^2=\sum_{i=1}^n\left(\frac{\partial\phi}{\partial
X_i}\right)^2\frac1{w_{x_i}}~~; \\
\label{eq:prpsX}
&& (\hat{\sigma}_{\hat{b}_{\rm X}}^\prime)^2=\sum_{i=1}^n\left(\frac{\partial
\psi}{\partial X_i}\right)^2\frac1{w_{x_i}}~~;
\end{lefteqnarray}
and the substitution of Eqs.\,(\ref{eq:pdfiXa}), 
%(\ref{eq:pdZ}),
(\ref{eq:pdfsX}), and 
(\ref{eq:dppXa}) into (\ref{eq:prphX}) and
(\ref{eq:prpsX}), respectively, yields:
\begin{lefteqnarray}
\label{eq:prphX1}
&& (\hat{\sigma}_{\hat{a}_{\rm X}}^\prime)^2=\frac{(\hat{a}_{\rm X})^4}
{(\widetilde{w_x})_{02}}
~~; \\
\label{eq:prpsX1}
&& (\hat{\sigma}_{\hat{b}_{\rm X}}^\prime)^2=\frac{(\hat{\sigma}_{\hat{a}_
{\rm X}}^\prime)^2}{(\hat{a}_{\rm X})^2}\left[\frac{(\widetilde{w_x})_{02}}
{(\widetilde{w_x})_{00}}+(\hat{a}_{\rm X})^2(\widetilde{X})^2\right]~~;
\end{lefteqnarray}
where Eqs.\,(\ref{seq:wZ}) and (\ref{eq:wgc})
have also been used.

The squared residual trace,
expressed by Eq.\,(\ref{eq:sR23}), in
the case under consideration via
Eqs.\,(\ref{eq:wgc}) and (\ref{eq:WiX})
reduces to:
\begin{lefteqnarray}
\label{eq:sR2X}
&& T_{\widetilde{R}}=(\widetilde{w_x})_{20}+(\hat{a}_{\rm X})^
{-2} (\widetilde{w_x})_{02}-2(\hat{a}_{\rm X})^{-1}(\widetilde{w_x})_{11}~~;
\end{lefteqnarray}
and concerning the regression line
slope and variance estimators,
$(\hat{\sigma}_{\hat{a}_{\rm X}})^2$ and
$(\hat{\sigma}_{\hat{b}_{\rm X}})^2$,
with due account taken of the scatter
of the data points, ${\sf P}_i\equiv
(X_i,Y_i)$, about the regression line,
in the case under discussion
Eqs.\,(\ref{eq:vara}) and (\ref{eq:varb}),
by use of (\ref{eq:WiX}), (\ref{eq:prphX1}),
(\ref{eq:prpsX1}) and (\ref{eq:sR2X})
reduce to:
\begin{lefteqnarray}
\label{eq:varaX}
&& (\hat{\sigma}_{\hat{a}_{\rm X}})^2=\frac{(\hat{a}_{\rm X})^2}{n-2}\left[
(\hat{a}_{\rm X})^2
\frac{(\widetilde{w_x})_{20}}{(\widetilde{w_x})_{02}}+1-2\hat{a}_{\rm X}\frac
{(\widetilde{w_x})_{11}}{(\widetilde{w_x})_{02}}\right]~~; \\
\label{eq:varbX}
&& (\hat{\sigma}_{\hat{b}_{\rm X}})^2=\frac{(\hat{\sigma}_{\hat{a}_{\rm X}})^
2}{(\hat{a}_{\rm X})^2}\left[
\frac{(\widetilde{w_x})_{02}}{(\widetilde{w_x})_{00}}+(\hat{a}_{\rm X})^2
(\widetilde{X})^2\right]~~;
\end{lefteqnarray}
where it can be seen that, in the case under
consideration, the exact expressions of the
regression line slope and intercept estimator, 
Eqs.\,(\ref{eq:varaX}) and (\ref{eq:varbX}),
respectively, do not coincide with the
approximate expression in the general case,
Eqs.\,(\ref{eq:vaap2}) and (\ref{eq:vabp2}),
respectively.

The substitution of Eq.\,(\ref{eq:aX})
into (\ref{eq:varaX}) yields:
\begin{lefteqnarray}
\label{eq:varaX2}
&& (\hat{\sigma}_{\hat{a}_{\rm X}})^2=\frac{(\hat{a}_{\rm X})^2}{n-2}\frac
{D_{w_x}}{[(\widetilde{w_x})_{11}]^2}~~; \\
\label{eq:Dwx}
&& D_{w_x}=(\widetilde{w_x})_{02}(\widetilde{w_x})_{20}-[(\widetilde{w_x})_
{11}]^2~~;
\end{lefteqnarray}
where $D_{w_x}$ is a weighted deviation
determinant.

In the special case of homoscedastic data,
$w_{x_i}=w_x$, $1\le i\le n$, which
implies $Q_i=Q$, $Q=W,V,U,P$, via
Eqs.\,(\ref{eq:Wi}), (\ref{eq:Vi}),
(\ref{eq:Ui}), (\ref{eq:Pi}),
and Eqs.\,(\ref{seq:Zp}) and
(\ref{eq:Qun}) hold.   Accordingly,
Eqs.\,(\ref{eq:aX}), (\ref{eq:bX}),
(\ref{eq:varaX}), (\ref{eq:varbX}),
and (\ref{eq:varaX2})
%and (\ref{eq:Dwx})
reduce to:
\begin{lefteqnarray}
\label{eq:aXu}
&& \hat{a}_{\rm X}=\frac{S_{02}}{S_{11}}~~; \\
\label{eq:bXu}
&& \hat{b}_{\rm X}=\overline{Y}-\hat{a}_{\rm X}\overline{X}~~; \\
\label{eq:vaXu}
&& (\hat{\sigma}_{\hat{a}_{\rm X}})^2=\frac{(\hat{a}_{\rm X})^2}{n-2}\left[
(\hat{a}_{\rm X})^2\frac{S_{20}}{S_{02}}+1-2\hat{a}_{\rm X}\frac{S_{11}}
{S_{02}}\right]~~; \\
\label{eq:vbXu}
&& (\hat{\sigma}_{\hat{b}_{\rm X}})^2=\frac{(\hat{\sigma}_{\hat{a}_{\rm X}})^
2}{(\hat{a}_{\rm X})^2}\left[\frac1nS_{02}+(\hat{a}_{\rm X})^2(\overline{X})^
2\right]~~; \\
\label{eq:vaXu2}
&& (\hat{\sigma}_{\hat{a}_{\rm X}})^2=\frac{(\hat{a}_{\rm X})^2}{n-2}\frac
{D_{\rm S}}{(S_{11})^2}~~;
\end{lefteqnarray}
where $D_{\rm S}$ is the deviation
determinant, Eq.\,(\ref{eq:DS}).

The expression of the regression
line slope variance estimator,
Eq.\,(\ref{eq:vaXu}), is different
from its counterpart calculated in
an earlier attempt (FB92), due to
the lack of an additional term
which is negligible for $D_{\rm S}/
(S_{11})^2\ll1$ and/or
$n\gg1$.   For a
formal demonstration, see
Appendix \ref{a:eqFB}.   Then the
expressions of the regression
line slope and intercept variance
estimators, $(\hat{\sigma}_{\hat{a}_{\rm X}})^2$
and $(\hat{\sigma}_{\hat{b}_{\rm X}})^2$,
reported above, hold provided
the residuals obey a Gaussian
distribution, as expected for
functional models (Y66; Y69),
with the caveat due to the
above mentioned discrepancy.

\subsection{Generalized orthogonal regression}
%Errors in $X$ equally weighted and
%equally correlated with respect to errors in $Y$}
\label{YXc2}

\noindent\noindent

In the limit of constant $y$ to $x$ variance
ratios and constant correlation coefficients,
%
%errors in $X$ equally weighted and equally
%correlated with respect to errors in $Y$,
%
the following relations hold:
\begin{lefteqnarray}
\label{eq:xyc2}
&& \frac{(\sigma_{yy})_i}{(\sigma_{xx})_i}=c^2~;~~\frac{w_{x_i}}
{w_{y_i}}=\Omega_i^{-2}=c^2~;~~\frac{(\sigma_{xy})_i}{(\sigma_{xx})_i}=r_ic=
rc~~;~~1\le i\le n~~;\qquad
%
%r_i=\frac{(\sigma_{xy})_i}{[(\sigma_{xx})_i(\sigma_{yy})_i]^{1/2}}=r;
%
\end{lefteqnarray}
where the weights are assumed to be inversely
proportional to related variances,
$w_{z_i}\propto1/(\sigma_{zz})_i$, $z=x,y$,
as usually done (e.g., FB92).   It is worth
noticing that Eq.\,(\ref{eq:xyc2}) holds for
both homoscedastic and heteroscedastic data.
It can be seen that the lines of adjustment
are oriented along the same direction (York,
1967) but are perpendicular
to the regression line only in the special case,
$c^2=1$, which is the genuine orthogonal regression
(e.g., Carroll et al., 2006, Chap.\,3, \S 3.4.2).

Earlier formulations of the model with respect
to the parent paper (Y66) may be found in
several attempts (e.g., Kummell, 1879;
Koopmans, 1937; Deming, 1943; Tintner, 1945;
Lindley, 1947; Anderson, 1951; Madansky, 1959)
as well as later investigations (e.g., Barnett,
1967; Moran, 1971; Kendall and Stuart, 1979,
Chap.\,29; Fuller, 1980, 1987, Chap.\,1,
Sect.\,1.3).

Taking into due account Eqs.\,(\ref{eq:Omi})
and (\ref{eq:xyc2}), Eqs.\,(\ref{eq:Wi}),
(\ref{eq:Vi}), (\ref{eq:Ui}) and (\ref{eq:Pi})
reduce to:
\begin{lefteqnarray}
\label{eq:Wic2}
&& W_i=\frac{w_{x_i}}{a^2+c^2-2rac}~~;\qquad1\le i\le n~~; \\
\label{eq:Vic2}
&& V_i=\frac{w_{x_i}}{(a^2+c^2-2rac)^2}~~;\qquad1\le i\le n~~; \\
\label{eq:Uic2}
&& U_i=\frac{rcw_{x_i}}{(a^2+c^2-2rac)^2}~~;\qquad1\le i\le n~~; \\
\label{eq:Pic2}
&& P_i=\frac{c^2w_{x_i}}{(a^2+c^2-2rac)^2}~~;\qquad1\le i\le n~~;
\end{lefteqnarray}
which, owing to Eq.\,(\ref{eq:wgc}), implies
the following:
\begin{equation}
\label{eq:Qtc2}
\widetilde{Q}_{pq}=k_{\rm Q}(\widetilde{w_x})_{pq}~~;\qquad Q=W,V,U,P~~;
\end{equation}
where $k_{\rm Q}=Q_i/w_{x_i}$ maintains constant.

Accordingly, the pseudo cubic, Eq.\,(\ref{eq:pcc}),
and the pseudo quadratic,\linebreak Eq.\,(\ref{eq:pqc}),
reduce to:
\begin{lefteqnarray}
\label{eq:pqcc}
&& [rc(\widetilde{w_x})_{20}-(\widetilde{w_x})_{11}]a^2-[c^2(\widetilde{w_x})
_{20}-(\widetilde{w_x})_{02}]a-[rc(\widetilde{w_x})_{02}-c^2(\widetilde{w_x})
_{11}]=0~\qquad
\end{lefteqnarray}
where $r=\sgn(r)$ in the limit of completely
correlated errors in $X$ and in $Y$.

In the special cases, $c^2\to+\infty$,
$c^2\to0$, Eq.\,(\ref{eq:pqcc}) reduces
to (\ref{eq:pqcY}) and (\ref{eq:pqcX}),
respectively, as expected.   The
regression line slope and intercept
estimators, $\hat{a}$ and $\hat{b}$,
can be derived from Eqs.\,(\ref{eq:pqcc})
and (\ref{eq:bar}), respectively,
where the parasite solution of the
pseudo quadratic equation, in the
former case, must be dismissed.

With regard to the regression line slope and
intercept variance estimators, $(\hat{\sigma}_
{\hat{a}_{\rm C}}^\prime)^2$ and $(\hat{\sigma}_
{\hat{b}_{\rm C}}^\prime)^2$, with no
account taken of the scatter of the
data points, ${\sf P}_i\equiv(X_i, Y_i)$,
about the regression line, in the case
under discussion Eqs.\,(\ref{eq:prph})
and (\ref{eq:prps}) reduce to:
\begin{lefteqnarray}
\label{eq:prphc}
&& \left(\frac{\partial\phi}{\partial\hat{a}_{\rm C}}\right)^2(\hat{\sigma}_
{\hat{a}_{\rm C}}^\prime)^2=\sum_{i=1}^n\left[\left(\frac{\partial\phi}
{\partial X_i}\right)^2+c^2\left(\frac{\partial\phi}{\partial Y_i}\right)^2+
2rc\frac{\partial\phi}{\partial X_i}\frac{\partial\phi}{\partial Y_i}
\right]\frac1{w_{x_i}}~~; \\
\label{eq:prpsc}
&& (\hat{\sigma}_{\hat{b}_{\rm C}}^\prime)^2=\sum_{i=1}^n\left[\left(\frac
{\partial\psi}{\partial X_i}\right)^2
+c^2\left(\frac{\partial\psi}{\partial Y_i}\right)^2+
2rc\frac{\partial\psi}{\partial X_i}\frac{\partial\psi}{\partial Y_i}
\right]\frac1{w_{x_i}}~~;
\end{lefteqnarray}
and the substitution of Eqs.\,(\ref{seq:pdfic}),
(\ref{eq:pdfsc}), and
(\ref{seq:dppc}), into (\ref{eq:prphc}) and
(\ref{eq:prpsc}), respectively, yields
cumbersome expressions of the regression
line slope and intercept variance
estimators, which shall not be explicitly
written here.

The squared residual trace,
expressed by Eq.\,(\ref{eq:sR23}), in
the case under consideration via
Eqs.\,(\ref{eq:wgc}) and (\ref{eq:Wic2})
reduces to:
\begin{lefteqnarray}
\label{eq:sR2c}
&& T_{\widetilde{R}}=\frac{(\hat{a}_{\rm C})^2(\widetilde
{w_x})_{20}+(\widetilde{w_x})_{02}-2\hat{a}_{\rm C}(\widetilde{w_x})_{11}}
{(\hat{a}_{\rm C})^2+c^2-2rc\hat{a}_{\rm C}}~~;
\end{lefteqnarray}
and concerning the regression line
slope and intercept variance estimators,
$(\hat{\sigma}_{\hat{a}_{\rm C}})^2$ and
$(\hat{\sigma}_{\hat{b}_{\rm C}})^2$,
with due account taken of the scatter
of the data points, ${\sf P}_i\equiv
(X_i,Y_i)$, about the regression line,
in the case under discussion
Eqs.\,(\ref{eq:vara}) and (\ref{eq:varb}),
by use of
%(\ref{eq:Wic2}), (\ref{seq:prphc1}), (\ref{seq:prpsc1}) and
(\ref{eq:sR2c})
may be cast into a more explicit form.

In the special case of uncorrelated errors,
$r\to0$, Eq.\,(\ref{eq:pqcc}) reduces to:
\begin{lefteqnarray}
\label{eq:pqc0}
&& (\widetilde{w_x})_{11}a^2+[c^2(\widetilde{w_x})_{20}-(\widetilde{w_x})_
{02}]a-c^2(\widetilde{w_x})_{11}=0~~;
\end{lefteqnarray}
which has the solutions (Deming, 1943):
\begin{lefteqnarray}
\label{eq:acu}
&& \hat{a}_{\rm C}=\frac{(\widetilde{w_x})_{02}-c^2(\widetilde{w_x})_{20}}
{2(\widetilde{w_x})_{11}}\left\{1\mp\left[1+c^2\left(\frac{(\widetilde{w_x})_
{02}-c^2(\widetilde{w_x})_{20}}{2(\widetilde{w_x})_{11}}\right)^{-2}\right]^
{1/2}\right\}~~;\qquad
\end{lefteqnarray}
and the regression line slope estimator
is obtained disregarding the parasite
solution.   Then the substitution of
Eq.\,(\ref{eq:acu}) into (\ref{eq:bar})
yields the regression line intercept
estimator, as:
\begin{equation}
\label{eq:bcu}
\hat{b}_{\rm C}=\widetilde{Y}-\hat{a}_{\rm C}\widetilde{X}~~;
\end{equation}
where the index, C, denotes the case
under discussion, with normal residuals
(FB92).

The squared regression line slope estimator,
via Eq.\,(\ref{eq:acu}), reads:
\begin{lefteqnarray}
\label{eq:acu2}
&& (\hat{a}_{\rm C})^2=\left\{\frac{(\widetilde{w_x})_{02}-c^2(\widetilde
{w_x})_{20}}{2(\widetilde{w_x})_{11}}\mp c\left[\frac1{c^2}\left(\frac
{(\widetilde{w_x})_{02}-c^2(\widetilde{w_x})_{20}}{2(\widetilde{w_x})_{11}}
\right)^2+1\right]^{1/2}\right\}^2;\qquad
\end{lefteqnarray}
where the square root, if sufficiently close
to unity, may be developed in binomial series
and the terms of higher order neglected.
The result is:
\begin{leftsubeqnarray}
\slabel{eq:acu2aa}
&& (\hat{a}_{\rm C})^2=\left\{\frac{(\widetilde{w_x})_{02}-c^2(\widetilde
{w_x})_{20}}{2(\widetilde{w_x})_{11}}\mp c\left[1+\frac12\frac1{c^2}\left(
\frac{(\widetilde{w_x})_{02}-c^2(\widetilde{w_x})_{20}}{2(\widetilde{w_x})_
{11}}\right)^2\right]\right\}^2;\qquad \\
\slabel{eq:acu2ab}
&& \frac1{c^2}\left(\frac{(\widetilde{w_x})_{02}-c^2(\widetilde{w_x})_{20}}
{2(\widetilde{w_x})_{11}}\right)^2\ll1~~;
\label{seq:acu2a}
\end{leftsubeqnarray}
which, performing some algebra and neglecting
the terms of higher order, takes the expression:
\begin{lefteqnarray}
\label{eq:acu2b}
&& (\hat{a}_{\rm C})^2=c^2\mp2c\frac{(\widetilde{w_x})_{02}-c^2(\widetilde
{w_x})_{20}}{2(\widetilde{w_x})_{11}}~;\quad
\frac1{c^2}\left(\frac{(\widetilde{w_x})_{02}-c^2(\widetilde{w_x})_{20}}
{2(\widetilde{w_x})_{11}}\right)^2\ll1~;\qquad
\end{lefteqnarray}
and additional algebra shows that the following
relation holds:
\begin{lefteqnarray}
\label{eq:acu2c}
&& \frac{(\hat{a}_{\rm C})^2-(\widetilde{w_x})_{02}/(\widetilde{w_x})_{20}}
{(\hat{a}_{\rm C})^2-c^2}=1\pm\frac1c\frac{(\widetilde{w_x})_{11}}
{(\widetilde{w_x})_{20}};~~
\frac1{c^2}\left(\frac{(\widetilde{w_x})_{02}-c^2(\widetilde{w_x})_{20}}
{2(\widetilde{w_x})_{11}}\right)^2\ll1;\qquad
\end{lefteqnarray}
where further attention has to be devoted
to the special case:
\begin{lefteqnarray}
\label{eq:acu2d}
&& \frac{(\hat{a}_{\rm C})^2-(\widetilde{w_x})_{02}/(\widetilde{w_x})_{20}}
{(\hat{a}_{\rm C})^2-c^2}=1\pm\lambda_{w_x}=1-\sgn[(\widetilde{w_x})_{11}]
\lambda_{w_x}~~; \\
\label{eq:alawx}
&& \lambda_{w_x}=\frac{(\widetilde{w_x})_{11}}{[(\widetilde{w_x})_{20}
(\widetilde{w_x})_{02}]^{1/2}}~~;\qquad c^2=\frac{(\widetilde{w_x})_{02}}
{(\widetilde{w_x})_{20}}~~;
\end{lefteqnarray}
expressed in terms of the regression
line correlation coefficient, $\lambda_{w_x}$.

With regard to the regression line slope and
intercept variance estimators, $(\hat{\sigma}_
{\hat{a}_{\rm C}}^\prime)^2$ and $(\hat{\sigma}_
{\hat{b}_{\rm C}}^\prime)^2$, the substitution
of Eqs.\,(\ref{seq:pdfcS}), (\ref{eq:pdacS2}),
and (\ref{seq:dpcS2}) into (\ref{eq:prphc}) and
(\ref{eq:prpsc}) particularized to uncorrelated
errors, $r=0$, yields after some algebra:
\begin{lefteqnarray}
\label{eq:prphc2}
&& (\hat{\sigma}_{\hat{a}_{\rm C}}^\prime)^2=(\hat{a}_{\rm C})^2\frac
{(\widetilde{w_x})_{02}+c^2(\widetilde{w_x})_{20}}
{[(\widetilde{w_x})_{11}]^2}~~; \\
\label{eq:prpsc2}
&& (\hat{\sigma}_{\hat{b}_{\rm C}}^\prime)^2=\frac{(\hat{a}_{\rm C})^2+c^2}
{(\widetilde{w_x})_{00}}+(\hat{a}_{\rm C})^2(\widetilde{X})^2\frac
{(\widetilde{w_x})_{02}+c^2(\widetilde{w_x})_{20}}
{[(\widetilde{w_x})_{11}]^2}~~;\quad
\end{lefteqnarray}
where Eq.\,(\ref{eq:wgc})
has also been used.

The squared residual trace,
expressed by Eq.\,(\ref{eq:sR2c}), in
the case under consideration reduces to:
\begin{lefteqnarray}
\label{eq:sR2c2}
&& T_{\widetilde{R}}=\frac{(\hat{a}_{\rm C})^2(\widetilde{w_x})
_{20}+(\widetilde{w_x})_{02}-2\hat{a}_{\rm C}(\widetilde{w_x})_{11}}
{(\hat{a}_{\rm C})^2+c^2}
%\nonumber \\
%&& \phantom{\sum_{i=1}^n(\widetilde{R_i})^2}
=\frac{(\hat{a}_{\rm C})^2(\widetilde{w_x})_{20}-
(\widetilde{w_x})_{02}}{(\hat{a}_{\rm C})^2-c^2}~~;
\end{lefteqnarray}
where Eq.\,(\ref{eq:pqc01}) has
also been used.
Concerning the regression line
slope and intercept variance estimators,
$(\hat{\sigma}_{\hat{a}_{\rm C}})^2$ and
$(\hat{\sigma}_{\hat{b}_{\rm C}})^2$,
%with due account taken of the scatter
%of the data points, ${\sf P}_i\equiv
%(X_i,Y_i)$, about the regression line,
%in the case under discussion
Eqs.\,(\ref{eq:vara}) and (\ref{eq:varb}),
by use of (\ref{eq:prphc2}),
(\ref{eq:prpsc2}) and (\ref{eq:sR2c2}),
after a lot of algebra reduce to:
\begin{lefteqnarray}
\label{eq:varaC}
&& (\hat{\sigma}_{\hat{a}_{\rm C}})^2=\frac1{n-2}\frac{(\hat{a}_{\rm C})^2}
{(\hat{a}_{\rm C})^2-c^2}
\frac{[(\widetilde{w_x})_{02}+c^2(\widetilde{w_x})_{20}][(\hat{a}_{\rm C})^2
(\widetilde{w_x})_{20}-(\widetilde{w_x})_{02}]}
{[(\widetilde{w_x})_{11}]^2}~~;\qquad \\
\label{eq:varbC}
&& (\hat{\sigma}_{\hat{b}_{\rm C}})^2=\frac1{n-2}\frac1{(\widetilde{w_x})_
{00}}\frac{(\hat{a}_{\rm C})^2+c^2}{(\hat{a}_{\rm C})^2-c^2}\left
[(\hat{a}_{\rm C})^2(\widetilde{w_x})_{20}-(\widetilde{w_x})_{02}\right]+
(\widetilde{X})^2(\hat{\sigma}_{\hat{a}_{\rm C}})^2~~;\qquad~
\end{lefteqnarray}
where further attention has to be devoted
to the special case, $c\to\hat{a}_{\rm C}$.

In the limit of errors in $X$ negligible
with respect to errors in $Y$, $c\to+\infty$,
$w_{x_i}/c^2\to w_{y_i}$, $\hat{a}_{\rm C}\to
\hat{a}_{\rm Y}$, it can be seen that
Eqs.\,(\ref{eq:varaC}) and (\ref{eq:varbC})
reduce to (\ref{eq:varaY}) and (\ref{eq:varbY}),
respectively, as expected.

In the limit of errors in $Y$ negligible
with respect to errors in $X$, $c\to0$,
$c^2w_{y_i}\to w_{x_i}$, $\hat{a}_{\rm C}\to
\hat{a}_{\rm X}$, it can be seen that
Eqs.\,(\ref{eq:varaC}) and (\ref{eq:varbC})
reduce to (\ref{eq:varaX}) and (\ref{eq:varbX}),
respectively, as expected.

In the limit of the special case  considered above,
$c^2\to(\widetilde{w_x})_{02}/(\widetilde{w_x})_{20}$,
the combination of
Eqs.\,(\ref{eq:acu2d}) and (\ref{eq:varbC}) yields:
\begin{lefteqnarray}
\label{eq:varbCd}
&& (\hat{\sigma}_{\hat{b}_{\rm C}})^2=\frac1{n-2}\frac{(\widetilde{w_x})_{20}}
{(\widetilde{w_x})_{00}}[(\hat{a}_{\rm C})^2+c^2]\{1-
\sgn[(\widetilde{w_x})_{11}]\lambda_{w_x}\}+(\widetilde{X})^2
(\hat{\sigma}_{\hat{a}_{\rm C}})^2~;\qquad~
\end{lefteqnarray}
in terms of the regression
line correlation coefficient, $\lambda_{w_x}$.

The parameter, $c^2$, appearing in
Eqs.\,(\ref{eq:varaC}) and (\ref{eq:varbC}),
may be eliminated via Eq.\,(\ref{eq:pqc01}).
The result is:
\begin{lefteqnarray}
\label{eq:c2}
&& c^2=\hat{a}_{\rm C}\frac{(\widetilde{w_x})_{02}-\hat{a}_{\rm C}
(\widetilde{w_x})_{11}}{\hat{a}_{\rm C}(\widetilde{w_x})_{20}-
(\widetilde{w_x})_{11}}~~; \\
\label{eq:ac2c2}
&& (\hat{a}_{\rm C})^2-c^2=\frac{\hat{a}_{\rm C}[(\hat{a}_{\rm C})^2
(\widetilde{w_x})_{20}-(\widetilde{w_x})_{02}]}{\hat{a}_{\rm C}
(\widetilde{w_x})_{20}-(\widetilde{w_x})_{11}}~~;
\end{lefteqnarray}
and the substitution of 
Eqs.\,(\ref{eq:c2}) and (\ref{eq:ac2c2}) 
into (\ref{eq:varaC}) yields after some algebra:
\begin{lefteqnarray}
\label{eq:varac2}
&& (\hat{\sigma}_{\hat{a}_{\rm C}})^2=\frac{(\hat{a}_{\rm C})^2}{n-2}\left\{
\frac{D_{w_x}}{[(\widetilde{w_x})_{11}]^2}\right. \nonumber \\
&& \phantom{(\hat{\sigma}_{\hat{a}_{\rm C}})^2=}\left.
+\left[\frac{D_{w_x}}{[(\widetilde
{w_x})_{11}]^2}+2-\frac{(\widetilde{w_x})_{02}+(\hat{a}_{\rm C})^2(\widetilde
{w_x})_{20}}{\hat{a}_{\rm C}(\widetilde{w_x})_{11}}\right]\right\}~~;
\end{lefteqnarray}
where $D_{w_x}$ is a weighted deviation
determinant, Eq.\,(\ref{eq:Dwx}).

In the special case of homoscedastic data,
$w_{x_i}=w_x$, $1\le i\le n$, which
implies $Q_i=Q$, $Q=W,V,U,P$, via
Eqs.\,(\ref{eq:Wic2})-(\ref{eq:Pic2}),
and Eqs.\,(\ref{seq:Zp}) and
(\ref{eq:Qun}) hold.   Accordingly,
Eqs.\,(\ref{eq:acu}), (\ref{eq:bcu}),
(\ref{eq:varaC}), (\ref{eq:varbC}),
(\ref{eq:varbCd}) and (\ref{eq:varac2})
reduce to:
\begin{lefteqnarray}
\label{eq:acS}
&& \hat{a}_{\rm C}=\frac{S_{02}-c^2S_{20}}{2S_{11}}\left\{1\mp\left[1+c^2
\left(\frac{S_{02}-c^2S_{20}}{2S_{11}}\right)^{-2}\right]^{1/2}\right\}~~; \\
\label{eq:bcS}
&& \hat{b}_{\rm C}=\overline{Y}-\hat{a}_{\rm C}\overline{X}~~; \\
\label{eq:vaacS}
&& (\hat{\sigma}_{\hat{a}_{\rm C}})^2=\frac1{n-2}\frac{(\hat{a}_{\rm C})^2}
{(\hat{a}_{\rm C})^2-c^2}\frac{[S_{02}+c^2S_{20}]
[(\hat{a}_{\rm C})^2S_{20}-S_{02}]}{(S_{11})^2}~~;\qquad \\
\label{eq:vabcS}
&& (\hat{\sigma}_{\hat{b}_{\rm C}})^2=\frac1{n-2}\frac1n\frac
{(\hat{a}_{\rm C})^2+c^2}{(\hat{a}_{\rm C})^2-c^2}\left[(\hat{a}_{\rm C})^2
S_{20}-S_{02}\right]+(\overline{X})^2(\hat{\sigma}_{\hat{a}_{\rm C}})^2~~;
\qquad~ \\
\label{eq:vabcd}
&& (\hat{\sigma}_{\hat{b}_{\rm C}})^2=\frac1{n-2}\frac1n
[(\hat{a}_{\rm C})^2+c^2][1-\sgn(S_{11})\lambda_{\rm S}]+
(\overline{X})^2(\hat{\sigma}_{\hat{a}_{\rm C}})^2~;\qquad~ \\
\label{eq:vacS}
&& (\hat{\sigma}_{\hat{a}_{\rm C}})^2=\frac{(\hat{a}_{\rm C})^2}{n-2}\left[
\frac{D_{\rm S}}{(S_{11})^2}+\left(\frac{D_{\rm S}}{(S_{11})^2}+2-\frac
{S_{02}+(\hat{a}_{\rm C})^2S_{20}}{\hat{a}_{\rm C}S_{11}}\right)\right]~~;
\qquad
\end{lefteqnarray}
where $D_{\rm S}$ is the deviation
determinant, Eq.\,(\ref{eq:DS}), and
\begin{equation}
\label{eq:alaS}
\lambda_{\rm S}=\frac{S_{11}}{[S_{20}S_{02}]^{1/2}}~~;\qquad
c^2=\frac{S_{02}}{S_{20}}~~;
\end{equation}
is the regression line correlation
coefficient, and the above value of
the parameter, $c^2$, relates to
Eq.\,(\ref{eq:vabcd}).

The expression of the regression
line slope estimator,
Eq.\,(\ref{eq:acS}), coincides
with its counterpart determined
using the method of moments estimators
[e.g., Fuller, 1987, Chap.\,1, \S
1.3.2, Eq.\,(1.3.7) therein].
More specifically, the method of moments
estimators and the least square
estimators of $a_{\rm C}$ are the
same (e.g., Fuller, 1987, Chap.\,1, \S
1.3.3).

The expression of the regression
line slope variance estimator,
Eq.\,(\ref{eq:vacS}), is different
from its counterpart calculated in
an earlier attempt (FB92), due to
a different term within round brackets
on the right-hand side of
Eq.\,(\ref{eq:vacS}).   For a
formal demonstration, see
Appendix \ref{a:eqFB}.   After long
and strong work, it can be seen that
the expression of both the regression
line slope variance estimator determined in an
earlier attempt (FB92) and intercept variance
estimator expressed by Eq.\,(\ref
{eq:vabcS}), coincide with their
counterparts reported in specific
textbooks for structural models and
uncorrelated errors (e.g., Fuller,
1987, Chap.\,1, Sect.\,1.3).

In the special case, $c^2=1$, the
above results reduce to their
counterparts related to genuine
orthogonal regression, where the
lines of adjustment are perpendicular
to the regression line (Adcock, 1877,
1878; Pearson, 1901; Jones, 1937;
Teissier, 1948; Kermack and Haldane,
1950).

Turning back to the general case of
weighted residuals, but restricting
to the special case, $c^2=(\widetilde
{w_x})_{02}/(\widetilde{w_x})_{20}$,
the pseudo quadratic, Eq.\,(\ref{eq:pqc0}),
reduces to:
\begin{equation}
\label{eq:pqm}
a^2-c^2=0~~;
\end{equation}
which has the solutions (Kermack and Haldane,
1950; Y66):
\begin{equation}
\label{eq:amu}
\hat{a}_{\rm R}=\mp\left[\frac{(\widetilde{w_x})_{02}}{(\widetilde{w_x})_{20}}
\right]^{1/2}~~;
\end{equation}
and the regression line slope estimator is
obtained disregarding the parasite solution.
Then the substitution of Eq.\,(\ref{eq:amu})
into (\ref{eq:bar}) yields the regression
line intercept estimator, as:
\begin{equation}
\label{eq:bmu}
\hat{b}_{\rm R}=\widetilde{Y}-\hat{a}_{\rm R}\widetilde{X}~~;
\end{equation}
where the index, R, denotes the case under
discussion, with normal residuals (Y66).

The regression line slope and intercept variance
estimators are obtained by substitution of
Eq.\,(\ref{eq:amu}) into (\ref{eq:varaC}),
(\ref{eq:varbCd}) and (\ref{eq:varac2}).
After some algebra, the result is:
\begin{lefteqnarray}
\label{eq:varam}
&& (\hat{\sigma}_{\hat{a}_{\rm R}})^2=\frac2{n-2}\frac{(\widetilde{w_x})_{02}}
{(\widetilde{w_x})_{20}}~~; \\
\label{eq:varbm}
&& (\hat{\sigma}_{\hat{b}_{\rm R}})^2=\frac2{n-2}\frac{(\widetilde{w_x})_{02}}
{(\widetilde{w_x})_{00}}\{1-\sgn[(\widetilde{w_x})_{11}]\lambda_{w_x}\}+
(\widetilde{X})^2(\hat{\sigma}_{\hat{a}_{\rm R}})^2~~; \\
\label{eq:varam2}
&& (\hat{\sigma}_{\hat{a}_{\rm R}})^2=\frac{(\hat{a}_{\rm R})^2}{n-2}\left\{
\frac{D_{w_x}}{[(\widetilde{w_x})_{11}]^2}
+\left[\frac{D_{w_x}}{[(\widetilde
{w_x})_{11}]^2}+2-2\hat{a}_{\rm R}\frac{(\widetilde{w_x})_{20}}
{(\widetilde{w_x})_{11}}\right]\right\}~~;\qquad
\end{lefteqnarray}
where $D_{w_x}$ is a weighted deviation
determinant, Eq.\,(\ref{eq:Dwx}).

In the special case of homoscedastic data,
$w_{x_i}=w_x$, $1\le i\le n$, which
implies $Q_i=Q$, $Q=W,V,U,P$, via
Eqs.\,(\ref{eq:Wic2})-(\ref{eq:Pic2}),
%(\ref{eq:Uic2}), (\ref{eq:Pic2}),
and Eqs.\,(\ref{seq:Zp}) and
(\ref{eq:Qun}) hold.   Accordingly,
Eqs.\,(\ref{eq:amu})-(\ref{eq:varam2})
reduce to:
\begin{lefteqnarray}
\label{eq:amS}
&& \hat{a}_{\rm R}=\mp\left(\frac{S_{02}}{S_{20}}\right)^{1/2}~~; \\
\label{eq:bmS}
&& \hat{b}_{\rm R}=\overline{Y}-\hat{a}_{\rm R}\overline{X}~~; \\
\label{eq:vaamS}
&& (\hat{\sigma}_{\hat{a}_{\rm R}})^2=\frac2{n-2}\frac{S_{02}}{S_{20}}~~; \\
\label{eq:vabmS}
&& (\hat{\sigma}_{\hat{b}_{\rm R}})^2=\frac2{n-2}\frac{S_{02}}n[1-\sgn
(S_{11})\lambda_{\rm S}]+
(\overline{X})^2(\hat{\sigma}_{\hat{a}_{\rm R}})^2~~;\qquad~ \\
\label{eq:vamS}
&& (\hat{\sigma}_{\hat{a}_{\rm R}})^2=\frac{(\hat{a}_{\rm R})^2}{n-2}\left\{
\frac{D_{\rm S}}{(S_{11})^2}+\left[\frac{D_{\rm S}}{(S_{11})^2}+2-2
\hat{a}_{\rm R}\frac{S_{20}}{S_{11}}\right]\right\}~~;\qquad
\end{lefteqnarray}
where $D_{\rm S}$ is the deviation
determinant, Eq.\,(\ref{eq:DS}).

The expression of the regression
line slope variance estimator,
Eq.\,(\ref{eq:vamS}), is different
from its counterpart calculated in
an earlier attempt (FB92), due to
a different term within square brackets
on the right-hand side of
Eq.\,(\ref{eq:vamS}).   For a
formal demonstration, see
Appendix \ref{a:eqFB}.

\subsection{Extension to structural models}
\label{exts}

A nontrivial question is to what extent the
above results, valid for functional models,
can be extended to structural models.   In
general, assumptions related to structural
models are different from their counterparts
related to functional models (e.g., Buonaccorsi,
2006; 2010, Chap.\,6, \S 6.4.5) but, on the
other hand, they could coincide for a special
subclass.   In any case, whatever different
assumptions and models can be made with regard
to structural and functional models, results
from the former are expected to tend to their
counterparts from the latter when the intrinsic
scatter is negligible with respect to the
instrumental scatter.
It is worth noticing that most
work on linear regression by astronomers
involves the situation where both intrinsic
scatter and heteroscedastic data are present
(e.g., AB96; Tremaine et al., 2002; Kelly,
2007).

In structural models, a true 
point, ${\sf P}_i^\ast(x_i^\ast,y_i^\ast)$, lying on the
(true) regression line, is shifted by intrinsic
scatter to an actual point, ${\sf P}_{{\rm S}i}^\ast
(x_{{\rm S}i},y_{{\rm S}i})$, outside the
regression line.   This last, in turn, is
shifted by instrumental scatter to an observed
point, ${\sf P}_i(X_i,Y_i)$.

The coordinates of observed and actual points,
according to Eq.\,(\ref{seq:erm}), and the
coordinates of actual and true points, are
assumed to be related as:
\begin{lefteqnarray}
\label{eq:csiF}
&& (\xi_{{\rm F}_z})_i=Z_i-z_{\rm{S}i}~~;\quad Z=X,Y~~;\quad1\le i\le n~~; \\
\label{eq:csiS}
&& (\xi_{{\rm S}_z})_i=z_{{\rm S}i}-
z_i^\ast~~;\quad z=x,y;\quad1\le i\le n~~;
\end{lefteqnarray}
where the random variables, $(\xi_{{\rm F}_z})_i$,
$(\xi_{{\rm S}_z})_i$, obey the distributions,
$f_{{\rm F}_zi}[(\xi_{{\rm F}_z})_i]$,
$f_{{\rm S}_zi}[(\xi_{{\rm S}_z})_i]$, respectively.
While the former distribution is necessarily Gaussian,
depending only on mesaurements processes, the
latter distribution may be different, depending
on a larger variety of processes.

With regard to true points, the substitution of
Eq.\,(\ref{eq:csiS}) into (\ref{eq:riS}) yields:
\begin{equation}
\label{eq:riSc}
y_i^\ast+(\xi_{{\rm S}_y})_i=a[x_i^\ast+(\xi_{{\rm S}_x})_i]+b+\epsilon_i~~;
\end{equation}
which, owing to Eq.\,(\ref{eq:rlf}), reduces to:
\begin{equation}
\label{eq:epi}
\epsilon_i=(\xi_{{\rm S}_y})_i-a(\xi_{{\rm S}_x})_i~~;
\end{equation}
where the contribution of each variable to
the intrinsic scatter is explicitly expressed.
For the true regression line i.e. fixed slope,
$a$, a null expectation value of $\epsilon_i$
necessarily implies null expectation values of
$(\xi_{{\rm S}_x})_i$ and $(\xi_{{\rm S}_y})_i$,
$1\le i\le n$, and vice versa.

With regard to observed points, the substitution of
Eq.\,(\ref{eq:csiF}) into (\ref{eq:riS}) yields:
\begin{equation}
\label{eq:riFc}
Y_i-(\xi_{{\rm F}_y})_i=a[X_i-(\xi_{{\rm F}_x})_i]+b+\epsilon_i~~;
\end{equation}
which, using Eq.\,(\ref{eq:epi}), takes the
form:
\begin{equation}
\label{eq:riFd}
Y_i=aX_i+b+[(\xi_{{\rm S}_y})_i+(\xi_{{\rm F}_y})_i]-a
[(\xi_{{\rm S}_x})_i+(\xi_{{\rm F}_x})_i]~~;
\end{equation}
where the contribution of each variable to
the intrinsic scatter is explicitly expressed.

A special subclass of structural models is
defined according to the following assumptions.
\begin{description}
\item[(a)]
The random variables, $(\xi_{{\rm F}_z})_i$,
$(\xi_{{\rm S}_z})_i$, $z=x,y$, $1\le i\le n$,
are independent.
\item[(b)]
The distributions, $f_{{\rm S}_zi}[(\xi_{{\rm S}_z})_i]$, 
$z=x,y$, $1\le i\le n$, related to the
intrinsic scatter, are Gaussian and corresponding
expectation values are null.
\item[(c)]
The sum of variances related to the distributions,
$f_{{\rm F}_zi}[(\xi_{{\rm F}_z})_i]$,
$f_{{\rm S}_zi}[(\xi_{{\rm S}_z})_i]$, $z=x,y$,
$1\le i\le n$, maintains constant as:
\begin{equation}
\label{eq:sUZi}
[(\sigma_{{\rm F}_z})_i]^2+[(\sigma_{{\rm S}_z})_i]^2=
[(\sigma_z)_i]^2=({\rm const}_z)_i;~z=x,y;~1\le i\le n.
\end{equation}
\end{description}
Owing to assumption (b), the related
distributions read:
\begin{lefteqnarray}
\label{eq:sUZi}
&& f_{{\rm U}_zi}[(\xi_{{\rm U}_z})_i]=\frac1{\sqrt{2\pi}(\sigma_{{\rm U}_z})_i}
\exp\left\{-\frac{[(\xi_{{\rm U}_z})_i]^2}{2[(\sigma_{{\rm U}_z})_i]^2}\right
\}~~; \nonumber \\
&& %\phantom{f_{{\rm U}i}[(\xi_{{\rm U}_z})_i]=}
{\rm U}={\rm F}, {\rm S}~~;\qquad z=x,y~~;\qquad1\le i\le n~~;
\end{lefteqnarray}
with regard to the random variables, $(\xi_{{\rm U}_z})_i$.

The random variables:
\begin{lefteqnarray}
\label{eq:csi}
&& (\xi_z)_i=(\xi_{{\rm F}_z})_i+(\xi_{{\rm S}_z})_i=Z_i-z_i^\ast;~
Z=X,Y;~z=x,y;~1\le i\le n; 
\end{lefteqnarray}
via assumption (a) obey the distribution:
\begin{equation}
\label{eq:fzi}
f_{zi}[(\xi_z)_i]=f_{{\rm F}_zi}[(\xi_{{\rm F}_z})_i]f_{{\rm S}_zi}
[(\xi_{{\rm S}_z})_i]~~;
\end{equation}
which, using a theorem of statistics,
via assumptions
(a), (b), is also Gaussian, expressed as:
\begin{lefteqnarray}
\label{eq:fzi2}
&& f_{z_i}[(\xi_z)_i]=\frac1{\sqrt{2\pi}(\sigma_z)_i}\exp\left\{-\frac
{[(\xi_z)_i]^2}{2[(\sigma_z)_i]^2}\right\}~~; \\
\label{eq:szi}
&& [(\sigma_z)_i]^2=[(\sigma_{{\rm F}_z})_i]^2+[(\sigma_{{\rm S}_z})_i]^2~~;
\quad~~z=x,y~~;\quad1\le i\le n~~;
\end{lefteqnarray}
accordingly, residuals obey a Gaussian distribution
and the weights, $w_{x_i}$, $w_{y_i}$, $1\le i\le n$,
remain unchanged via assumption (c).   Then, for
a selected regression estimator, the regression line slope
and intercept variance estimators are independent of the
amount of instrumental and intrinsic scatter, including
the limit of null intrinsic scatter (functional models)
and null instrumental scatter (extreme structural models).
In this view, the whole subclass of structural models
under consideration could be related to functional
modelling (Carroll et al., 2006, Chap.\,2, \S 2.1).

\section{An example of astronomical application}
\label{apfm}
\subsection{Astronomical introduction}
\label{asin}

\noindent\noindent

With regard to stellar populations,
the dependence of oxygen abundance
on iron abundance, or [O/H]-[Fe/H]
relation%
\footnote{For a generic nuclide, N,
the logarithmic number abundance is
defined as [N/H$]=\log($N/H$)-\log
($N/H$)_\odot$, normalized to hydrogen,
H, and to solar abundance, (N/H$)_\odot$.
The related mass abundance is defined as
$\phi_{\rm N}=Z_{\rm N}/(Z_{\rm N})_
\odot$.   The relation: $\log\phi_
{\rm N}=[$N/H] holds to a good extent
(Pagel, 1989; Malinie et al., 1993;
Rocha-Pinto and Maciel, 1996; Caimmi,
2007).},
has been deeply investigated during
the last decade (e.g., Carretta et al.,
2000; Israelian et al., 2001a,b;
Barbuy et al. (eds.), 2001; Jonsell
et al., 2005; Fulbright et al., 2005;
Garcia Perez et al., 2006; Melendez
et al., 2006; Fabbian et al., 2009,
hereafter quoted as Fal09; Rich and
Boesgaard, 2009, hereafter quoted
as RB09; Schmidt et al., 2009, hereafter
quoted as Sal09).

It has been
realized that the [O/H]-[Fe/H] relation
is strongly dependent on both the
selection of the spectroscopic oxygen
lines and the choice of the atmosphere
model.
The discrepancy due to using different
methods and different models remains
large, and no general consensus on the
best choice still exists.   For further
details refer to an earlier attempt (Caimmi,
2010).

Oxygen is the most abundant metal%
\footnote{In astrophysical language,
all elements heavier than helium are
called ``metals''.}
in the universe, but it is more difficult
than iron to detect.   The population
of available samples where oxygen
abundances are directly determined,
does not exceed a few hundreds at
most (e.g., Ramirez et al., 2007;
Melendez et al., 2008; RB09; Fal09;
Sal09).   Oxygen abundance in larger
samples may be deduced by use of an
inferred [O/H]-[Fe/H] relation.

According to the stellar evolution
theory, oxygen is produced only via
type II supernovae (SnII), characterized
by massive $(m\appgeq8$m$_\odot)$
progenitors.   On the contrary,
iron is produced also via type
Ia supernovae (SnIa), where
a white dwarf $(m<m_{\rm C})$
attains the Chandrasekhar
limit $(m_{\rm C}\approx1.4$m$_\odot)$
due to mass accretion from
a close red giant companion,
where white dwarfs are related
to low-mass $(m\appleq8$m$_\odot)$
progenitors.   During the lifetime
of primeval SnII progenitors,
$\tau\appleq0.1$Gyr, a linear
[O/H]-[Fe/H] relation is expected,
while the contrary holds at later
times unless SnII contribution to
iron production remains dominant.
Three samples shall be used for
bivariate least squares linear
regression, namely RB09,
Fal09, Sal09, where the denomination
comes from related parent papers.

The RB09 sample $(N=49)$ is made of a homogeneous
subsample $(N=24)$ of metal-poor ($-3.5<[$Fe/H$]<
-2.2$) stars, and a non homogeneous subsample
$(N=25)$ of higher-metallicity ($-3.1<[$Fe/H$]<
-0.5$) stars.   In both cases, the stellar
population remains unspecified and oxygen abundance
has been determined using standard local thermodynamical
equilibrium (LTE) one-dimensional hydrostatic model
atmospheres.
Standard deviations are provided for each star, where
typical values
are $\sigma_{\rm[Fe/H]}=\sigma_{\rm[O/H]}=0.15$.
For further details refer to the parent paper (RB09).

The Fal09 sample $(N=44)$ is made of halo stars ($-3.3<
[$Fe/H$]<-1.0$) where oxygen abundance has been determined
using three different methods involving (a) LTE one-dimensional
hydrostatic model atmospheres; (b) three-dimensional
hydrostatic model atmospheres in absence of LTE with
no account taken of the inelastic collisions via neutral
H atoms $(S_{\rm H}=0)$; (c)
three-dimensional hydrostatic model atmospheres in absence
of LTE with due account taken of the inelastic collisions
via neutral H atoms $(S_{\rm H}=1)$.
Standard deviations are not reported for each star,
but typical values are mentioned to be
$\sigma_{\rm[Fe/H]}=\sigma_{\rm[O/H]}=0.15$.
For further details refer to the parent paper (Fal09).

The RB09 and Fal09 samples have $N=11$ (necessarily
halo) stars in common, where the values assumed for
effective temperature and surface gravity have been
determined using different methods, yielding different
values for each star.   For further details refer to
an earlier attempt (Caimmi, 2010).

The Sal09 sample $(N=63)$ is made of cool (late K
and M) dwarfs $(-1.8<[$Fe/H$]<+0.2)$ where oxygen
abundance has been determined by use of the
\linebreak
$\gamma$
$R_2$ 0|0 TiO band at 7054 \AA~ combined with previously
derived abundances of Ti and Fe.
Standard deviations are provided for each star, where
typical values may be estimated
as $\sigma_{\rm[Fe/H]}=\sigma_{\rm[O/H]}=0.15$.
A single star, LHS 185, is mentioned (Sal09, Table 1
therein) but excluded from further analysis.
For further details refer to the parent paper (Sal09).

In conclusion, the Fal09 sample is made of
homoscedastic data, while the remaining
RB09 and Sal09 samples are made of
heteroscedastic data.   To a first extent,
the latter may be considered as made of
homoscedastic data where standard deviations
are approximated to related typical values.
Under the further assumption that intrinsic
scatter is negligible with respect to
instrumental scatter, i.e. functional models,
the general results of section
\ref{fumo} may be particularized
to the case under discussion, where errors
in [Fe/H] and [O/H] may be considered as
uncorrelated $(r_i=0$, $1\le i\le n)$ to
a good extent.

\subsection{Statistical results}
\label{stre}

\indent\indent

The [O/H]-[Fe/H] empirical relations are
interpolated using the regression models,
G, Y, X, O, R, for heteroscedastic data
(RB09 and Sal09 samples) and Y, X, O, R,
for homoscedastic data (Fal09 sample,
cases LTE, SH0, SH1) and heteroscedastic
data where instrumental scatters are
taken equal to related typical values,
$\sigma_{\rm [Fe/H]}=0.15$, $\sigma_{\rm
[O/H]}=0.15$, for both FB09 and Sal09
samples.   Slope and intercept estimators
together with related dispersion estimators
are listed in Tables \ref{t:fog} and
\ref{t:fo} for heteroscedastic and
homoscedastic data, respectively.   Also
listed are values of slope and intercept
dispersion estimators by earlier attempts
(Y66; FB92) for comparison with their
counterparts calculated in the current
paper (CRS).

Owing to high difficulties
intrinsic to the determination of slope
and intercept dispersion estimators in
the general case, related calculations
were not performed in dealing with G models
and only approximate expressions (Y66),
Eqs.\,(\ref{eq:vaap2}) and (\ref{eq:vabp2}),
were used.
\begin{table}
\caption{Regression line slope and intercept
estimators, $\hat{a}$ and $\hat{b}$, and
related dispersion estimators, $\hat{\sigma}_
{\hat{a}}$, and $\hat{\sigma}_{\hat{b}}$,
for heteroscedastic models, G, Y, X, O, R, applied
to the [O/H]-[Fe/H] empirical relation
deduced from the following samples (from up to
down): RB09, Sal09. 
Values related to different slope and intercept
dispersion estimators (Y66) are also reported
for comparison with current results
(CRS).   For G models, slope and intercept
dispersion estimators were not evaluated
in the present attempt.   For Y models, different slope or
intercept dispersion estimators yield coinciding values,
as expected.}
\label{t:fog}
\begin{center}
\begin{tabular}{cccccccc}
\hline
%\multicolumn{1}{c|}{model} &
%\multicolumn{4}{c|}{RN09} &
%\multicolumn{6}{c|}{Fal09} \\
% & & & & & & & & \\
\multicolumn{1}{c}{$m$} &
\multicolumn{1}{c}{$\hat{a}$} &
\multicolumn{2}{c}{$\hat{\sigma}_{\hat{a}}$} &
%\multicolumn{1}{c}{$(\hat{\sigma}_{\hat{a}})_{\rm Y66}$} &
\multicolumn{1}{c}{$\hat{b}$} &
\multicolumn{2}{c}{$\hat{\sigma}_{\hat{b}}$} &
%\multicolumn{1}{c}{$(\hat{\sigma}_{\hat{b}})_{\rm Y66}$}  \\
\multicolumn{1}{c}{sample} \\
  &        &   CRS  &  Y66   &           &   CRS  &  Y66   \\

\hline

  &        &        &        &           &        &        &       \\
G & 0.7279 &        & 0.0294 & $+$0.0043 &        & 0.0672 & RB09  \\
Y & 0.6714 & 0.0314 & 0.0314 & $-$0.1121 & 0.0675 & 0.0675 &       \\
X & 0.7305 & 0.0290 & 0.0279 & $+$0.0316 & 0.0735 & 0.0712 &       \\
O & 0.6964 & 0.0278 & 0.0271 & $-$0.0512 & 0.0707 & 0.0689 &       \\
R & 0.7050 & 0.0282 & 0.0272 & $-$0.0305 & 0.0725 & 0.0693 &       \\
  &        &        &        &           &        &        &       \\
G & 0.6383 &        & 0.0435 & $+$0.0619 &        & 0.0251 & Sal09 \\
Y & 0.6167 & 0.0398 & 0.0398 & $+$0.0439 & 0.0198 & 0.0198 &       \\
X & 0.8652 & 0.0829 & 0.0664 & $+$0.3080 & 0.0673 & 0.0575 &       \\
O & 0.6355 & 0.0637 & 0.0541 & $+$0.1461 & 0.0525 & 0.0469 &       \\
R & 0.6927 & 0.0700 & 0.0560 & $+$0.1864 & 0.0549 & 0.0485 &       \\
\hline                            
\end{tabular}                     
\end{center}                      
\end{table}                       
\begin{table}
\caption{Regression line slope and intercept
estimators, $\hat{a}$ and $\hat{b}$, and
related dispersion estimators, $\hat{\sigma}_
{\hat{a}}$, and $\hat{\sigma}_{\hat{b}}$,
for homoscedastic models, Y, X, O, R, applied
to the [O/H]-[Fe/H] empirical relation
deduced from the following samples (from up 
to down): RB09, Sal09, Fal09, cases LTE, SH0, 
SH1.   Values related to different slope and 
intercept dispersion estimators (Y66, FB92) are 
also reported for comparison with current results
(CRS).   For Y models, different slope or
intercept dispersion estimators yield coinciding
values, as expected.}
\label{t:fo}
\begin{center}
\begin{tabular}{ccccccccc} \hline
%\multicolumn{1}{c|}{model} &
%\multicolumn{4}{c|}{RN09} &
%\multicolumn{6}{c|}{Fal09} \\
% & & & & & & & & \\
\multicolumn{1}{c}{$m$} &
\multicolumn{1}{c}{$\hat{a}$} &
\multicolumn{3}{c}{$\hat{\sigma}_{\hat{a}}$} &
%\multicolumn{1}{c}{$(\hat{\sigma}_{\hat{a}})_{\rm FB92}$} &
%\multicolumn{1}{c}{$(\hat{\sigma}_{\hat{a}})_{\rm Y66}$} &
\multicolumn{1}{c}{$\hat{b}$} &
\multicolumn{2}{c}{$\hat{\sigma}_{\hat{b}}$} &
%\multicolumn{1}{c}{$(\hat{\sigma}_{\hat{b}})_{\rm Y66}$}  \\
\multicolumn{1}{c}{sample} \\
% &        &        &        &        &           &        &                \\
  &        &  CRS   &  FB92  &  Y66   &           &  CRS   &  Y66   &       \\
% &        &        &        &        &           &        &                \\
\hline
  &        &        &        &        &           &        &        &       \\
Y & 0.6917 & 0.0317 & 0.0317 & 0.0317 & $-$0.0766 & 0.0737 & 0.0737 & FB09  \\
X & 0.7600 & 0.0348 & 0.0349 & 0.0332 & $+$0.0742 & 0.0806 & 0.0773 &       \\
O & 0.7143 & 0.0331 & 0.0327 & 0.0319 & $-$0.0268 & 0.0766 & 0.0741 &       \\
R & 0.7251 & 0.0336 & 0.0332 & 0.0321 & $-$0.0030 & 0.0778 & 0.0746 &       \\
  &        &        &        &        &           &        &        &       \\
Y & 0.5868 & 0.0461 & 0.0461 & 0.0461 & $+$0.0908 & 0.0338 & 0.0338 & Sal09 \\
X & 0.8077 & 0.0635 & 0.0637 & 0.0541 & $+$0.2011 & 0.0430 & 0.0397 &       \\
O & 0.6476 & 0.0526 & 0.0509 & 0.0468 & $+$0.1212 & 0.0363 & 0.0343 &       \\
R & 0.6885 & 0.0562 & 0.0541 & 0.0479 & $+$0.1416 & 0.0381 & 0.0352 &       \\
  &        &        &        &        &           &        &        &       \\
Y & 0.8961 & 0.0303 & 0.0303 & 0.0303 & $+$0.5476 & 0.0663 & 0.0663 & Fal09 \\ 
X & 0.9381 & 0.0317 & 0.0318 & 0.0310 & $+$0.6366 & 0.0693 & 0.0678 & (LTE) \\ 
O & 0.9150 & 0.0311 & 0.0310 & 0.0305 & $+$0.5877 & 0.0680 & 0.0666 &       \\ 
R & 0.9168 & 0.0312 & 0.0310 & 0.0305 & $+$0.5916 & 0.0681 & 0.0667 &       \\ 
  &        &        &        &        &           &        &        &       \\
Y & 1.2261 & 0.0432 & 0.0432 & 0.0432 & $+$0.8717 & 0.0945 & 0.0945 & Fal09 \\
X & 1.2884 & 0.0454 & 0.0454 & 0.0443 & $+$1.0037 & 0.0991 & 0.0968 & (SH0) \\
O & 1.2640 & 0.0448 & 0.0445 & 0.0436 & $+$0.9519 & 0.0978 & 0.0953 &       \\
R & 1.2569 & 0.0445 & 0.0443 & 0.0435 & $+$0.9369 & 0.0973 & 0.0950 &       \\
  &        &        &        &        &           &        &        &       \\
Y & 1.0492 & 0.0341 & 0.0341 & 0.0341 & $+$0.6518 & 0.0745 & 0.0745 & Fal09 \\
X & 1.0946 & 0.0356 & 0.0356 & 0.0348 & $+$0.7479 & 0.0777 & 0.0761 & (SH1) \\
O & 1.0732 & 0.0350 & 0.0349 & 0.0343 & $+$0.7027 & 0.0765 & 0.0750 &       \\
R & 1.0716 & 0.0350 & 0.0348 & 0.0343 & $+$0.6993 & 0.0764 & 0.0750 &       \\
\hline                            
\end{tabular}                     
\end{center}                      
\end{table}                       
The regression line slope and intercept
estimators are calculated using
Eqs.\,(\ref{eq:pcc}) and (\ref{eq:bar}),
respectively.   For the remaining models,
the regression line slope and intercept
estimators and related dispersion estimators
are calculated using
Eqs.\,(\ref{eq:aY}), (\ref{eq:bY}),
(\ref{eq:varaY}), (\ref{eq:varbY}), and
(\ref{eq:aYu}), (\ref{eq:bYu}),
(\ref{eq:vaYu}), (\ref{eq:vbYu}), case Y,
heteroscedastic and homoscedastic data,
respectively;
Eqs.\,(\ref{eq:aX}), (\ref{eq:bX}),
(\ref{eq:varaX}), (\ref{eq:varbX}), and
(\ref{eq:aXu}), (\ref{eq:bXu}),
(\ref{eq:vaXu}), (\ref{eq:vbXu}), case X,
heteroscedastic and homoscedastic data,
respectively;
Eqs.\,(\ref{eq:acu}), (\ref{eq:bcu}),
(\ref{eq:varaC}), (\ref{eq:varbC}), and
(\ref{eq:acS}), (\ref{eq:bcS}),
(\ref{eq:vaacS}), (\ref{eq:vabcS}), all
related to the special value, $c^2=1$
(genuine orthogonal regression), case O,
heteroscedastic and homoscedastic data,
respectively;
Eqs.\,(\ref{eq:amu}), (\ref{eq:bmu}),
(\ref{eq:varam}), (\ref{eq:varbm}), and
(\ref{eq:amS}), (\ref{eq:bmS}),
(\ref{eq:vaamS}), (\ref{eq:vabmS}), case R,
heteroscedastic and homoscedastic data,
respectively.

\begin{figure*}[t]
\begin{center}      
\includegraphics[scale=0.8]{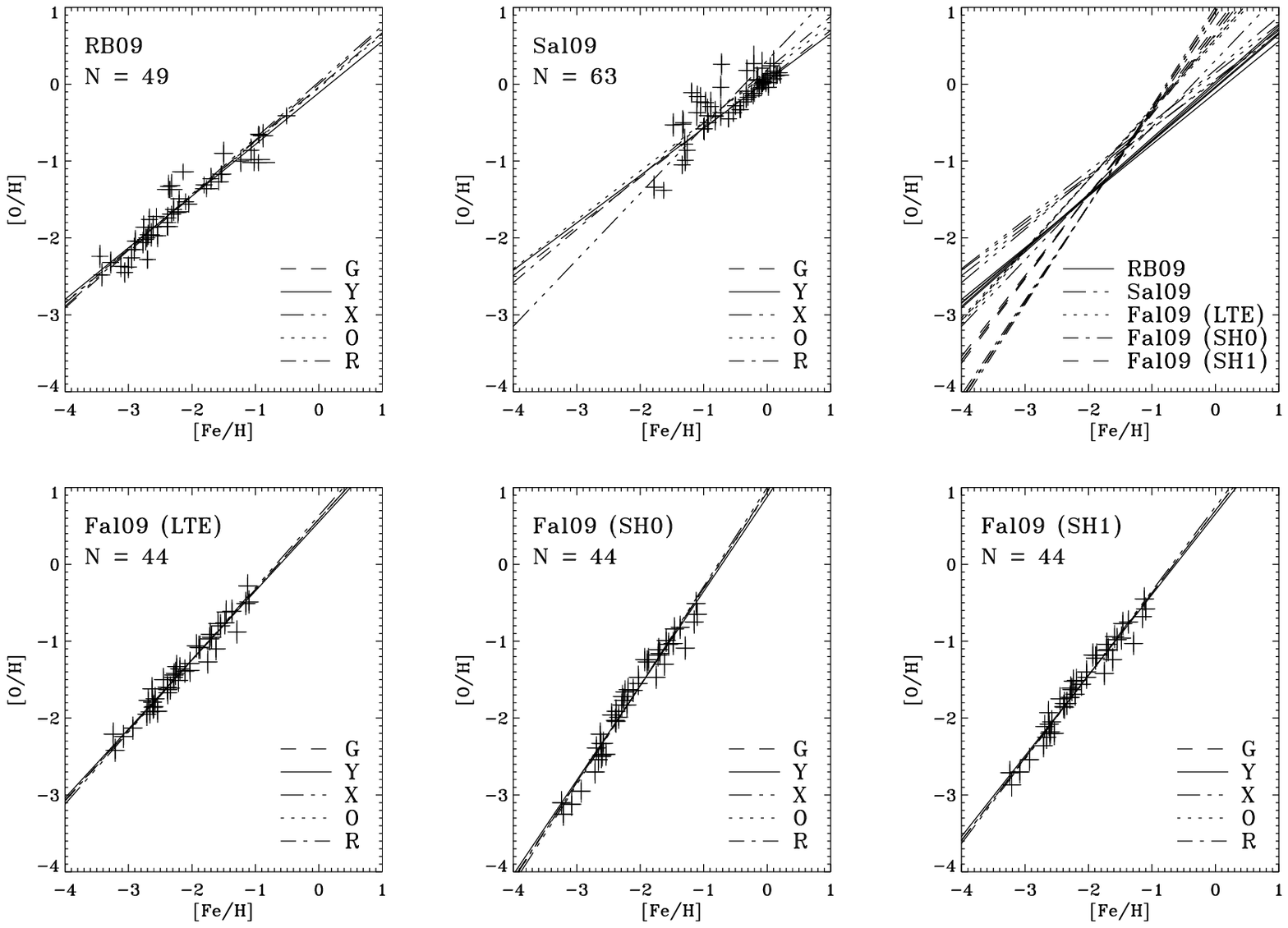}                      
\caption{Regression lines related to
[O/H]-[Fe/H] empirical relations
deduced from two samples with heteroscedastic
data, RB09 and Sal09, and three samples with
homoscedastic data (using the computer code
for heteroscedastic data), Fal09, cases LTE, SH0,
and SH1, indicated on each panel together
with related population and model captions.
The regression lines
related to five different methods are shown for
each sample on the top right panel.   For further
details refer to the text.}
\label{f:fog}
\end{center}       
\end{figure*}

\begin{figure*}[t]
\begin{center}      
\includegraphics[scale=0.8]{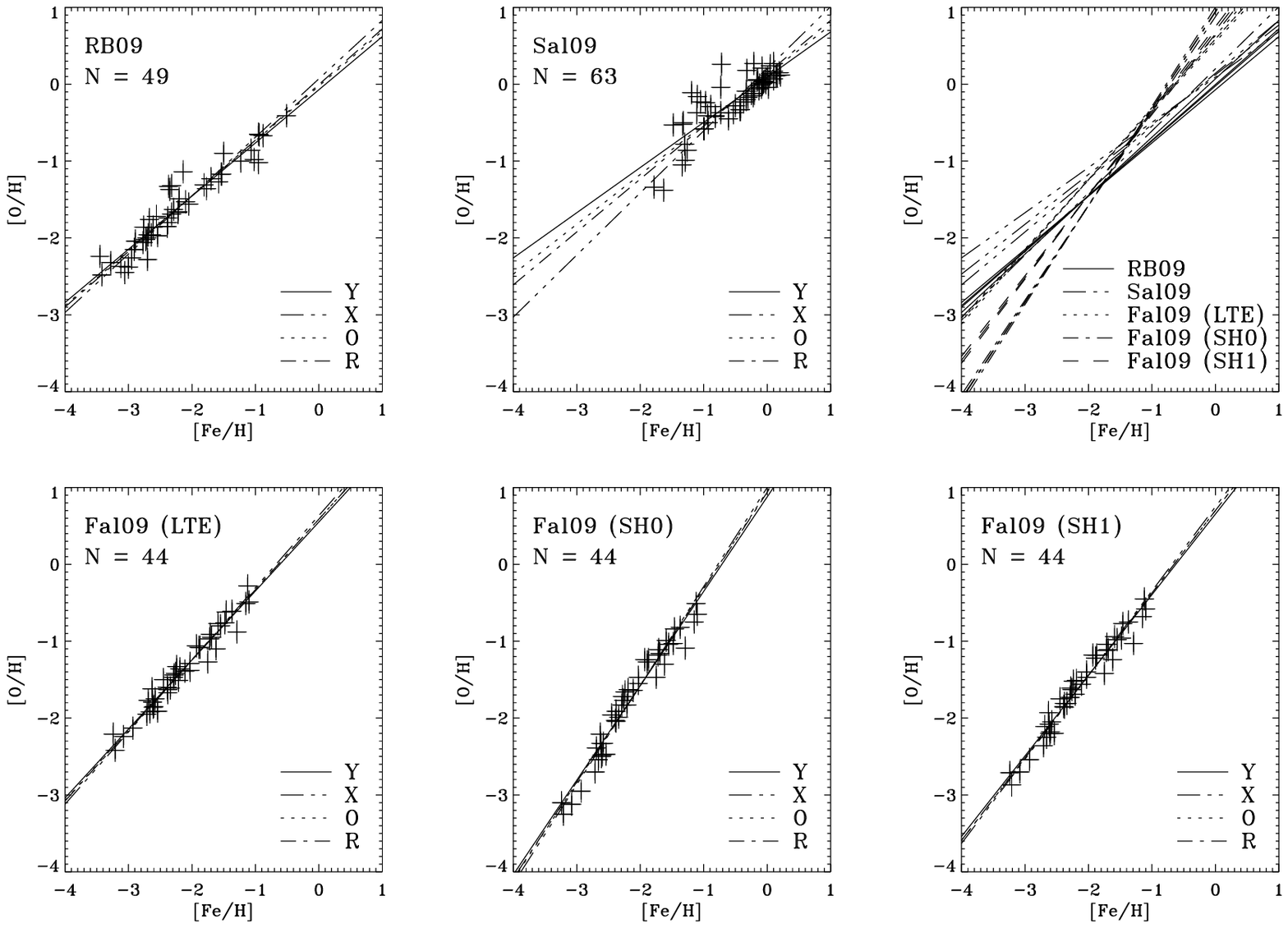}                      
\caption{Regression lines related to
[O/H]-[Fe/H] empirical relations
deduced from two samples with heteroscedastic
data (with instrumental scatters taken equal
to related typical values), RB09 and Sal09,
and three samples with
homoscedastic data, Fal09, cases LTE, SH0,
and SH1, indicated on each panel together
with related population and model captions.
The regression lines
related to four different methods are shown for
each sample on the top right panel.   For further
details refer to the text.}
\label{f:fo}
\end{center}       
\end{figure*}

The regression lines determined by
use of the above mentioned methods
are plotted in Figs.\,\ref{f:fog}
and \ref{f:fo} for heteroscedastic
and homoscedastic data, respectively,
where sample denomination and
population are indicated on each panel
together with model captions.
Homoscedastic data
are conceived as a special case of
heteroscedastic data in Fig.\,\ref
{f:fog} to test the computer code, which
is different for heteroscedastic and
homoscedastic data.   It can be seen
that lower panels of Figs.\,\ref{f:fog}
and \ref{f:fo} coincide, and the regression
lines related to models G and O in
lower panels of Figs.\,\ref{f:fog} and
\ref{f:fo} also coincide, as expected.
The whole set of regression lines for
all methods and all samples is shown
in the upper right panel of Figs.\,\ref
{f:fog} and \ref{f:fo}.

An inspection of Tables \ref{t:fog}-\ref{t:fo}
and Figs.\,\ref{f:fog}-\ref{f:fo} discloses
the following.
\begin{description}
\item[(1)]
Either of the inequalities (Ial90):
\begin{leftsubeqnarray}
\slabel{eq:disa}
&& \hat{a}_{\rm Y}<\hat{a}_{\rm O}<\hat{a}_{\rm R}<1<\hat{a}_{\rm X}~~;
\qquad S_{11}>0~~; \\
\slabel{eq:disb}
&& \hat{a}_{\rm Y}<1<\hat{a}_{\rm R}<\hat{a}_{\rm O}<\hat{a}_{\rm X}~~;
\qquad S_{11}>0~~;
\label{seq:dis}
\end{leftsubeqnarray}
holds for both heteroscedastic and
homoscedastic data.   In addition,
$\hat{a}_{\rm Y}<\hat{a}_{\rm G}<
\hat{a}_{\rm X}$
for heteroscedastic data, but a
counterexample is provided in an
earlier attempt (Y66).
\item[(2)]
Slope and intercept estimators by different
methods are consistent within $\mp\sigma$
for samples with lower dispersion (FB09,
Fal09), while the contrary holds for samples
with higher dispersion (Sal09), with regard
to both heteroscedastic and homoscedastic
data.
\item[(3)]
Slope and intercept dispersion estimators
coincide with their counterparts related 
to earlier attempts for both heteroscedastic
(Y66) and homoscedastic (Y66; FB92) data,
in the special case of Y models.  For the
other models, the approximations exploited
in an earlier attempt (Y66) make lower
limits with respect to current results,
while an alternative expression of the
slope dispersion estimator (FB92) yields
slightly different results.
\item[(4)]
Systematic variations due to different sample data
are dominant with respect to the instrumental
scatter.
\end{description}
In conclusion, regression lines deduced
from different sample data represent
correct (from the standpoint of regression
models considered in the current attempt)
[O/H]-[Fe/H] relations, but
no definitive choice can be made until
systematic errors due to different methods
and/or spectral lines in determining oxygen
abundance, are alleviated.

\section{Discussion}\label{disc}

\noindent\noindent

For an assigned sample, structural models
belonging to a special subclass
are indistinguishable from functional models
provided restrictions (a)-(c) hold as
outlined in subsection \ref{exts}.
Accordingly, the results of the
current paper also apply to structural
models of the kind considered.   The
expression of regression line slope and
intercept estimators and related variance
estimators in terms of weighted
deviation traces, for heteroscedastic
and homoscedastic data, makes a first step
towards a unified formalism of 
bivariate least squares linear regression.
The bisector method has not
been dealt with in earlier attempts related
to functional models (Y66; Y69), but only
in later investigations involving extreme structural
models (Ial90; FB92).   For this reason, the
bisector method has not been considered in the
current paper.

Exact expressions of regression line slope
and intercept estimators and related variance
estimators have been determined from general
formulae (Y69) in the limit of generalized
%
%errors in $X$ uncorrelated and equally weighted
%with respect to errors in $Y$
%
orthogonal regression i.e. $(\sigma_{yy})_i/
(\sigma_{xx})_i=c^2$, $1\le i\le n$.
It is noteworthy
that a constant variance ratio, $c^2$, for
all data points, does not
necessarily imply equal variances,
$(\sigma_{xx})_i=\sigma_{xx}=$ const,
$(\sigma_{yy})_i=\sigma_{yy}=$ const,
$1\le i\le n$.   While regression line
slope and intercept estimators attain
a coinciding expression in different
attempts (Y66; Y69; Ial90; FB92) with
regard to a fixed model, the results
of the current paper show that the
contrary holds for related variance
estimators.

Approximate expressions provided in
earlier attempts (Y66; Y69) make (at
least in computed cases) a lower limit
to their exact counterparts, as shown
in Tables \ref{t:fog}-\ref{t:fo}.   On
the other hand, alternative expressions
given in a later investigation, restricted
to regression line slope variance estimators,
yield different results (FB92).

The above mentioned discrepancy could be
explained in different ways, namely: (1)
calculations performed in the current
paper were checked and repeated twice or
more, but something wrong cannot be
excluded; (2) the expression of the
regression line slope variance estimator
determined in the current paper, is
approximate intead of exact, contrary to
what reported in the parent paper (Y69);
(3) the expression of the regression line
slope variance estimator used in an
earlier attempt [FB92, Eq.\,(14) therein]
for structural models with normal residuals
and dominant intrinsic scatter, is
approximate instead of exact; (4) for
%
%errors in $X$ equally weighted and equally
%correlated with respect to errors in $Y$,
%
generalized orthogonal regression,
the expression of the regression line slope
variance estimator is different for
functional (Y69) and structural (FB92)
models even if, in the latter case,
residuals obey a Gaussian distribution
and the intrinsic scatter is dominant with
respect to the instrumental scatter; (5)
with regard to the regression line slope
variance estimator, the method of partial
differentiation, used in an earlier attempt
(Y69) and in the current paper, yields
different results with respect
to the method of moments estimators
[e.g., Fuller, 1987, Chap.\,1, \S
1.3.2, Eq.\,(1.3.7) therein].

It is well known that the regression line
slope and intercept estimators for Y models,
Eqs.\,(\ref{eq:aYu}) and (\ref{eq:bYu}),
are biased (e.g., Fuller, 1987, Chap.\,1,
\S 1.1.1; Carroll et al., 2006, Chap.\,3,
\S 3.2; Buonaccorsi, 2010, Chap.\,4, \S 4.4).
Biases can be explicitly expressed in the
special case of homoscedastic models where
random variables obey Gaussian distributions.
More specifically, the condition $1-\rho_{20}
\ll1$ ensures bias effects are negligible,
where $\rho_{20}$ is the reliability ratio:
\begin{equation}
\label{eq:rho20}
\rho_{20}=\frac{S_{20}}{S_{20}+(n-1)\sigma_{xx}}~~;
\end{equation}
which implies $0\le\rho_{20}\le1$.   For
further details refer to specific monographies
(e.g., Fuller, 1987, Chap.\,1, \S 1.1.1;
Carroll et al., 2006, Chap.\,3, \S 3.2.1;
Buonaccorsi, 2010, Chap.\,4, \S 4.4).
Accordingly, the reliability ratio cannot
exceed unity, or in other words the regression
line slope estimator, Eq.\,(\ref{eq:aYu}),
is biased towards zero, as clearly shown in
current literature (e.g., Kelly, 2007).
Following a similar line of thought with
regard to the regression line slope estimator
for X models, Eq.\,(\ref{eq:aXu}), discloses
the last is biased towards infinity, in the
sense that the true slope is overestimated.

The regression line slope estimator for O
models (genuine orthogonal regression) and
R models (major-axis regression) lie between
their counterparts related to Y and X models,
according to Eq.\,(\ref{seq:dis}), which
implies bias corrections (e.g., Carroll et al.,
2006, Chap.\,3, \S 3.4.2).   Though there is
skepticism about an indiscriminate use of
generalized orthogonal regression estimators,
still it is accepted the method is viable
provided both instrumental and intrinsic
scatter are known (e.g., Carroll et al.,
2006, Chap.\,3, \S 3.4.2; Buonaccorsi,
2010, Chap.\,4, \S 4.5).

With regard to heteroscedastic data,
an inspection of Tables \ref{t:fog}-\ref{t:fo}
shows that for lower data dispersion
(RB09 sample) the values of regression
line slope and intercept estimators,
deduced for heteroscedastic (Table \ref{t:fog})
and homoscedastic (Table \ref{t:fo}) data,
are systematically smaller in the former
case with respect to the latter, but
remain consistent within $\mp\sigma$.
For larger dispersion data (Sal09 sample)
no systematic trend of the kind considered
appears, but the values of regression line
slope and intercept estimators are still
consistent within $\mp\sigma$ in different
alternatives.   It may be a general property
of the regression models considered in the
current attempt or, more realistically,
intrinsic to the samples selected for the
application performed in section \ref{apfm}.

The reliability ratio, Eq.\,(\ref{eq:rho20}),
has been calculated for all sample data
together with its counterpart for X models:
\begin{equation}
\label{eq:rho02}
\rho_{02}=\frac{S_{02}}{S_{02}+(n-1)\sigma_{yy}}~~;
\end{equation}
and the inequalities, $\rho_{20}>0.92$,
$\rho_{02}>0.91$, hold in any case except
$\rho_{02}>0.86$ for the Sal09 sample,
which implies poorly biased regression line
slope and intercept estimators for the samples
considered using Y and X models and, a fortiori,
using O and R models.

\section{Conclusion}
\label{conc}

\noindent\noindent

From the standpoint of a unified analytic formalism of
bivariate least squares linear regression, functional
models have been conceived as structural
models where the intrinsic scatter is
negligible (ideally null) with respect to
the instrumental scatter.

Within the framework of classical error models,
the dependent variable has been related to the
independent variable according to the well
known additive
model (e.g., Carroll et al., 2006, Chap.\,1, \S 1.2,
Chap.\,3, \S 3.2.1; Buonaccorsi, 2010, Chap.\,4,
\S 4.3).  Then the classical
approach pursued in earlier papers (Y66;
Y69) has been reviewed using
a new formalism in terms
of weighted deviation traces
which, for homoscedastic data, reduce to usual
quantities, leaving aside an
unessential (but dimensional) multiplicative
factor.

Regression line slope and intercept
estimators, and related variance estimators,
have been expressed in the general case of
correlated errors in $X$ and in $Y$ for
heteroscedastic data, and in the opposite limiting
situations of (i) uncorrelated errors in
$X$ and in $Y$, and (ii) completely
correlated errors in $X$ and in $Y$.
The special case of (C) generalized 
%
%errors in $X$
%equally weighted and equally correlated
%with respect to errors in $Y$ i.e.
%constant variance ratio and correlation
%coefficient ratio for all data points,
%
orthogonal regression
has been considered in detail together with
well known subcases, namely: (Y) errors
in $X$ negligible (ideally null) with
respect to errors in $Y$; (X) errors
in $Y$ negligible (ideally null) with
respect to errors in $X$; (O) genuine orthogonal
regression; (R) reduced major-axis regression.

In the limit of homoscedastic data, the
results determined for functional models
have been compared with their counterparts
related to extreme structural models i.e. the
instrumental scatter is negligible
(ideally null) with respect to the
intrinsic scatter (Ial90; FB92).
While regression line slope and intercept
estimators for functional and structural
models have necessarily been found to
coincide, the contrary has been shown
for related variance estimators
even if the residuals obey a Gaussian
distribution, with the exception of Y
models.

An example of astronomical
application has been considered, concerning
the [O/H]-[Fe/H] empirical relations
deduced from five samples related to
different stars and/or different methods
of oxygen abundance determination.
For selected samples and assigned
methods, different regression models
have been found to
yield consistent results within the
errors $(\mp\sigma)$ for both
heteroscedastic and homoscedastic data.
Conversely, it has been shown that
samples related to different
methods produce discrepant results,
due to the presence of (still undetected)
systematic errors, which implies no
definitive statement can be made at
present.   A comparison has also been made
between different expressions of
regression line slope and intercept
variance estimators, where fractional
discrepancies were found to be not
exceeding a few percent, which grows
up to about 20\% in presence of large
dispersion data.

An extension of the
results to structural models has
been left to a forthcoming paper.

\section{Note added in proof}

The author is indebted with G.J. Babu and
E.D. Feigelson for providing an earlier
version of the erratum of their quoted paper
(FB92) before publication (Feigelson and Babu,
2011).   The quotation FB92 throughout the
text has to be intended as including the
original paper and the erratum (Feigelson
and Babu, 1992, 2011).

\section{Addendum}

The slope variance estimator for generalized
orthogonal regression
with normal residuals, expressed in an earlier
attempt [FB92, Eq.\,(4) therein, hereafter quoted
as Eq.\,(FB4ev)] has been revised
[FB92, erratum 2011, first equation therein,
hereafter quoted as Eq.\,(FB4lv)].
Accordingly, the derivation of
Eqs.\,(\ref{eq:vac2b}), (\ref{eq:vacX1}),
(\ref{eq:vacc2}), which started from
Eq.\,(FB4ev), should be repeated
using Eq.\,(FB4lv).   It can be seen the
latter is closer to Eq.\,(\ref{eq:vacS})
than the former.   The difference is small
for the numerical application
shown in Sect.\,\ref{apfm}.

\section*{Acknowledgements}
The author is grateful to an anonymous
referee for enlightening comments which
made substantial improvement to an earlier
version of the current attempt.
Thanks are due to G.J. Babu, E.D. Feigelson,
M.A. Bershady, I. Lavagnini, S.J. Schmidt for fruitful
e-mail correspondance on their 
quoted papers (FB92; AB96; Lavagnini
and Magno, 2007; Sal09;
respectively), and to C. Toniolo,
C. Ghetti, S. Zoletto, V. Nascimbeni, for providing
references on a consistent number of (non astronomical)
statistical investigations quoted throughout the text.

\appendix
\section*{Appendix}

\section{Solutions of cubic equations}
\label{a:psec}

\noindent\noindent

Without loss of generality, a
cubic equation may be
cast under the standard form:
\begin{equation}
\label{eq:z3}
z^3+a_1z^2+a_2z+a_3=0~~;
\end{equation}
where the discriminant reads:
\begin{lefteqnarray}
\label{eq:D}
&& D=Q^3+R^2~~; \\
\label{eq:Q}
&& Q=\frac{3a_2-a_1^2}9~~; \\
\label{eq:R}
&& R=\frac{9a_1a_2-27a_3-2a_1^3}{54}~~;
\end{lefteqnarray}
and one (the other two being
complex coniugate) or three
real solutions exist, according
if $D>0$ or $D\le0$, respectively.
The related expressions are:
\begin{equation}
\label{eq:solrc}
z_1=(R+D^{1/2})^{1/3}+(R-D^{1/2})^{1/3}-\frac13a_1~~;
\end{equation}
for $D>0$ and:
\begin{lefteqnarray}
\label{eq:solr1}
&& z_1=2\sqrt{-Q}\cos\left(\frac\theta3+\frac{0\pi}3\right)-\frac13a_1~~; \\
\label{eq:solr2}
&& z_2=2\sqrt{-Q}\cos\left(\frac\theta3+\frac{2\pi}3\right)-\frac13a_1~~; \\
\label{eq:solr3}
&& z_3=2\sqrt{-Q}\cos\left(\frac\theta3+\frac{4\pi}3\right)-\frac13a_1~~; \\
\label{eq:theta}
&& \theta=\arctan\frac{\sqrt{-D}}R~~;
\end{lefteqnarray}
for $D\le0$, where $\theta$ has to be replaced
by $\theta\mp\pi$ as appropriate, when a
discontinuity of the $\arctan$ function occurs.
A null factor has been put into
Eq.\,(\ref{eq:solr1}) to save aesthetics.

\section{Determination of partial derivatives}
\label{a:pade}

\noindent\noindent

The explicit expressions of the regression
line slope and intercept variance estimators,
imply the calculation of the partial derivatives
appearing in Eqs.\,(\ref{eq:prph}) and
(\ref{eq:prps}), respectively.   With regard
to the regression line slope variance estimator,
$(\hat{\sigma}_{\hat{a}}^\prime)^2$, the slope
estimator, $\hat{a}$, has to be considered as
an independent variable in performing partial
derivatives, $\partial/\partial Z_i$, $Z=X,Y$.
Using Eqs.\,(\ref{eq:Wi}), (\ref{seq:wZ}) and
(\ref{eq:wgc}),
the partial derivatives of the deviation from
the weighted mean, with respect to the coordinates,
read:
\begin{lefteqnarray}
\label{eq:pdZ}
&& \frac{\partial}{\partial Z_i}(Z_\ell-\widetilde{Z})=\delta_{i\ell}-W_i^
\prime~~; \\
\label{eq:Wip}
&& W_i^\prime=\frac{W_i}{\widetilde{W}_{00}}~~;\qquad\widetilde{W}_{00}=\sum_
{i=1}^nW_i=n\overline{W}~~;
\end{lefteqnarray}
where $\delta_{i\ell}$ is the Kronecker
symbol and the prime does not mean first
derivation.

Using Eqs.\,(\ref{eq:wgc}) and (\ref{eq:pdZ}),
the partial derivatives of the weighted deviation
traces, with respect to the
coordinates, read:
\begin{leftsubeqnarray}
\slabel{eq:pdQa}
&& \frac{\partial\widetilde{Q}_{pq}}{\partial X_i}=pQ_i(X_i-\widetilde{X})^
{p-1}(Y_i-\widetilde{Y})^q-pW_i^\prime\widetilde{Q}_{p-1,q}~~;\quad p>0~~; \\
\slabel{eq:pdQb}
&& \frac{\partial\widetilde{Q}_{pq}}{\partial Y_i}=qQ_i(X_i-\widetilde{X})^p
(Y_i-\widetilde{Y})^{q-1}-qW_i^\prime\widetilde{Q}_{p,q-1}~~;\quad q>0~~; \\
\slabel{eq:pdQc}
&& \frac{\partial\widetilde{Q}_{0q}}{\partial X_i}=0~~;\qquad
\frac{\partial\widetilde{Q}_{p0}}{\partial Y_i}=0~~;
\label{seq:pdQ}
\end{leftsubeqnarray}
where the cases of interest are
$\widetilde{Q}_{pq}=
\widetilde{V}_{11}, \widetilde{U}_{20},
\widetilde{P}_{20}, \widetilde{V}_{02},
\widetilde{P}_{11}, \widetilde{U}_{02}$.

Using Eqs.\,(\ref{eq:Wi}), (\ref{eq:Vi}),
(\ref{eq:Ui}) and (\ref{eq:Pi}), the
partial derivatives of the weights, $W_i$,
with respect to the regression line slope
estimator, $\hat{a}$, read:
\begin{equation}
\label{eq:pdW}
\frac{\partial W_i}{\partial\hat{a}}=2(U_i-\hat{a}V_i)~~;
\end{equation}
and, in addition:
\begin{lefteqnarray}
\label{eq:pdV}
&& \frac{\partial V_i}{\partial\hat{a}}=4(U_i^\prime-\hat{a}V_i^\prime)~~; \\
\label{eq:Upi}
&& U_i^\prime=\frac{W_i^3r_i}{w_{x_i}^2\Omega_i}~~; \\
\label{eq:Vpi}
&& V_i^\prime=\frac{W_i^3}{w_{x_i}^2}~~; \\
\label{eq:pdU}
&& \frac{\partial U_i}{\partial\hat{a}}=4(U_i^\pprime-\hat{a}U_i^\prime)~~; \\
\label{eq:Usi}
&& U_i^\pprime=\frac{W_i^3r_i^2}{w_{x_i}^2\Omega_i^2}~~; \\
\label{eq:pdP}
&& \frac{\partial P_i}{\partial\hat{a}}=4(U_i^\ppprime-\hat{a}P_i^\prime)~~;\\
\label{eq:Uti}
&& U_i^\ppprime=\frac{W_i^3r_i}{w_{x_i}^2\Omega_i^3}~~; \\
\label{eq:Ppi}
&& P_i^\prime=\frac{W_i^3}{w_{x_i}^2\Omega_i^2}~~;
\end{lefteqnarray}
where
%Eq.\,(\ref{eq:Pi}) has also been considered and
the prime, the second, and the third, do not mean
first, second, and third derivation, respectively.

Using Eqs.\,(\ref{eq:wgc}) and (\ref{eq:pdV})-(\ref{eq:Ppi}),
the partial derivatives of related weighted deviation
traces, with respect to the regression
line slope estimator, read:
\begin{lefteqnarray}
\label{eq:pdVt}
&& \frac{\partial\widetilde{V}_{pq}}{\partial\hat{a}}=4(\widetilde{U}_{pq}^
\prime-\hat{a}\widetilde{V}_{pq}^\prime)~~; \\
\label{eq:pdUt}
&& \frac{\partial\widetilde{U}_{pq}}{\partial\hat{a}}=4(\widetilde{U}_{pq}^
\pprime-\hat{a}\widetilde{U}_{pq}^\prime)~~; \\
\label{eq:pdPt}
&& \frac{\partial\widetilde{P}_{pq}}{\partial\hat{a}}=4(\widetilde{U}_{pq}^
\ppprime-\hat{a}\widetilde{P}_{pq}^\prime)~~;
\end{lefteqnarray}
which, together with Eq.\,(\ref{eq:pdZ}),
complete the set of results needed for an
explicit expression of the partial
derivatives of the function,
$\phi(X_i,Y_i,\hat{a})$, defined by
Eq.\,(\ref{seq:phi}).

Using Eqs.\,(\ref{seq:phi}) and (\ref{seq:pdQ}),
the explicit expression of the partial derivatives
with respect to the coordinates, read:
\begin{leftsubeqnarray}
\slabel{eq:pdfia}
&& \frac{\partial\phi}{\partial X_i}=\left\{\left[V_i(Y_i-\widetilde{Y})-W_i^
\prime\widetilde{V}_{01}\right]-2\left[U_i(X_i-\widetilde{X})-W_i^\prime
\widetilde{U}_{10}\right]\right\}(\hat{a})^2 \nonumber \\
&& \phantom{\frac{\partial\phi}{\partial X_i}=}+
2\left[P_i(X_i-\widetilde{X})-W_i^\prime\widetilde{P}_{10}\right]\hat{a}-
\left[P_i(Y_i-\widetilde{Y})-W_i^\prime\widetilde{P}_{01}\right]~~; \\
\slabel{eq:pdfib}
&& \frac{\partial\phi}{\partial Y_i}=\left[V_i(X_i-\widetilde{X})-W_i^
\prime\widetilde{V}_{10}\right](\hat{a})^2-2\left[V_i(Y_i-\widetilde{Y})-W_i^
\prime\widetilde{V}_{01}\right]\hat{a} \nonumber \\
&& \phantom{\frac{\partial\phi}{\partial X_i}=}-
\left\{\left[P_i(X_i-\widetilde{X})-W_i^\prime\widetilde{P}_{10}\right]-2
\left[U_i(Y_i-\widetilde{Y})-W_i^\prime\widetilde{U}_{01}\right]\right\}~~;
\qquad
\label{seq:pdfi}
\end{leftsubeqnarray}
and using Eqs.\,(\ref{seq:phi}), (\ref{eq:pdVt}),
(\ref{eq:pdUt}), (\ref{eq:pdPt}), the explicit
expression of the partial derivative with respect
to the regression line slope estimator reads:
\begin{lefteqnarray}
\label{eq:pdfs}
&& \frac{\partial\phi}{\partial\hat{a}}=4(\widetilde{U}_{20}^
\prime-\widetilde{V}_{11}^\prime)(\hat{a})^3+4(\widetilde{U}_{11}^\prime-
\widetilde{U}_{20}^\pprime-\widetilde{P}_{20}^\prime+\widetilde{V}_{02}^\prime)
(\hat{a})^2 \nonumber \\
&& \phantom{\frac{\partial\widetilde{\phi}}{\partial\hat{a}}=}+
2(2\widetilde{U}_{20}^\ppprime-2\widetilde{U}_{02}^\prime+2\widetilde{P}_{11}
^\prime-2\widetilde{U}_{02}^\prime+\widetilde{V}_{11}-\widetilde{U}_{20})
\hat{a} \nonumber \\
&& \phantom{\frac{\partial\widetilde{\phi}}{\partial\hat{a}}=}+
(4\widetilde{U}_{02}^\pprime-4\widetilde{U}_{11}^\ppprime+
\widetilde{P}_{20}-\widetilde{V}_{02})~~;
\end{lefteqnarray}
finally, the substitution of
Eqs.\,(\ref{seq:pdfi}) and (\ref{eq:pdfs})
into (\ref{eq:prph}) yields an explicit
expression of the regression line slope
variance estimator, $(\hat{\sigma}_{\hat
{a}}^\prime)^2$, with no account taken
of the scatter of the data points, ${\sf
P}_i\equiv(X_i,Y_i)$, about the regression
line (Y69).

With regard to the regression line intercept
variance estimator, $(\hat{\sigma}_{\hat{b}}
^\prime)^2$, the regression line slope estimator,
$\hat{a}$, has to be considered as a function
of the coordinates, $\hat{a}=F_a(X_i,Y_i)$,
$1\le i\le n$, in performing partial derivatives,
$\partial/\partial Z_i$, $Z=X,Y$.   Accordingly,
$\partial W_i/\partial Z_i\ne0$.    Using
Eq.\,(\ref{seq:wZ}), the
%(\ref{eq:Wi}) and
partial derivatives of the weighted mean,
with respect to the coordinates,
after some algebra read:
\begin{equation}
\label{eq:dpZtb}
\frac{\partial\widetilde{Z}}{\partial Z_i}=\frac1{\widetilde{W}_{00}}\left[
W_i+\sum_{\ell=1}^n\frac{\partial W_\ell}{\partial Z_i}\left(Z_\ell-
\widetilde{Z}\right)\right]~~;\quad Z=X,Y~~;
\end{equation}
which, using Eqs.\,(\ref{eq:wgc}) and
(\ref{eq:pdW}), together with a theorem
on the partial derivatives of the function
of a function:
\begin{equation}
\label{eq:tff}
\frac{\partial\phi}{\partial Z_i}=\frac{\partial\phi}{\partial\hat{a}}\frac
{\partial\hat{a}}{\partial Z_i}~~;\qquad
\frac{\partial W_\ell}{\partial Z_i}=\frac{\partial W_\ell}{\partial\hat{a}}
\frac{\partial\hat{a}}{\partial Z_i}~~;\qquad1\le i\le n~~;
\end{equation}
after additional algebra takes the explicit form:
\begin{leftsubeqnarray}
\slabel{eq:dpZba}
&& \frac{\partial\widetilde{X}}{\partial X_i}=\frac1{\widetilde{W}_{00}}
\left[W_i+2\left(\widetilde{U}_{10}-\hat{a}\widetilde{V}_{10}\right)\frac
{\partial\phi/\partial X_i}{\partial\phi/\partial\hat{a}}\right]~~; \\
\slabel{eq:dpZbb}
&& \frac{\partial\widetilde{Y}}{\partial Y_i}=\frac1{\widetilde{W}_{00}}
\left[W_i+2\left(\widetilde{U}_{01}-\hat{a}\widetilde{V}_{01}\right)\frac
{\partial\phi/\partial Y_i}{\partial\phi/\partial\hat{a}}\right]~~;
\label{seq:dpZb}
\end{leftsubeqnarray}
and using Eqs.\,(\ref{eq:stb}) and (\ref{seq:dpZb}),
the explicit expressions of the partial derivatives
with respect to the coordinates read:
\begin{leftsubeqnarray}
\slabel{eq:dppa}
&& \frac{\partial\psi}{\partial X_i}=-\hat{a}\frac{\partial\widetilde{X}}
{\partial X_i}-\widetilde{X}\frac{\partial\hat{a}}{\partial X_i} \nonumber \\
&& \phantom{\frac{\partial\psi}{\partial X_i}}=
-\frac{\hat{a}W_i}{\widetilde{W}_{00}}-\left[\frac{2\hat{a}}{\widetilde{W}
_{00}}\left(\widetilde{U}_{10}-\hat{a}\widetilde{V}_{10}\right)+\widetilde{X}
\right]\frac{\partial\phi/\partial X_i}{\partial\phi/\partial\hat{a}}~~; \\
\slabel{eq:dppb}
&& \frac{\partial\psi}{\partial Y_i}=\frac{\partial\widetilde{Y}}
{\partial Y_i}-\widetilde{X}\frac{\partial\hat{a}}{\partial Y_i} \nonumber \\
&& \phantom{\frac{\partial\psi}{\partial Y_i}}=
\frac{W_i}{\widetilde{W}_{00}}+\left[\frac2{\widetilde{W}
_{00}}\left(\widetilde{U}_{01}-\hat{a}\widetilde{V}_{01}\right)-\widetilde{X}
\right]\frac{\partial\phi/\partial Y_i}{\partial\phi/\partial\hat{a}}~~;
\label{seq:dpp}
\end{leftsubeqnarray}
and the substitution of Eqs.\,(\ref{seq:dpp})
into (\ref{eq:prps}) yields an explicit
expression of the regression line intercept
variance estimator, $(\hat{\sigma}_{\hat{b}^
\prime})^2$, with no account taken of the
scatter of the data points, ${\sf P}_i\equiv
(X_i,Y_i)$, about the regression line (Y69).

In the special case of errors in $X$
negligible with respect to errors in
$Y$, analysed in subsection \ref{XnrY},
Eqs.\,(\ref{eq:Wip}), (\ref{eq:Upi}),
(\ref{eq:Vpi}), (\ref{eq:Usi}),
(\ref{eq:Uti}), (\ref{eq:Ppi}), via
Eqs.\,(\ref{eq:WiY})-(\ref{eq:PiY})
reduce to:
\begin{lefteqnarray}
\label{eq:WpiY}
&& W_i^\prime=\frac{w_{y_i}}{(\widetilde{w_{y}})_{00}}~~;\qquad
(\widetilde{w_{y}})_{00}=\sum_{i=1}^nw_{y_i}=n\overline{w_y}~~; \\
\label{eq:UpiY}
&& U_i^\prime=0~~; \\
\label{eq:VpiY}
&& V_i^\prime=0~~; \\
\label{eq:UsiY}
&& U_i^\pprime=0~~; \\
\label{eq:UtiY}
&& U_i^\ppprime=0~~; \\
\label{eq:PpiY}
&& P_i^\prime=0~~;
\end{lefteqnarray}
and the particularization of 
Eq.\,(\ref{eq:wgc}) to the
case under discussion via
Eqs.\,(\ref{eq:WiY})-(\ref{eq:PiY}),
selecting $Q=V, U, P;$ $p=0,1$;
$q=1,0$; and using Eq.\,(\ref{seq:wZ}),
yields:
\begin{lefteqnarray}
\label{eq:P10Y}
&& \widetilde{Q}_{10}=0~~; \\
\label{eq:P01Y}
&& \widetilde{Q}_{01}=0~~;
\end{lefteqnarray}
finally, the particularization of
Eqs.\,(\ref{eq:wgc}), (\ref{seq:pdfi}),
(\ref{eq:pdfs}),
(\ref{seq:dpp}), to the case under
discussion via Eqs.\,(\ref{eq:WiY})-(\ref{eq:PiY})
and (\ref{eq:WpiY})-(\ref{eq:P01Y}) produces:
\begin{leftsubeqnarray}
\slabel{eq:pdfiYa}
&& \frac{\partial\phi}{\partial X_i}=w_{y_i}[2\hat{a}_{\rm Y}(X_i-\widetilde
{X})-(Y_i-\widetilde{Y})]~~; \\
\slabel{eq:pdfiYb}
&& \frac{\partial\phi}{\partial Y_i}=-w_{y_i}(X_i-\widetilde{X})~~;
\label{seq:pdfiY}
\end{leftsubeqnarray}
\begin{equation}
\label{eq:pdfsY}
\frac{\partial\phi}{\partial\hat{a}_{\rm Y}}=(\widetilde{w_y})_{20}~~;
\end{equation}
\begin{leftsubeqnarray}
\slabel{eq:dppYa}
&& \frac{\partial\psi}{\partial X_i}=-w_{y_i}\left\{\frac{\hat{a}_{\rm Y}}
{(\widetilde{w_{y}})_{00}}+\frac{\widetilde{X}}{(\widetilde{w_{y}})_{20}}
\left[2\hat{a}_{\rm Y}(X_i-\widetilde{X})-(Y_i-\widetilde{Y})\right]\right\}
~~; \\
\slabel{eq:dppYb}
&& \frac{\partial\psi}{\partial Y_i}=w_{y_i}\left[\frac1{(\widetilde{w_{y}})_
{00}}+\frac{\widetilde{X}}{(\widetilde{w_{y}})_{20}}(X_i-\widetilde{X})\right]
~~;
\label{seq:dppY}
\end{leftsubeqnarray}
and the substitution of Eqs.\,(\ref{eq:pdfiYb}),
(\ref{eq:pdfsY}), and (\ref{eq:dppYb})
into (\ref{eq:prphY})
and (\ref{eq:prpsY}) yields an explicit
expression of the regression line slope
and intercept variance estimator, $(\hat{\sigma}_ 
{\hat{a}_{\rm Y}}^\prime)^2$ and $(\hat{\sigma}_
{\hat{b}_{\rm Y}}^\prime)^2$, respectively, with
no account taken of the scatter of the
data points, ${\sf P}_i\equiv(X_i,Y_i)$,
about the regression line, Eqs.\,(\ref{eq:prphY1})
and (\ref{eq:prpsY1}).

In the special case of errors in $Y$
negligible with respect to errors in
$X$, analysed in subsection \ref{YnrX},
Eqs.\,(\ref{eq:Wip}), (\ref{eq:Upi}),
(\ref{eq:Vpi}), (\ref{eq:Usi}),
(\ref{eq:Uti}), (\ref{eq:Ppi}), via
Eqs.\,(\ref{eq:WiX})-(\ref{eq:PiX})
reduce to:
\begin{lefteqnarray}
\label{eq:WpiX}
&& W_i^\prime=\frac{w_{x_i}}{(\widetilde{w_{x}})_{00}}~~;\qquad
(\widetilde{w_{x}})_{00}=\sum_{i=1}^nw_{x_i}=n\overline{w_x}~~; \\
\label{eq:UpiX}
&& U_i^\prime=0~~; \\
\label{eq:VpiX}
&& V_i^\prime=\frac{w_{x_i}}{(\hat{a}_{\rm X})^6}~~; \\
\label{eq:UsiX}
&& U_i^\pprime=0~~; \\
\label{eq:UtiX}
&& U_i^\ppprime=0~~; \\
\label{eq:PpiX}
&& P_i^\prime=0~~;
\end{lefteqnarray}
and the particularization of 
Eq.\,(\ref{eq:wgc}) to the
case under discussion via
Eqs.\,(\ref{eq:WiX})-(\ref{eq:PiX}),
selecting $Q=V, U, P;$ $p=0,1$;
$q=1,0$; and using Eq.\,(\ref{seq:wZ}),
yields again Eqs.\,(\ref{eq:P10Y}) and
(\ref{eq:P01Y}).
Finally, the particularization of
Eqs.\,(\ref{seq:pdfi}), (\ref{eq:pdfs}),
(\ref{seq:dpp}), to the case under
discussion via Eqs.\,(\ref{eq:wgc}),
(\ref{eq:WiX})-(\ref{eq:PiX}),
(\ref{eq:aX}), (\ref{eq:P10Y}),
(\ref{eq:P01Y}), and
(\ref{eq:WpiX})-(\ref{eq:PpiX}) produces:
\begin{leftsubeqnarray}
\slabel{eq:pdfiXa}
&& \frac{\partial\phi}{\partial X_i}=(\hat{a}_{\rm X})^{-2}w_{x_i}
(Y_i-\widetilde{Y})~~; \\
\slabel{eq:pdfiXb}
&& \frac{\partial\phi}{\partial Y_i}=(\hat{a}_{\rm X})^{-3}w_{x_i}
[\hat{a}_{\rm X}(X_i-\widetilde{X})-2(Y_i-\widetilde{Y})]~~;
\label{seq:pdfiX}
\end{leftsubeqnarray}
\begin{equation}
\label{eq:pdfsX}
\frac{\partial\phi}{\partial\hat{a}_{\rm X}}=(\hat{a}_{\rm X})^{-4}
(\widetilde{w_x})_{02}~~;
\end{equation}
\begin{leftsubeqnarray}
\slabel{eq:dppXa}
&& \frac{\partial\psi}{\partial X_i}=-\hat{a}_{\rm X}w_{x_i}\left[\frac1
{(\widetilde{w_{x}})_{00}}+\frac{\hat{a}_{\rm X}\widetilde{X}}
{(\widetilde{w_{x}})_{02}}(Y_i-\widetilde{Y})\right]~~; \\
\slabel{eq:dppXb}
&& \frac{\partial\psi}{\partial Y_i}=w_{x_i}\left\{\frac1{(\widetilde{w_{x}})_
{00}}+\frac{\hat{a}_{\rm X}\widetilde{X}}{(\widetilde{w_{x}})_{02}}\left[2
(Y_i-\widetilde{Y})-\hat{a}_{\rm X}(X_i-\widetilde{X})\right]\right\}~~;
\label{seq:dppX}
\end{leftsubeqnarray}
and the substitution of Eqs.\,(\ref{eq:pdfiXa}),
(\ref{eq:pdfsX}), and (\ref{eq:dppXa})
into (\ref{eq:prphX})
and (\ref{eq:prpsX}) yields an explicit
expression of the regression line slope
and intercept variance estimator, $(\hat{\sigma}_ 
{\hat{a}_{\rm X}}^\prime)^2$ and $(\hat{\sigma}_
{\hat{b}_{\rm X}}^\prime)^2$, respectively, with
no account taken of the scatter of the
data points, ${\sf P}_i\equiv(X_i,Y_i)$,
about the regression line, Eqs.\,(\ref{eq:prphX1})
and (\ref{eq:prpsX1}).

In the special case of generalized
orthogonal regression,
%
%errors in $X$
%equally weighted and equally correlated
%with respect to errors in $Y$,
%
analysed in subsection \ref{YXc2},
Eqs.\,(\ref{eq:Wip}), (\ref{eq:Upi}),
(\ref{eq:Vpi}), (\ref{eq:Usi}),
(\ref{eq:Uti}), (\ref{eq:Ppi}), via
Eqs.\,(\ref{eq:Wic2})-(\ref{eq:Pic2})
reduce to:
\begin{lefteqnarray}
\label{eq:Wpic}
&& W_i^\prime=\frac{w_{x_i}}{(\widetilde{w_{x}})_{00}}~~;\qquad
(\widetilde{w_{x}})_{00}=\sum_{i=1}^nw_{x_i}=n\overline{w_x}~~; \\
\label{eq:Upic}
&& U_i^\prime=\frac{rcw_{x_i}}{(a^2+c^2-2rca)^3}~~; \\
\label{eq:Vpic}
&& V_i^\prime=\frac{w_{x_i}}{(a^2+c^2-2rca)^3}~~; \\
\label{eq:Usic}
&& U_i^\pprime=\frac{r^2c^2w_{x_i}}{(a^2+c^2-2rca)^3}~~; \\
\label{eq:Utic}
&& U_i^\ppprime=\frac{rc^3w_{x_i}}{(a^2+c^2-2rca)^3}~~; \\
\label{eq:Ppic}
&& P_i^\prime=\frac{c^2w_{x_i}}{(a^2+c^2-2rca)^3}~~;
\end{lefteqnarray}
and the particularization of 
Eq.\,(\ref{eq:wgc}) to the
case under discussion via
Eqs.\,(\ref{eq:Wic2})-(\ref{eq:Pic2}),
selecting $Q=V, U, P;$ $p=0,1$;
$q=1,0$; and using Eq.\,(\ref{seq:wZ}),
yields again Eqs.\,(\ref{eq:P10Y}) and
(\ref{eq:P01Y}).
Finally, the particularization of
Eqs.\,(\ref{seq:pdfi}), (\ref{eq:pdfs}),
(\ref{seq:dpp}), to the case under
discussion via Eqs.\,(\ref{eq:Wic2})-(\ref{eq:Pic2}),
(\ref{eq:P10Y}), (\ref{eq:P01Y}), 
and (\ref{eq:Wpic})-(\ref{eq:Ppic}) produces:
\begin{leftsubeqnarray}
\slabel{eq:pdfica}
&& \frac{\partial\phi}{\partial X_i}=\frac
{w_{x_i}[(\hat{a}_{\rm C})^2-c^2](Y_i-\widetilde{Y})+2
w_{x_i}c\hat{a}_{\rm C}(c-r\hat{a}_{\rm C})(X_i-\widetilde{X})}
{[(\hat{a}_{\rm C})^2+c^2-2rc\hat{a}_{\rm C}]^2}~~; \\
\slabel{eq:pdficb}
&& \frac{\partial\phi}{\partial Y_i}=\frac
{w_{x_i}[(\hat{a}_{\rm C})^2-c^2](X_i-\widetilde{X})-2
w_{x_i}(\hat{a}_{\rm C}-rc)(Y_i-\widetilde{Y})}
{[(\hat{a}_{\rm C})^2+c^2-2rc\hat{a}_{\rm C}]^2}~~;
\label{seq:pdfic}
\end{leftsubeqnarray}
\begin{lefteqnarray}
\label{eq:pdfsc}
&& \frac{\partial\phi}{\partial\hat{a}_{\rm C}}=
[(\hat{a}_{\rm C})^2+c^2-2rc\hat{a}_{\rm C}]^{-3}
\{2[rc(\widetilde{w_{x}})_{20}-(\widetilde{w_{x}})_{11}](\hat{a}_{\rm C})^3
\nonumber \\
&& \phantom{\frac{\partial\phi}{\partial\hat{a}_{\rm C}}=
[(\hat{a}_{\rm C})^2+c^2-2rc\hat{a}_{\rm C}]^{-3}\{}+
[(\widetilde{w_{x}})_{02}-c^2(\widetilde{w_{x}})_{20}][3(\hat{a}_{\rm C})^2-
c^2] \nonumber \\
&& \phantom{\frac{\partial\phi}{\partial\hat{a}_{\rm C}}=
[(\hat{a}_{\rm C})^2+c^2-2rc\hat{a}_{\rm C}]^{-3}\{}+
2[c^2(\widetilde{w_{x}})_{11}-rc(\widetilde{w_{x}})_{02}][3\hat{a}_
{\rm C}-2rc]\}~~;\qquad
\end{lefteqnarray}
\begin{leftsubeqnarray}
\slabel{eq:dppca}
&& \frac{\partial\psi}{\partial X_i}=-\frac{\hat{a}_{\rm C}w_{x_i}}
{(\widetilde{w_{x}})_{00}}-\frac{\widetilde{X}w_{x_i}}{[(\hat{a}_{\rm C})^2+
c^2-2rc\hat{a}_{\rm C}]^2}\left(\frac{\partial\phi}{\partial\hat{a}_{\rm C}}
\right)^{-1}
\nonumber \\
&& \phantom{\frac{\partial\psi}{\partial X_i}=}\times
\{[(\hat{a}_{\rm C})^2-c^2](Y_i-\widetilde{Y})-2c\hat{a}_{\rm C}
[r\hat{a}_{\rm C}-c](X_i-\widetilde{X})\}~~;\qquad \\
\slabel{eq:dppcb}
&& \frac{\partial\psi}{\partial Y_i}=\frac{w_{x_i}}{(\widetilde{w_{x}})_{00}}
-\frac{\widetilde{X}w_{x_i}}{[(\hat{a}_{\rm C})^2+c^2-2rc\hat{a}_{\rm C}]^2}
\left(\frac{\partial\phi}{\partial\hat{a}_{\rm C}}\right)^{-1} \nonumber \\
&& \phantom{\frac{\partial\psi}{\partial Y_i}=}\times
\{[(\hat{a}_{\rm C})^2-c^2](X_i-\widetilde{X})-2
[\hat{a}_{\rm C}-rc](Y_i-\widetilde{Y})\}~~;
\label{seq:dppc}
\end{leftsubeqnarray}
and the substitution of Eqs.\,(\ref{seq:pdfic}),
(\ref{eq:pdfsc}) and
(\ref{seq:dppc}) into (\ref{eq:prphc})
and (\ref{eq:prpsc}) yields an explicit
expression of the regression line slope
and intercept variance estimator, $(\hat{\sigma}_ 
{\hat{a}_{\rm C}}^\prime)^2$ and $(\hat{\sigma}_
{\hat{b}_{\rm C}}^\prime)^2$, respectively, with
no account taken of the scatter of the
data points, ${\sf P}_i\equiv(X_i,Y_i)$,
about the regression line.
%, Eqs.\,(\ref{eq:prphc2}) and (\ref{eq:prpsc2}).

In the limit of uncorrelated errors, $r\to0$,
Eqs.\,(\ref{seq:pdfic}), (\ref{eq:pdfsc}) and
(\ref{seq:dppc}) reduce to:
\begin{leftsubeqnarray}
\slabel{eq:pdfcSa}
&& \frac{\partial\phi}{\partial X_i}=\frac
{w_{x_i}}{[(\hat{a}_{\rm C})^2+c^2]^2}
\{[(\hat{a}_{\rm C})^2-c^2](Y_i-\widetilde{Y})+2
\hat{a}_{\rm C}c^2(X_i-\widetilde{X})\}~~; \\
\slabel{eq:pdfcSb}
&& \frac{\partial\phi}{\partial Y_i}=\frac
{w_{x_i}}{[(\hat{a}_{\rm C})^2+c^2]^2}
\{[(\hat{a}_{\rm C})^2-c^2](X_i-\widetilde{X})-2
\hat{a}_{\rm C}(Y_i-\widetilde{Y})\}~~;
\label{seq:pdfcS}
\end{leftsubeqnarray}
\begin{lefteqnarray}
\label{eq:pdacS}
&& \frac{\partial\phi}{\partial\hat{a}_{\rm C}}=\frac
{[3(\hat{a}_{\rm C})^2-c^2]
[(\widetilde{w_{x}})_{02}-c^2(\widetilde{w_{x}})_{20}]-2\hat{a}_{\rm C}
[(\hat{a}_{\rm C})^2-3c^2](\widetilde{w_{x}})_{11}}
{[(\hat{a}_{\rm C})^2+c^2]^3}~~;\qquad
\end{lefteqnarray}
\begin{leftsubeqnarray}
\slabel{eq:dpcSa}
&& \frac{\partial\psi}{\partial X_i}=-\frac{\hat{a}_{\rm C}w_{x_i}}
{(\widetilde{w_{x}})_{00}}-\widetilde{X}[(\hat{a}_{\rm C})^2+c^2] \nonumber \\
&& \phantom{\frac{\partial\psi}{\partial X_i}=}\times
\frac{[(\hat{a}_{\rm C})^2-c^2]w_{x_i}(Y_i-\widetilde{Y})+2\hat{a}_{\rm C}c^2
w_{x_i}(X_i-\widetilde{X})}{[3(\hat{a}_{\rm C})^2-c^2]
[(\widetilde{w_{x}})_{02}-c^2(\widetilde{w_{x}})_{20}]-2\hat{a}_{\rm C}
[(\hat{a}_{\rm C})^2-3c^2](\widetilde{w_{x}})_{11}};\qquad \\
\slabel{eq:dpcSb}
&& \frac{\partial\psi}{\partial Y_i}=\frac{w_{x_i}}{(\widetilde{w_{x}})_{00}}
-\widetilde{X}[(\hat{a}_{\rm C})^2+c^2] \nonumber \\
&& \phantom{\frac{\partial\psi}{\partial X_i}=}\times
\frac{[(\hat{a}_{\rm C})^2-c^2]w_{x_i}(X_i-\widetilde{X})-2\hat{a}_{\rm C}
w_{x_i}(Y_i-\widetilde{Y})}{[3(\hat{a}_{\rm C})^2-c^2]
[(\widetilde{w_{x}})_{02}-c^2(\widetilde{w_{x}})_{20}]-2\hat{a}_{\rm C}
[(\hat{a}_{\rm C})^2-3c^2](\widetilde{w_{x}})_{11}};\qquad
\label{seq:dpcS}
\end{leftsubeqnarray}
on the other hand,  Eq.\,(\ref{eq:pqc0})
may be cast under the equivalent form:
\begin{lefteqnarray}
\label{eq:pqc01}
&& [(\hat{a}_{\rm C})^2-c^2](\widetilde{w_{x}})_{11}=\hat{a}_{\rm C}
[(\widetilde{w_{x}})_{02}-c^2(\widetilde{w_{x}})_{20}]~~;
\end{lefteqnarray}
and the substitution of Eq.\,(\ref{eq:pqc01}) into
(\ref{eq:pdacS}) and (\ref{seq:dpcS}) yields:
\begin{lefteqnarray}
\label{eq:pdacS2}
&& \frac{\partial\phi}{\partial\hat{a}_{\rm C}}=\frac{(\widetilde{w_{x}})_
{11}}{\hat{a}_{\rm C}[(\hat{a}_{\rm C})^2+c^2]}~~;
\end{lefteqnarray}
\begin{leftsubeqnarray}
\slabel{eq:dpcS2a}
&& \frac{\partial\psi}{\partial X_i}=-\frac{\hat{a}_{\rm C}w_{x_i}}
{(\widetilde{w_{x}})_{00}}-\frac{\widetilde{X}\hat{a}_{\rm C}}
{(\hat{a}_{\rm C})^2+c^2} \nonumber \\
&& \phantom{\frac{\partial\psi}{\partial X_i}=}\times
\frac{[(\hat{a}_{\rm C})^2-c^2]w_{x_i}(Y_i-\widetilde{Y})+2\hat{a}_{\rm C}c^2
w_{x_i}(X_i-\widetilde{X})}{(\widetilde{w_{x}})_{11}}~~; \\
\slabel{eq:dpcS2b}
&& \frac{\partial\psi}{\partial Y_i}=\frac{w_{x_i}}
{(\widetilde{w_{x}})_{00}}-\frac{\widetilde{X}\hat{a}_{\rm C}}
{(\hat{a}_{\rm C})^2+c^2} \nonumber \\
&& \phantom{\frac{\partial\psi}{\partial X_i}=}\times
\frac{[(\hat{a}_{\rm C})^2-c^2]w_{x_i}(X_i-\widetilde{X})-2\hat{a}_{\rm C}
w_{x_i}(Y_i-\widetilde{Y})}{(\widetilde{w_{x}})_{11}}~~;
\label{seq:dpcS2}
\end{leftsubeqnarray}
finally, the substitution of Eqs.\,(\ref{seq:pdfcS}),
(\ref{eq:pdacS2}) and (\ref{seq:dpcS2}) into
(\ref{eq:prphc}) and (\ref{eq:prpsc})
particularized to uncorrelated errors,
$r=0$, yields an explicit
expression of the regression line slope
and intercept variance estimator, $(\hat{\sigma}_ 
{\hat{a}_{\rm C}}^\prime)^2$ and $(\hat{\sigma}_
{\hat{b}_{\rm C}}^\prime)^2$, respectively, with
no account taken of the scatter of the
data points, ${\sf P}_i\equiv(X_i,Y_i)$,
about the regression line,
Eqs.\,(\ref{eq:prphc2}) and (\ref{eq:prpsc2}).

\section{Regression line slope and intercept
variance estimators}
\label{a:resu}

\noindent\noindent

According to Eqs.\,(\ref{eq:vaap1}) and
(\ref{eq:vabp1}), an explicit expression
of the squared residual trace implies
an explicit expression of the regression
line slope and intercept variance estimators,
$(\hat{\sigma}_{\hat{a}})^2$ and
$(\hat{\sigma}_{\hat{b}})^2$, respectively (Y69).
The former may be obtained by substitution of
Eqs.\,(\ref{seq:XY}) and (\ref{eq:lai}) into
(\ref{eq:ssR}).

After some algebra, the result is:
\begin{lefteqnarray}
\label{eq:sR21}
&& T_{\widetilde{R}}=\sum_{i=1}^n\frac1{1-r_i^2}\frac{W_i^2
(\hat{a}X_i+\hat{b}-Y_i)^2}{w_{x_i}}\frac1{\Omega_i^2} \nonumber \\
&& \phantom{F=}\times
\left[(\hat{a}\Omega_i-r_i)^2+(\hat{a}\Omega_ir_i-1)^2-2r_i(\hat{a}\Omega_i-
r_i)(\hat{a}\Omega_ir_i-1)\right]~~;
\end{lefteqnarray}
where, in addition, the following relation holds:
\begin{lefteqnarray}
\label{eq:ide}
&& (\hat{a}\Omega_i-r_i)^2+(\hat{a}\Omega_ir_i-1)^2-2r_i(\hat{a}\Omega_i-
r_i)(\hat{a}\Omega_ir_i-1) \nonumber \\
&& =[(\hat{a})^2\Omega_i^2-2\hat{a}\Omega_ir_i+1](1-r_i^2)=w_{x_i}\Omega_i^2
W_i^{-1}(1-r_i^2)~~;
\end{lefteqnarray}
where the former equality makes an identity
and the latter is owing to Eq.\,(\ref{eq:Wi}).

The substitution of Eq.\,(\ref{eq:ide})
into (\ref{eq:sR21}) yields:
\begin{lefteqnarray}
\label{eq:sR22}
&& T_{\widetilde{R}}=\sum_{i=1}^nW_i(\hat{a}X_i+\hat{b}-Y_i)^2~~;
\end{lefteqnarray}
and the combination of Eqs.\,(\ref{eq:DXY}),
(\ref{eq:wgc}) and (\ref{eq:sR22}) produces:
\begin{lefteqnarray}
\label{eq:sR23}
&& T_{\widetilde{R}}=(\hat{a})^2\widetilde{W}_{20}+
\widetilde{W}_{02}-2\hat{a}\widetilde{W}_{11}~~;
\end{lefteqnarray}
which is the result of interest.

The substitution Eqs.\,(\ref{eq:prph}),
(\ref{seq:pdfi}), (\ref{eq:pdfs}) and
(\ref{eq:sR23}) into (\ref{eq:vara})
yields an explicit expression of the
regression line slope variance estimator,
$(\hat{\sigma}_{\hat{a}})^2$, with due
account taken of the scatter of the data
points, ${\sf P}_i\equiv(X_i,Y_i)$, about
the regression line (Y69).

The substitution of Eqs.\,(\ref{eq:prps}),
(\ref{seq:dpp}), and
(\ref{eq:sR23}) into (\ref{eq:varb})
yields an explicit expression of the
regression line intercept variance estimator,
$(\hat{\sigma}_{\hat{b}})^2$, with due
account taken of the scatter of the data
points, ${\sf P}_i\equiv(X_i,Y_i)$, about
the regression line (Y69).

If Eqs.\,(\ref{eq:vara}) and (\ref{eq:varb})
are approximated by Eqs.\,(\ref{eq:vaap1})
and (\ref{eq:vabp1}), respectively, the
result is expressed by Eqs.\,(\ref{eq:vaap2})
and (\ref{eq:vabp2}), respectively.

In the special case of unweighted residuals,
$W_i=W$, $\widetilde{W}_{pq}=WS_{pq}$,
Eq.\,(\ref{eq:sR23}) reduces to:
\begin{lefteqnarray}
\label{eq:FS}
&& T_{\widetilde{R}}=W\left[(\hat{a})^2S_{20}+
S_{02}-2\hat{a}S_{11}\right]~~;
\end{lefteqnarray}
in terms of deviation traces.

\section{Equivalence between earlier and current
formulation}
\label{a:eqFB}

\noindent\noindent

In the special case of homoscedastic functional
or extreme structural models, where the ratio of instrumental
or intrinsic dispersions in the two variables,
$Y$ and $X$, maintains constant, $(\sigma_{yy})_i
/(\sigma_{xx})_i=c^2$, $1\le i\le n$, and the
errors in $Y$ and in $X$ are uncorrelated, $r_i=0$,
$1\le i\le n$, the regression
line slope estimator and the related dispersion
estimator read:
\begin{lefteqnarray}
\label{eq:ac2}
&& \hat{a}=\frac{S_{02}-c^2S_{20}}
{2S_{11}}\left\{1\mp\left[1+c^2\left(\frac{S_{02}-
c^2S_{20}}{2S_{11}}\right)^{-2}\right]^{1/2}\right\}~~; \\
\label{eq:vac2a}
&& (\hat{\sigma}_{\hat{a}})^2=\frac{(\hat{a})^2}{(S_{11})^2}\left[
\left(\frac{S_{11}}{\hat{a}}+\frac{S_{02}-\hat{a}
S_{11}}{c^2}\right)R-\frac{(\hat{a})^2}{n-1}\left(\frac{S_{02}-\hat{a}
S_{11}}{c^2}\right)^2\right]~~;\qquad \\
\label{eq:vac2b}
&& R=\frac1{n-2}\sum_{i=1}^n\left[(Y_i-\overline{Y})-\hat{a}(X_i-\overline{X})
\right]^2~~;
\end{lefteqnarray}
under the further assumptions of unweighted
normal
residuals and large samples (FB92). With
regard to Eq.\,(\ref{eq:ac2}), the plus
instead of the double sign appears in
the parent paper (FB92) but the latter
is mentioned in an earlier paper (Ial90).
In this case, the (physically meaningless)
parasite solution must be disregarded.

The substitution of Eqs.\,(\ref{eq:Spq})
into (\ref{eq:vac2b}) yields after some
algebra:
\begin{lefteqnarray}
\label{eq:Rpr}
&& R=\frac1{n-2}\left[S_{02}+(\hat{a})^2S_{20}-2
\hat{a}S_{11}\right]~~; 
\end{lefteqnarray}
in terms of deviation traces.

The substitution of Eq.\,(\ref{eq:FS})
into (\ref{eq:Rpr}) yields:
\begin{equation}
\label{eq:RFW}
R=\frac1{n-2}\frac{T_{\widetilde{R}}}W~~;
\end{equation}
which shows the relation between different
sums of squared residuals, $R$ and $T_{\widetilde{R}}$, via
the dimensional constant, $1/[(n-2)W]$.  A
dimensionless counterpart to Eq.\,(\ref{eq:RFW})
reads:
\begin{equation}
\label{eq:RFa}
\frac R{\hat{a}S_{11}}=\frac1{n-2}\frac{T_{\widetilde{R}}}{\hat{a}
\widetilde{W}_{11}}~~;
\end{equation}
where $\widetilde{W}_{11}=WS_{11}$
in the case under consideration.

In the limit of errors in $X$ negligible with
respect to errors in $Y$, $c^2\to+\infty$,
the square root on the right-hand
side of Eq.\,(\ref{eq:ac2}) may be developed
in binomial series with the terms of higher
order neglected.   The result is:
\begin{equation}
\label{eq:acY}
\hat{a}_{\rm Y}=\frac{S_{11}}{S_{20}}~~;
\end{equation}
where the plus in the double sign on the
right-hand side of Eq.\,(\ref{eq:ac2})
corresponds to a (physically meaningless)
infinite value and for this reason has
been disregarded, while the minus has
been considered.

The combination of Eqs.\,(\ref{eq:vac2a})
and (\ref{eq:Rpr}), after neglecting the
terms of higher order with respect to
unity, yields:
\begin{equation}
\label{eq:vacY1}
(\hat{\sigma}_{\hat{a}_{\rm Y}})^2=\frac1{n-2}\frac{\hat{a}_{\rm Y}}
{S_{11}}\left[S_{02}+(\hat{a}_{\rm Y})^2
S_{20}-2\hat{a}_{\rm Y}S_{11}\right]~~;
\end{equation}
and the substitution of
Eq.\,(\ref{eq:acY}) into
(\ref{eq:vacY1}) produces:
\begin{equation}
\label{eq:vacY2}
(\hat{\sigma}_{\hat{a}_{\rm Y}})^2=\frac{(\hat{a}_{\rm Y})^2}{n-2}\frac{D_
{\rm S}}{(S_{11})^2}~~;
\end{equation}
in terms of the deviation determinant,
expressed by Eq.\,(\ref{eq:DS}).   It
can be seen that Eqs.\,(\ref{eq:acY})
and (\ref{eq:vacY2}) coincide with
(\ref{eq:aYu}) and (\ref{eq:vaYu2}),
respectively, which implies the
equivalence between earlier (FB92)
and current (this paper) formulation,
in the special case under discussion
$(c^2\to+\infty)$.

In the limit of errors in $Y$ negligible with
respect to errors in $X$, $c^2\to0$,
the term in square brackets on the right-hand
side of Eq.\,(\ref{eq:ac2}) tends to unity.
The result is:
\begin{equation}
\label{eq:acX}
\hat{a}_{\rm X}=\frac{S_{02}}{S_{11}}~~;
\end{equation}
where the minus in the double sign on the
right-hand side of Eq.\,(\ref{eq:ac2})
corresponds to a (physically meaningless)
null value and for this reason has
been disregarded, while the plus has
been considered.

With this restriction,  the
square root on the right-hand
side of Eq.\,(\ref{eq:ac2}) may be
developed in binomial series with
the terms of higher order neglected.
After some algebra, the result is:
\begin{equation}
\label{eq:DSX}
\frac{S_{02}-\hat{a}_{\rm X}S_{11}}{c^2}=
\frac{D_{\rm S}}{S_{02}}~~;
\end{equation}
in terms of the deviation determinant,
Eq.\,(\ref{eq:DS}).

The substitution of Eq.\,(\ref{eq:DSX})
into (\ref{eq:vac2a}) yields:
\begin{equation}
\label{eq:vacX1}
(\hat{\sigma}_{\hat{a}_{\rm X}})^2=\frac{(\hat{a}_{\rm X})^2}
{(S_{11})^2}\left[\frac{S_{11}R}{\hat{a}_
{\rm X}}+\frac{D_{\rm S}R}{S_{02}}-\frac{(\hat{a}_{\rm X})
^2}{n-1}\frac{(D_{\rm S})^2}{(S_{02})^2}\right]~~;
\end{equation}
and the substitution of
Eqs.\,(\ref{eq:Rpr}) and (\ref{eq:acX})
into (\ref{eq:vacX1}) produces:
\begin{equation}
\label{eq:vacX2}
(\hat{\sigma}_{\hat{a}_{\rm X}})^2=\frac{(\hat{a}_{\rm X})^2}{n-2}\frac
{D_{\rm S}}{(S_{11})^2}\left[1+\frac1{n-1}\frac{D_{\rm S}}
{(S_{11})^2}\right]~~;
\end{equation}
in terms of the deviation determinant,
expressed by Eq.\,(\ref{eq:DS}).   It
can be seen that Eq.\,(\ref{eq:acX})
coincides with Eq.\,(\ref{eq:aXu}), 
which implies the equivalence between
earlier (FB92)
and current (this paper) formulation,
in the special case under discussion
$(c^2\to0)$, concerning the regression
line slope estimator, $\hat{a}_{\rm X}$.
The contrary holds for the regression
line slope variance estimator,
$(\hat{\sigma}_{\hat{a}_{\rm X}})^2$,
where Eq.\,(\ref{eq:vacX2}) overstimates
Eq.\,(\ref{eq:vaXu2}), the difference
decreasing for increasing $n$ and/or
decreasing $D_{\rm S}/(S_{11})^2$,
and tends to be null in the limit,
$n\to+\infty$ and/or $D_{\rm S}/
(S_{11})^2\to0$.

Turning to the general case, it can be
seen that Eq.\,(\ref{eq:acu}) for
unweighted residuals, $w_{x_i}=w_x$,
$(\widetilde{w_x})_{pq}=w_xS_{pq}$,
reduces to Eq.\,(\ref{eq:ac2}),
which implies the validity of the
pseudo quadratic, Eq.\,(\ref{eq:pqc0}).
The substitution of Eqs.\,(\ref{eq:c2})
and Eq.\,(\ref{eq:Rpr})
into (\ref{eq:vac2a}) yields
after a lot of algebra:
\begin{equation}
\label{eq:vacc2}
(\hat{\sigma}_{\hat{a}})^2=\frac{(\hat{a})^2}{n-2}\left[\frac
{D_{\rm S}}{(S_{11})^2}+\frac1{n-1}\left(\frac{\hat{a}S_{20}}
{S_{11}}-1\right)^2\right]~~;
\end{equation}
where in the limit, $c\to+\infty$,
$c\to0$, Eq.\,(\ref{eq:vacc2})
reduce to (\ref{eq:vacY2}) and
(\ref{eq:vacX2}), respectively.
A comparison between 
Eqs.\,(\ref{eq:vacc2}) and
(\ref{eq:vacS}) shows that
earlier (FB92) and current
(this paper) formulation are
different due to the second
term within square brackets.

In the special case, $c^2=
S_{02}/S_{20}$, which implies
$\hat{a}=(S_{02}/S_{20})^{1/2}$,
a similar conclusion is attained
by comparison between
Eqs.\,(\ref{eq:vacc2}) and
(\ref{eq:vamS}).

\end{document}